\documentclass[12pt,letterpaper]{iopart}
\usepackage[utf8]{inputenc}
\usepackage[T1]{fontenc}
\usepackage[left=1.0in,right=1.0in,top=1in,bottom=1in,heightrounded]{geometry}
\usepackage{iopams}
\expandafter\let\csname equation*\endcsname=\relax
\expandafter\let\csname endequation*\endcsname=\relax 
\usepackage{amsmath}
\usepackage{amsfonts,amssymb}
\usepackage{mathabx}
\usepackage{lmodern}
\usepackage{siunitx}
\DeclareSIUnit\gauss{G}
\usepackage{physics}
\usepackage[hang]{footmisc}
\usepackage{enumitem}
\setlist[enumerate,1]{label = \arabic*.,ref = \arabic*}
\usepackage{adjustbox}
\usepackage[all]{nowidow}
\usepackage{rotating}
\usepackage{pdflscape}
\usepackage{multirow}
\usepackage{booktabs}
\usepackage[font=small,bf,labelsep=period]{caption}
\usepackage{microtype}
\usepackage{graphicx,color}
\usepackage{hyperref}
\usepackage{doi}
\usepackage{CJKutf8}
\usepackage{etoolbox}

\makeatletter
\long\def\@makefntext#1{\parindent 1em\noindent 
 \makebox[1em][l]{\footnotesize\rm$\m@th{\textsuperscript{\arabic{footnote}}}$}%
 \footnotesize\rm #1}
\def\@makefnmark{\hbox{\textsuperscript{${\arabic{footnote}}\m@th$}}}
\def\@thefnmark{\arabic{footnote}}
\makeatother

\makeatletter
\newrobustcmd{\fixappendix}{%
  \patchcmd{\l@section}{1.5em}{7em}{}{}%
  \patchcmd{\l@subsection}{2.3em}{7em}{}{}%
}
\makeatother

\usepackage[backend=biber, style=numeric, citestyle=numeric-comp, date=year,labeldate=year, sorting=none, giveninits=true, maxnames=10]{biblatex}
\AtBeginBibliography{\small}
\AtEveryBibitem{\clearfield{number}\clearfield{month}\clearfield{url}}
\AtEveryCitekey{\clearfield{number}\clearfield{month}\clearfield{url}}
\renewbibmacro{in:}{%
  \ifentrytype{article}{}{\printtext{\bibstring{in}\intitlepunct}}}
\DeclareFieldFormat{pages}{#1}
\DeclareFieldFormat[article]{volume}{\mkbibbold{#1}}

\renewbibmacro*{issue+date}{%
  \printtext[parens]{%
    \usebibmacro{date}}%
  \newunit}

\renewbibmacro*{journal+issuetitle}{%
  \usebibmacro{journal}%
  \setunit*{\addspace}%
  \iffieldundef{series}
    {}
    {\newunit
     \printfield{series}%
     \setunit{\addspace}}%
  \usebibmacro{volume+number+eid}%
  \setunit{\bibpagespunct}%
  \printfield{pages}%
  \setunit{\addspace}%
  \usebibmacro{issue+date}%
  \newunit}
\renewbibmacro*{note+pages}{%
  \printfield{note}%
  \newunit}

\begin{document}

\title[Matter-wave Atomic Gradiometer Interferometric Sensor (MAGIS-100)]{\vspace{-3em}\\Matter-wave Atomic Gradiometer Interferometric Sensor (MAGIS-100)}

\setlength{\mathindent}{45pt} %Default is 72pt
\begin{CJK*}{UTF8}{gbsn}
%% Do not edit directly. The author list is generated by a script 
\author{Mahiro Abe$^{1}$, Philip~Adamson$^{2}$, Marcel~Borcean$^{2}$, Daniela~Bortoletto$^{8}$, Kieran~Bridges$^{5}$, Samuel~P~Carman$^{1}$, Swapan~Chattopadhyay$^{2,7}$, Jonathon~Coleman$^{5}$, Noah~M~Curfman$^{2}$, Kenneth~DeRose$^{3}$, Tejas~Deshpande$^{3}$, Savas~Dimopoulos$^{1}$, Christopher~J~Foot$^{8}$, Josef~C~Frisch$^{9}$, Benjamin~E~Garber$^{1}$, Steve~Geer$^{2}$, Valerie~Gibson$^{6}$, Jonah~Glick$^{3}$, Peter~W~Graham$^{1}$, Steve~R~Hahn$^{2}$, Roni~Harnik$^{2}$, Leonie~Hawkins$^{5}$, Sam~Hindley$^{5}$, Jason~M~Hogan$^{1}$, Yijun~Jiang~\mbox{(姜一君)}$^{1}$, Mark~A~Kasevich$^{1}$, Ronald~J~Kellett$^{2}$, Mandy~Kiburg$^{2}$, Tim~Kovachy$^{3}$, Joseph~D~Lykken$^{2}$, John~March-Russell$^{8}$, Jeremiah~Mitchell$^{6,7}$, Martin~Murphy$^{2}$, Megan~Nantel$^{1}$, Lucy~E~Nobrega$^{2}$, Robert~K~Plunkett$^{2}$, Surjeet~Rajendran$^{4}$, Jan~Rudolph$^{1}$, Natasha~Sachdeva$^{3}$, Murtaza~Safdari$^{9}$, James~K~Santucci$^{2}$, Ariel~G~Schwartzman$^{9}$, Ian~Shipsey$^{8}$, Hunter~Swan$^{1}$, Linda~R~Valerio$^{2}$, Arvydas~Vasonis$^{2}$, Yiping~Wang$^{3}$, and Thomas~Wilkason$^{1}$}

\begin{indented}
\item[](\normalsize The MAGIS-100 Collaboration)
\end{indented}
\vspace{10pt}

\address{$^{1}$ Stanford University, Stanford, California 94305, USA}
\address{$^{2}$ Fermi National Accelerator Laboratory, Batavia, Illinois 60510, USA}
\address{$^{3}$ Center for Fundamental Physics, Northwestern University, Evanston, Illinois 60208, USA}  
\address{$^{4}$ The Johns Hopkins University, Baltimore, MD 21218, USA}
\address{$^{5}$ University of Liverpool, Department of Physics, Liverpool, United Kingdom}
\address{$^{6}$ Cavendish Laboratory, University of Cambridge, Cambridge, United Kingdom}
\address{$^{7}$ Department of Physics, Northern Illinois University, DeKalb, Illinois 60115, USA}
\address{$^{8}$ Clarendon Laboratory, University of Oxford, Oxford OX1 3PU United Kingdom}
\address{$^{9}$ SLAC National Accelerator Laboratory, 2575 Sand Hill Road, Menlo Park, CA 94025, USA}

\ead{hogan@stanford.edu}

\begin{abstract}
MAGIS-100 is a next-generation quantum sensor under construction at Fermilab that aims to explore fundamental physics with atom interferometry over a 100-meter baseline. This novel detector will search for ultralight dark matter, test quantum mechanics in new regimes, and serve as a technology pathfinder for future gravitational wave detectors in a previously unexplored frequency band. It combines techniques demonstrated in state-of-the-art 10-meter-scale atom interferometers with the latest technological advances of the world's best atomic clocks. MAGIS-100 will provide a development platform for a future kilometer-scale detector that would be sufficiently sensitive to detect gravitational waves from known sources. Here we present the science case for the MAGIS concept, review the operating principles of the detector, describe the instrument design, and study the detector systematics.

\end{abstract}

\maketitle

\end{CJK*}

\tableofcontents
\newpage
\markboth{Matter-wave Atomic Gradiometer Interferometric Sensor (MAGIS-100)}

\section{Introduction}

Long-baseline atom interferometry is a rapidly growing field with a variety of exciting fundamental physics applications.  Science opportunities include gravitational wave detection~\cite{dimopoulos2008atomic,hogan2011atomic,Yu2011,graham2013new,canuel2018MIGA,Canuel2019ELGAR,kolkowitz2016gravitational,ZAIGA2020,abou2020aedge,Badurina_2020,Graham:2016plp, graham2017mid}, searches for ultralight (wave-like) dark matter candidates~\cite{arvanitaki2018search,Graham:2015ifn} and for dark energy~\cite{Hamilton2015}, tests of gravity and searches for new fundamental interactions (``fifth forces'')~\cite{Rosi2014,Biedermann2015,Rosi2017b,fray2004atomic,Schlippert2014ep,Zhou2015ep,Barrett2016,kuhn2014bose,barrett2015correlative,PhysRevLett.113.023005,PhysRevA.88.043615,Hartwig2015,asenbaum2020atom,williams2016quantum,berge2019exploring}, precise tests of the Standard Model~\cite{Bouchendira2011,parker2018measurement}, and tests of quantum mechanics~\cite{Arndt2014,Bassi2013,Nimmrichter2013,Bassi2017,altamirano2018gravity,kovachy2015quantum, asenbaum2016phase,xu2019probing,zych2011quantum,Roura2020}.  Such experiments take advantage of the ongoing evolution of the precision and accuracy of atomic sensors.  Optical lattice clocks now regularly attain 18 digits of frequency resolution~\cite{hinkley2013atomic,bloom2014optical} and beyond~\cite{Marti2018image,mcgrew2018atomic}, while atom interferometers continue to improve both in inertial sensing applications~\cite{bongs2019taking} and in precision metrology, including measurements of Newton's gravitational constant~\cite{fixler2007atom,PhysRevLett.100.050801,Rosi2014}, the fine structure constant~\cite{Bouchendira2011,parker2018measurement}, and the Equivalence Principle~\cite{fray2004atomic,Schlippert2014ep,Zhou2015ep,Barrett2016,kuhn2014bose,barrett2015correlative,PhysRevLett.113.023005,PhysRevA.88.043615,Hartwig2015,asenbaum2020atom,williams2016quantum,berge2019exploring}. The broad scientific potential of long-baseline quantum sensor networks has been widely recognized~\cite{WalsworthReport,PreskillReport,battaglieri2017us}. These sensors are noted for their potential use in searching for new fundamental forces, dark matter, and other dark sector ingredients~\cite{WalsworthReport}. At the same time, they offer the possibility of testing quantum mechanics over record-breaking macroscopic distances and timescales, and of searching for gravitational waves in an unexplored frequency range. Here we describe the Matter-wave Atomic Gradiometer Interferometric Sensor (MAGIS) research program to develop long-baseline atom interferometers with the aim of realizing these scientific goals. In particular, we present the MAGIS-100 detector design, the pathfinder project in this multistage effort.

The MAGIS concept~\cite{graham2013new,graham2017mid} takes advantage of features of both clocks and atom interferometers to allow for a single-baseline gravitational wave detector~\cite{PhysRevD.78.042003,Yu2011,graham2013new}.  It aims to detect gravitational waves in the scientifically rich, so-far unexplored `mid-band' frequency range between $0.01~\text{Hz}$ and $3~\text{Hz}$. This band lies below the sensitivity range of existing terrestrial interferometers (LIGO/Virgo) and above the frequency band of the planned LISA satellite detector. Simultaneously, the MAGIS concept enables the exploration of new regions of dark-sector parameter-space~\cite{BRNreport} by being sensitive to proposed scalar- and vector-coupled dark matter candidates in the ultralight range ($10^{-15}$~eV -- $10^{-14}$~eV).

MAGIS-100 is the first detector facility in a family of proposed experiments based on the MAGIS concept. The instrument features a 100-meter vertical baseline and is now under construction at the Fermi National Accelerator Laboratory (Fermilab). State-of-the-art atom interferometers are currently operating at the 10-meter scale~\cite{Dickerson2013,kovachy2015quantum,asenbaum2016phase,asenbaum2020atom,Hartwig2015,zhou2011development}, while a kilometer-scale detector is likely required to detect gravitational waves from known sources. MAGIS-100 is the first step to push the limits of atom interferometry beyond the lab-scale and bridge the gap to future detectors. It is designed to be operated in the manner of a full-scale detector and aims to achieve the high up-time required from such a facility. MAGIS-100 will explore a wide variety of systematic errors and backgrounds to serve as a technology demonstrator for future gravitational wave detection with atom interferometry. Additionally, the detector is expected to be sensitive enough to search for potential ultralight dark matter couplings beyond current limits.  By operating in two distinct dark matter search modes, MAGIS-100 can look for both scalar-coupled and vector-coupled dark matter candidates in so-far unexcluded regions of parameter space. Finally, by extending the scale of matter-wave interferometers to a 100-meter baseline, MAGIS-100 has the opportunity to advance the frontier of quantum science and sensor technologies, including tests of the validity of quantum mechanics in a regime in which massive particles are delocalized over record-setting macroscopic time~\cite{Dickerson2013,xu2019probing} and length~\cite{kovachy2015quantum,asenbaum2016phase} scales.

Over the past several years, there has been widespread, growing international interest in pursuing long-baseline atomic sensors for gravitational wave detection~\cite{BertoldiGWOverview}. This has sparked a number of proposals for both space-based instruments and terrestrial detectors, some of which are already under construction today. In France, significant progress has been made towards the $200~\text{m}$ baseline underground gravitational wave detector prototype MIGA (Matter-wave laser based Interferometer Gravitation Antenna)~\cite{canuel2018MIGA}. A follow-on proposal has called for the construction of ELGAR (European Laboratory for Gravitation and Atom-interferometric Research)~\cite{Canuel2019ELGAR,Canuel2020ELGARtechnologies}, an underground detector with horizontal $32~\text{km}$ arms aiming to detect gravitational waves in the mid-band (infrasound) frequency range. In China, work has started to build ZAIGA (Zhaoshan long-baseline Atom Interferometer Gravitation Antenna)~\cite{ZAIGA2020}, a set of $300~\text{m}$ vertical shafts separated by kilometer-scale laser links that will use atomic clocks and atom interferometry for a wide range of research, including gravitational wave detection and tests of general relativity. In the UK, a broad collaboration of seven institutions has recently advanced the multi-stage program AION (Atom Interferometer Observatory and Network)~\cite{Badurina_2020}, which aims to progressively construct atom interferometers at the 10- and then 100-meter scale, in order to develop technologies for a full-scale kilometer-baseline instrument for both gravitational wave detection and dark matter searches. A variety of space-based gravitational wave detectors have also been proposed to access the lower frequency ranges inaccessible to terrestrial observatories. These proposals are based both on optical lattice atomic clocks~\cite{kolkowitz2016gravitational,ebisuzaki2020ino} and atom interferometers~\cite{graham2017mid,hogan2016atom,abou2020aedge,hogan2011atomic,loriani2019atomic}, two technologies that are in fact closely related~\cite{norcia2017role}.

The ambitious scope of these endeavors from around the world is evidence of the widespread enthusiasm for the scientific prospects of long-baseline atomic sensing. The numerous projects complement each other through the diversity of approaches, allowing for the development of alternate atomic sensing technologies in parallel (see \Tref{table:compareExps}). This is a key factor in collectively overcoming the technological challenges towards a full-scale gravitational wave detector. The ultimate synergy of this global effort would be to realize a network of observatories in the spirit of the the LIGO/Virgo/KAGRA collaboration. Correlating data collected simultaneously by several atomic sensor gravitational wave detectors operating in the mid-band frequency range would be a powerful way to improve background rejection and increase overall sensitivity~\cite{Badurina_2020}.

\begin{table}[t]
\centering
\small
\caption{Comparison of long-baseline atom interferometer projects. MAGIS-100 will pursue Sr clock atom interferometry (clock AI)~\cite{graham2013new} over a single, vertical baseline. The AION project is based on the same concept. MAGIS-100 additionally features a dark matter search mode using dual-isotope AI~\cite{overstreet2018effective, asenbaum2020atom} with two-photon Bragg transitions~\cite{muller2008, kovachy2015quantum}. MIGA will employ cavity-assisted Rb atom interferometry with Bragg transitions over two horizontal baselines. ZAIGA will employ three vertical baselines, connected via km-scale horizontal shafts. The proposed experimental techniques range from two-photon Raman~\cite{kasevich1991atomic} and Bragg transitions, to optical lattice clock (OLC) comparisons.}
\label{table:compareExps}
\begin{tabular}{lcccccc} 
 \toprule
 \multirow{2}{*}{Project} & Baseline & Number of & \multirow{2}{*}{Orientation} & \multirow{2}{*}{Atom} & \multirow{2}{*}{Atom Optics} & \multirow{2}{*}{Location} \tabularnewline
  & Length & Baselines & & & &  \tabularnewline
 \midrule
  MAGIS-100 & $100~\text{m}$ & 1 & Vertical   & Sr     & Clock AI, Bragg   & USA \tabularnewline
  AION~\cite{Badurina_2020}    & $100~\text{m}$ & 1 & Vertical   & Sr     & Clock AI          & UK \tabularnewline
  MIGA~\cite{canuel2018MIGA}   & $200~\text{m}$ & 2 & Horizontal & Rb     & Bragg             & France \tabularnewline
  ZAIGA~\cite{ZAIGA2020}       & $300~\text{m}$ & 3 & Vertical   & Rb, Sr & Raman, Bragg, OLC & China \tabularnewline
 \bottomrule
\end{tabular}
\end{table}

\section{Science Motivation}

In the following, we describe the scientific motivation for MAGIS-100 and anticipated follow-on detectors. The sensitivity of these instruments to many science signals of interest scales proportionally to the baseline length $L$, motivating the development of detectors at the $100~\text{m}$ scale, and eventually, at the kilometer scale.  Sensitivity can also be enhanced through the use of advanced large momentum transfer (LMT) atom optics, as well as by reducing phase noise in the matter-wave interference fringes (atom shot noise) through the use of both high flux atom sources and quantum entangled atoms.  In estimating the potential science impact of MAGIS-100 and future advanced MAGIS-style detectors, we consider sensitivities based both on the current state-of-the-art, as well as on targeted improvements in these various detector parameters.  The goals of this technology development path are summarized in \Tref{table:futurevision}.

\begin{table}
\centering
\footnotesize
\caption{Detector design parameter targets for MAGIS-100 and follow-on detectors.  The baseline length $L$ is the total end-to-end length of the detector. LMT atom optics of order $n$ refers to an $n\hbar k$ momentum splitting between the two arms of the atom interferometer (corresponding to $n$ photon recoil kicks).  The atom phase noise $\delta\phi$ listed is for a single atom source, and assumes improvements in atom flux and the use of spin-squeezed atomic states~\cite{hosten2016measurement,PhysRevLett.116.093602}. The multiple atom sources are assumed to be distributed uniformly along the baseline. MAGIS-100 (initial) corresponds to current state-of-the-art parameters, while (final) assumes atom optics operating at the projected physical limit for this baseline. The space-based configuration is discussed in greater detail in~\cite{graham2017mid}.
}
\label{table:futurevision}
\begin{tabular}{lccccc} 
 \toprule
   \multirow{2}{*}{Experiment} & \multirow{2}{*}{(Proposed) Site} & Baseline & LMT Atom   & Atom    &  Phase Noise  \tabularnewline
  &   &  $L$ (m)  & Optics $n$ & Sources & $\delta\phi$ ($\text{rad}/\sqrt{\text{Hz}}$)  \tabularnewline
  \midrule
 Sr prototype tower & Stanford & $10$ & $10^2$ & 2 & $10^{-3}$ \tabularnewline
 MAGIS-100 (initial) & Fermilab (MINOS shaft) & $100$ & $10^2$ & 3 & $10^{-3}$ \tabularnewline
 MAGIS-100 (final) & Fermilab (MINOS shaft) & $100$ & $4\times10^4$ & 3 & $10^{-5}$ \tabularnewline
 MAGIS-km & Homestake mine (SURF) & $2000$ & $4\times10^4$ & 40 & $10^{-5}$ \tabularnewline 
 MAGIS-Space & Medium Earth orbit (MEO) & $4\times 10^7$ & $10^3$ & 2 & $10^{-4}$ \tabularnewline
 \bottomrule
\end{tabular}
\end{table}

\subsection{Gravitational waves}

The recent discovery of gravitational waves by LIGO is a historic event of profound scientific significance~\cite{Abbott2016}.  It is a re-affirmation of general relativity, but its main significance is as the beginning of a new form of astronomy.  Gravitational waves provide an entirely new spectrum with which to view the universe and will be a major part of the future of astronomy, astrophysics, and cosmology.

Gravitational waves allow observations that are impossible with normal electromagnetic telescopes as they are not meaningfully attenuated by passage through intervening matter. Moreover, unlike electromagnetism, all forms of matter couple equally to gravity, so in particular no objects are neutral with respect to gravity. Thus, gravitational wave astronomy enables the observation of objects that do not emit light.  For example, black holes and other compact objects are probably best studied with gravitational waves.  Furthermore, the earliest observation of the universe possible with light comes from the time of Cosmic Microwave Background (CMB) formation, since prior to that all photons were thermalized. Gravitational waves do not thermalize, and can carry information about the earliest epochs in the universe, back to and even including inflation or whatever other process sets the initial conditions for the hot Big Bang.  The early universe was likely both extremely hot (with temperature $T\gg {\rm TeV}$, much more than the MeV temperatures indirectly probed by Big Bang Nucleosynthesis) and sometimes far-from-equilibrium, so observing such early times can teach us not just about the beginning of our universe, but also about the highest energy scales and most basic laws of nature, potentially far beyond the energies that can be probed in any particle collider.

\begin{figure}[t]
\begin{center}
\includegraphics[width=1\textwidth]{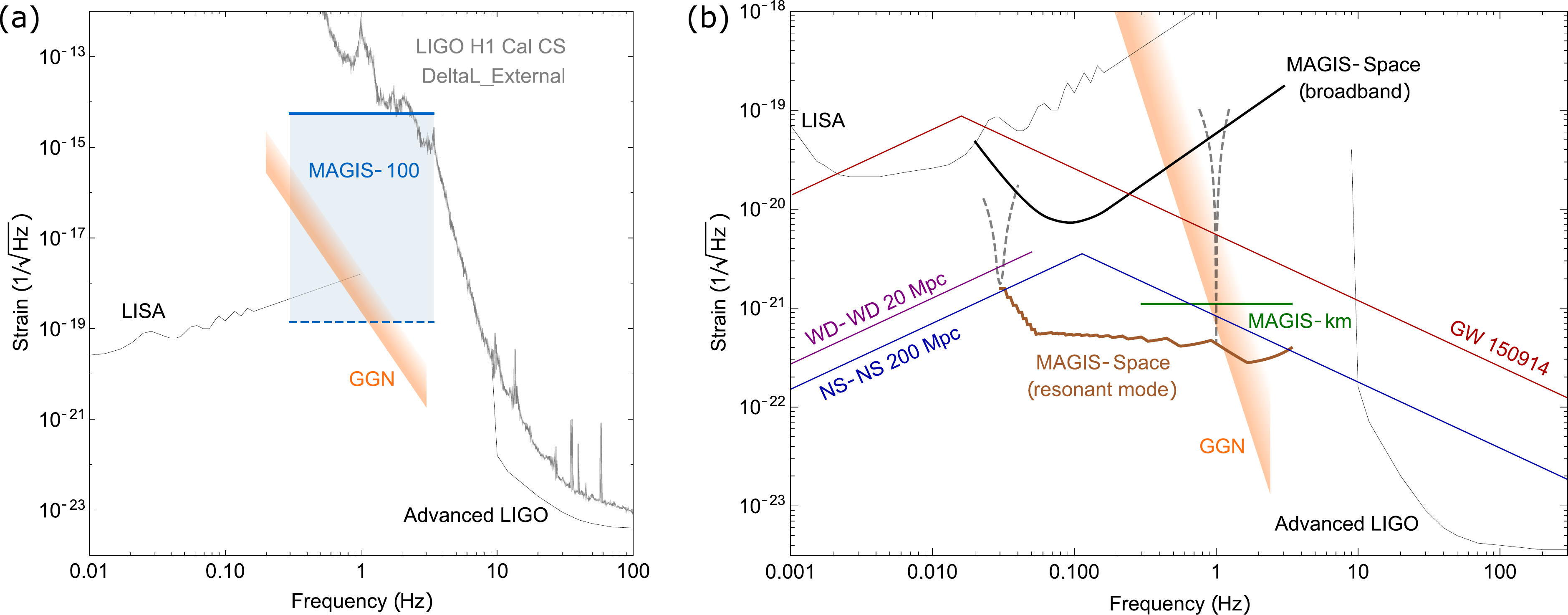}
\end{center}
    \caption{(a) Projected gravitational wave strain sensitivity for MAGIS-100 and follow-on detectors. The solid blue line shows initial performance using current state of the art parameters (\Tref{table:futurevision}, initial). The dashed line assumes parameters improved to their physical limits (\Tref{table:futurevision}, final). LIGO low frequency calibration data (gray) is shown as an estimate for the state-of-the-art performance in the mid-band frequency range~\cite{LIGOcalibration}. An estimate of gravity gradient noise (GGN) at the Fermilab site is shown as an orange band (see \Sref{sec:seismic-measurements}). (b) Estimated sensitivity of a future km-scale terrestrial detector (MAGIS-km, green) and satellite-based detector (MAGIS-Space, brown) using detector parameters from \Tref{table:futurevision}.  The detector can be switched between both broadband (black, solid) and narrow resonant modes (black, dashed).  The resonant enhancement $Q$ can be tuned by adjusting the pulse sequence~\cite{Graham:2016plp}. Two example resonant responses are shown targeting $0.03~\text{Hz}$ ($8\hbar k$ atom optics, $Q=9$) and $1~\text{Hz}$ ($1\hbar k$ atom optics, $Q=300$).  The brown curve is the envelope of the peak resonant responses, as could be reached by scanning the target frequency across the band.  Sensitivity curves for LIGO~\cite{AdvancedLIGO2015} and LISA~\cite{prince2002lisa} are shown for reference. Also shown are a selection of mid-band sources including neutron star (NS) and white dwarf (WD) binaries (blue and purple) as well as a black hole binary already detected by LIGO (red). The GGN band (orange) is a rough estimate based on seismic measurements at the SURF site~\cite{Harms_2010}.}
\label{fig:gw-sensitivity-100}
\end{figure}

Current gravitational wave observations use terrestrial laser interferometry techniques. LIGO and other ground-based laser interferometer designs are sensitive to gravitational waves between a few Hz and $10~\text{kHz}$, but are severely limited at lower frequencies due to coupling to seismic noise~\cite{Abbott2016,Abbott2017LIGOVirgo,Punturo2010}. However, frequencies below LIGO's range carry important information about the early universe and many other sources. At much lower frequencies, the planned space-based LISA detector is targeted at the $1~\text{mHz}$ -- $50~\text{mHz}$ range~\cite{bartolo2016science}.  In order to fully realize the potential of gravitational wave observations, it is important to cover as many different frequency bands as possible~\cite{Kawamura2006,Crowder2005,Ni2016}.  Atomic sensors appear promising for filling the gap between LIGO and LISA, observing gravitational waves in the mid-band, roughly $30~\text{mHz}$ to $3~\text{Hz}$. \Fref{fig:gw-sensitivity-100}(b) shows the projected strain sensitivity in the mid-band for future full-scale MAGIS-style detectors, both on the ground and in space.  Achieving this level of performance will be challenging, but the potential payoff is significant.  By serving as a technology demonstrator, MAGIS-100 will pave the way for these next generation detectors.

There are a number of compelling reasons to explore the mid-band.  Historically, each time a frequency band of electromagnetic radiation was explored for the first time it led to new unexpected discoveries.  Since gravitational waves are a fundamentally new way to observe the universe, perhaps the most important discovery could be something we do not expect.  For this reason alone it is important to build detectors in all frequency bands.

In addition, the mid-band may be optimal for observing signals of cosmological origin. This frequency range is above the white dwarf ``confusion noise'' but can still extend low enough in frequency to see certain cosmological sources~\cite{Graham:2016plp}.  This band can also be an excellent place to search for gravitational waves from inflation and reheating, and certain models such as axion inflation may give signals large enough to be detected by future versions of MAGIS~\cite{Graham:2016plp}. Furthermore, thermal phase transitions in the early universe at scales above the weak scale~\cite{randall2007gravitational,Caprini:2009fx,Caprini:2009yp,Konstandin:2011dr,Schwaller_2015,Hindmarsh:2017gnf,Cutting:2019zws,Breitbach_2019,Caprini:2019egz}, or quantum tunnelling transitions in cold hidden sectors~\cite{GarciaGarcia:2016xgv}, or networks of cosmic strings~\cite{depies2007stochastic}, or collapsing domain walls~\cite{Saikawa2017}, or axion dynamics in the early universe~\cite{Machado_2019,Machado_2020}, may produce detectable gravitational wave signals in this band. In general, gravitational waves are a direct way to observe the earliest times in the birth of the universe, and the mid-band may provide a valuable window into this high energy physics.

The mid-band possesses a potential two-fold advantage for the investigation of a stochastic gravitational wave background resulting from a first-order phase transition.  A transition occurring with nucleation temperature $T_n \simeq 1~{\rm TeV}$ typically leads to a stochastic spectrum with a peak occurring at frequency $f\simeq  2\times 10^{-2}~{\rm Hz}$~\cite{Caprini:2009fx}. (Here for definiteness we have assumed that the inverse duration, $\beta$, of the transition satisfies $\beta/H_n\sim 10^2$ as is often the case for electroweak scale and above transitions, where $H_n$ is the Hubble parameter at nucleation.)  So, for transitions corresponding to energy scales in the region above that already partially probed by the LHC ($T_n \gtrsim {\rm TeV}$) the peak of the gravitational wave spectrum commonly occurs in the mid-band region.  Secondly, as causality enforces (in the long-wavelength limit) a universal $f^3$ rise of the power spectrum of gravitational waves produced by local sub-horizon scale dynamics~\cite{Caprini:2009fx}, the information that distinguishes the underlying microphysics is usually contained in the peak and especially post-peak high-frequency range. The reason for this is that gravitational waves produced in thermal phase transitions primarily arise from a combination of long-lasting sound wave and turbulent bulk fluid motions of the plasma acting on far sub-horizon length scales~\cite{Caprini:2009yp,Caprini:2015zlo,Hindmarsh:2017gnf,Cutting:2019zws,Caprini:2019egz,Ellis:2020awk}. (In the case of runaway bubble walls, and small $\beta/H_n$, the motion of the underlying order parameter profile can also contribute significantly.)  These depend in turn on the bubble wall velocity and properties of the plasma, as well as the amount of super-cooling, all properties determined by the microphysics.  The information-rich post-peak frequency region is again typically contained within the mid-band for transitions corresponding to the so-far-unexplored TeV scale and above. 

The mid-band also provides excellent complementarity with the capabilities of LISA~\cite{Caprini:2019egz,Ellis:2020awk} 
for characterising the properties of a stochastic gravitational wave background arising from a thermal phase transition slightly above the electroweak scale. This is because the spectrum needs to be determined over a ${\cal O}(10^2)$ frequency range to reliably extract the most important information, and, in addition, there exist many extensions of the Standard Model not yet excluded by the LHC with transitions occurring at $T_n \gtrsim 500~{\rm GeV}$, so with peak frequency falling at the upper-end of the LISA sensitivity range.

Such speculative cosmological sources would clearly be a major discovery, but there are also important sources in the mid-band which are certain to exist, in some cases because they have already been observed by LIGO later in their lifetimes, when their frequencies have increased.  These include neutron star (NS) binaries as well as black hole binaries with masses around a few to tens of solar masses.  Additionally, the recent discovery by LIGO of a binary black hole merger with a total mass of 150 solar masses motivates further studies of intermediate mass black holes~\cite{LIGO2020Intermediate}.  Earlier observations of such binaries at lower frequencies (including mid-band frequencies) would provide valuable astrophysical insight.  Moreover, even heavier black holes than the LIGO observation, hundreds of solar masses and above, would merge in the mid-band.  Therefore, this would be the ideal band to discover such black holes, if they exist.

Importantly, many black hole or neutron star binaries that are observed in the mid-band can later be observed by LIGO once they evolve to higher frequencies.  Such joint observation would be a powerful new source of information.  This would allow an atomic sensor in the mid-band to give a prediction of the time and location of a merger event.  Since the lifetimes of many sources in the mid-frequency band are comparable to the orbital period of the Earth, the mid-band is ideal for sky localization and prediction of merger events, and in fact they can be localized even by a single-baseline detector~\cite{graham2018localizing}. Optical, x-ray, gamma ray, and other telescopes could use this forewarning to observe these sources as they merge.  A mid-band atomic detector would thus allow simultaneous observation of these merger events by gravitational wave and electromagnetic telescopes.  For example, in the NS-NS kilonova merger recently observed by LIGO~\cite{Abbott2017}, optical and other EM telescopes only started viewing the object roughly a day (or more) after merger, which is a very long time compared to the natural dynamical timescale of around a millisecond.  It would be very beneficial to have prior information on the timing and location of such an event so that the final merger could be observed.  Thus, observations in the mid-band have the potential to be a powerful complement to detection by LIGO, and can significantly enhance multi-messenger astronomy.

To better understand the nature and origin of black holes, including those being observed by LIGO, it would be beneficial to have a measurement of their initial spins and orbital eccentricities.  This can for example discriminate between different production mechanisms (e.g.~cosmological or astrophysical)~\cite{Mandel:2017pzd}. 
To accurately measure the initial spins, it is necessary to record many cycles before the merger, and this is aided by observing at lower frequencies below LIGO's band. Thus, the mid-band appears to be a promising band for measuring the spin of merging black holes.

There are also white dwarf binary mergers that can be detected in the mid-band, which are not observable at higher frequencies~\cite{schneider2001low}.  Such a merger could be a type IA supernova.  The question of the origin of type IA supernovae, whether from a single white dwarf or a white dwarf binary merger, has attracted significant interest and is clearly of major importance (see e.g.~\cite{Mandel:2017pzd} and references therein).  As just one example, the nature of type IA supernova clearly affects their use as standard candles for measuring the cosmological expansion rate and the properties of dark energy.

In addition to being interesting and important astrophysically, observing compact objects such as black holes, neutron stars, and white dwarfs may well also teach us about particle physics.  For example, supernovae and other such extreme astrophysical objects have already been used to set some of the best limits on axions and other light particles, and  gravitational wave observations may allow many more such tests for new physics.  As just one example, superradiance around black holes~\cite{Arvanitaki:2009fg, Arvanitaki:2010sy,Brito:2015oca} may allow us to constrain or even discover such particles with future gravitational wave observations. One way to use black hole superradiance to set
limits on axions or other very light bosonic particles requires that we have precise measurements of the black holes' initial spins.  As noted above, the mid-band appears quite promising for these black hole spin measurements.  In addition, there can be dramatic monochromatic gravitational wave signals via axion annihilations or level transitions in the axion (or very light particle) cloud that superradiance produces in the vicinity of the black hole~\cite{Arvanitaki:2009fg, Arvanitaki:2010sy,Arvanitaki:2014wva,Arvanitaki:2016qwi,Baryakhtar:2017ngi}.

Further, such nominally astrophysical observations can shed light on important cosmological questions.  It has long been known that a binary merger can be a gravitational wave ``standard siren''~\cite{Schutz:1986gp}. Such standard sirens allow us to measure the expansion rate of the universe with fewer systematic uncertainties because the gravitational merger signal is very clean.  Accurate angular localization is very important for this measurement, and thus mid-band observations could contribute significantly to this program by providing sky localization information for LIGO-Virgo and optical telescopes well before the merger event, as well as providing localization information for events without a direct optical counterpart such as binary black hole mergers (see e.g.~\cite{Chen:2016tys}). This would enable a better understanding of cosmology and, in particular, a better measurement of the Hubble constant and the dark energy equation of state.

As a technology demonstrator, MAGIS-100 is not expected to be sufficiently sensitive to detect known candidate sources of gravitational waves. Nevertheless, within its frequency range it aims to achieve a record sensitivity, improving on the current bounds~\cite{PhysRevD.95.082004} (\Fref{fig:gw-sensitivity-100}(a)). While there are not any known gravitational wave sources that MAGIS-100 will be sensitive enough to detect, whenever we probe far beyond current bounds there is always the possibility of an unexpected discovery.  A follow-up full-scale, kilometer-baseline terrestrial detector would have the potential to observe many of the sources discussed above in the mid-band.  In addition, MAGIS-100 will set the stage for a possible future satellite-based detector~\cite{graham2017mid}, which could access the complete mid-band range as shown in \Fref{fig:gw-sensitivity-100}(b).

\subsection{Dark matter}

\label{sec:DMIntro}

Cosmological and astronomical measurements have established that the energy budget of the universe is dominated by dark energy and dark matter, but their nature remains unknown.  Discovering the properties of these unknown constituents of the universe is therefore one of the most important scientific problems of our time.  Dark matter can lead to time-dependent signals in high precision quantum sensor networks, enabling a unique probe of its existence. In particular, these time-dependent signals can be caused by ultralight dark matter candidates. Observational bounds permit a 10\% fraction or more of the dark matter to have a mass as low as \SI{e-22}{eV}, whereas current experiments have focused on  dark matter in a narrow range of masses (e.g.  around \SI{100}{GeV} for WIMPs). Given the null results from the present generation of dark matter experiments, it is important to broaden the search to cover a variety of dark matter candidates. Well motivated theories indicate that the mass range from \SI{e-22}{eV} to \SI{e-3}{eV} is particularly interesting. Potential dark matter candidates within this range include, for spin-0, the QCD axion~\cite{Peccei:1977hh,Weinberg:1977ma,Wilczek:1977pj,Preskill:1982cy}, axion-like-particles and ultralight string moduli~\cite{Essig:2013lka}, the relaxion~\cite{Graham:2015cka},  and for spin 1, the dark photon~\cite{Essig:2013lka}. Dark matter in this mass range has a large number density and can be described as a classical field that oscillates at a frequency determined by the mass of the dark matter particle. This results in time-dependent effects that can be searched for using quantum sensors. These effects arise  because as the classical dark matter field oscillates, the properties of the sensor (such as the quantum energy level and spin) also change, leading to time-dependent signals. The fact that the dark matter signal oscillates at a frequency set by fundamental physics (the mass of the dark matter) serves as a powerful discriminant against a variety of noise sources, enabling high precision searches for the ultra-weak effects of dark matter.

Even though there are a wide variety of theoretical dark matter candidates, there are only four dominant experimental signatures of this oscillating classical field. The oscillating field can induce currents in circuits, exert accelerations on test masses, cause precession of spins, and change the values of fundamental constants~\cite{BRNreport}. Multiple experiments are currently searching for the first of these effects.  With its unique sensitivity to accelerations, spin and atomic energy levels, MAGIS-100 would be sensitive to the three other dominant effects of a component of dark matter in the mass range \SI{e-22}{eV}\,--\,\SI{e-15}{eV}~\cite{arvanitaki2018search, Graham:2017ivz, Graham:2015ifn}. In fact, three separate dark matter searches can be performed using this quantum sensor.

\begin{figure}
\centering
  \includegraphics[width=\textwidth]{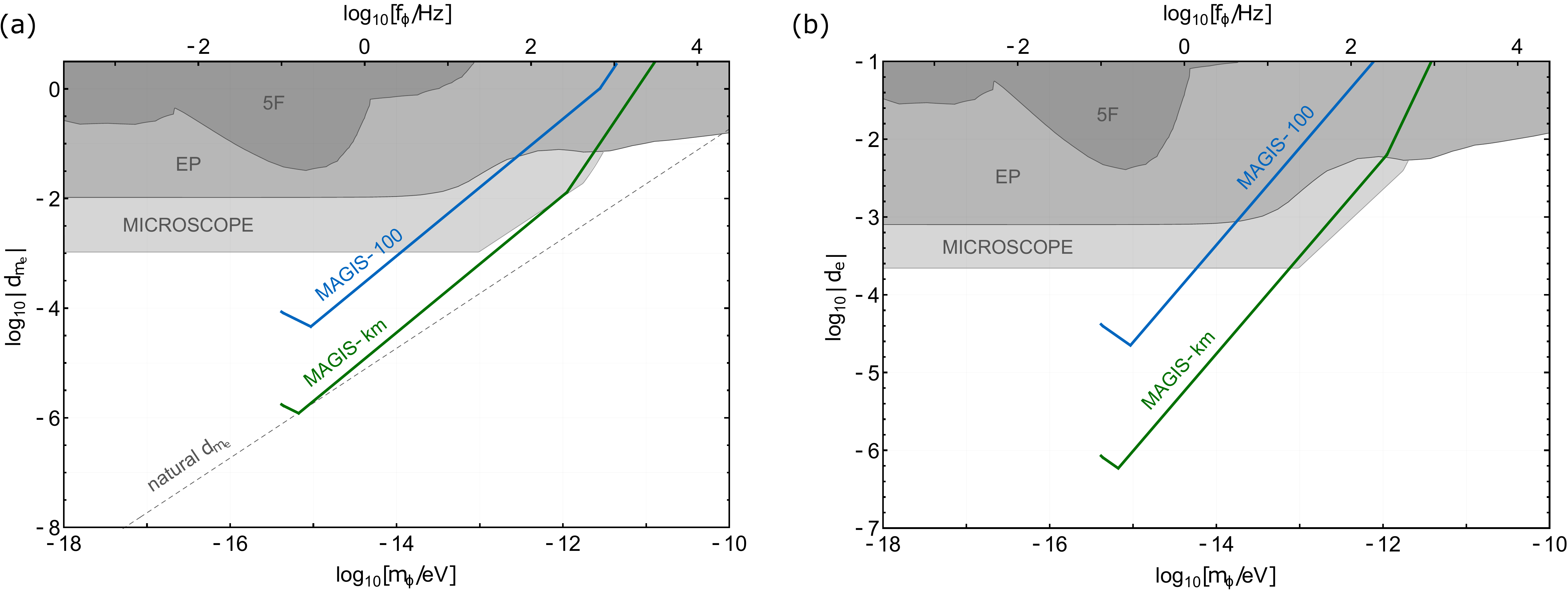}
    \caption{Sensitivity of MAGIS-100 to ultralight dark matter fields coupled to the electron mass with strength $d_{m_e}$ (a) and the fine structure constant with strength $d_e$ (b), shown as a function of the mass of the scalar field $m_\phi$ (or alternatively the frequency of the field - top scale)~\cite{arvanitaki2018search}. The blue sensitivity curve assumes a shot noise limited phase resolution and corresponds to 1~year of data acquisition ($1000~\hbar k$ atom optics, $10^{-4}~\text{rad}/\sqrt{\text{Hz}}$ phase resolution).  We assume a density of $0.3~\text{GeV}/\text{cm}^3$ for each candidate dark matter field.  The gray bands show existing bounds, derived from equivalence principle (EP) and fifth force (5F) tests~\cite{arvanitaki2018search}, as well as the MICROSCOPE satellite EP experiment~\cite{Touboul_2019microscope}. The green curve is the projected sensitivity of a future kilometer-scale detector.} 
\label{fig:dme-and-dm-sensitivity}
\end{figure}

First, dark matter that affects fundamental constants, such as the electron mass or the fine structure constant, will change the energy levels of the quantum states used in the interferometer, causing them to oscillate at the Compton frequency of the candidate dark matter particle.  This effect can be searched for by comparing two simultaneous atom interferometers separated along the MAGIS-100 baseline (see Section \ref{Sec:MAGISConcept}).  The sensitivity to several such dark matter candidates is shown in \Fref{fig:dme-and-dm-sensitivity}.

\begin{figure}[t]
\begin{center}
\includegraphics[width=0.55\textwidth]{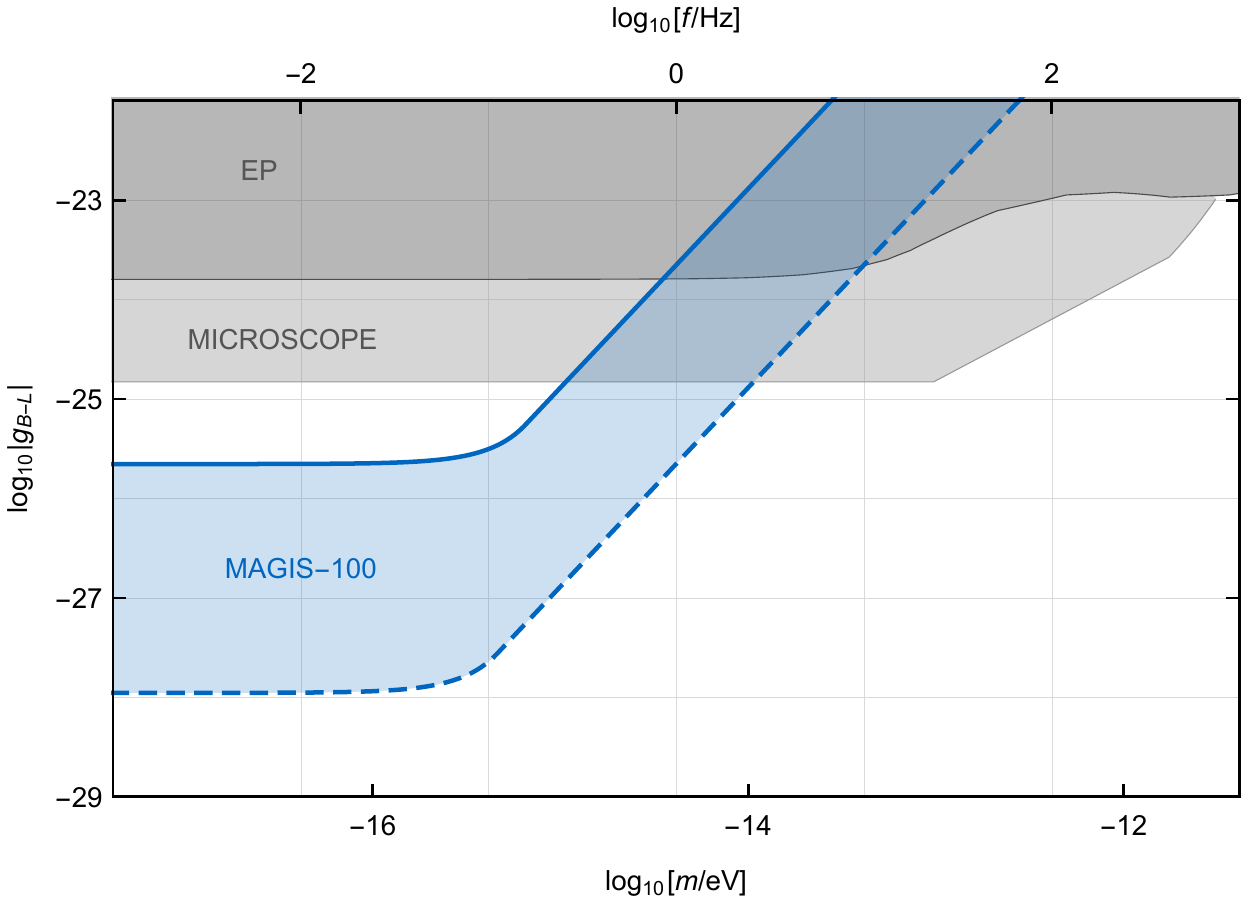}
    \caption{Dark matter sensitivity of MAGIS-100 for an ultralight vector field coupled to $B-L$.  The initial (solid) curve is for $10^{-14}g/\sqrt{\text{Hz}}$ acceleration sensitivity (assumes $50~\text{m}$ launch, $100~\hbar k$ atom optics, $10^6~\text{atoms/s}$ flux, shot noise limited), while the upgraded (dashed) curve is for $6\times10^{-17}g/\sqrt{\text{Hz}}$ (assumes $100~\text{m}$ launch, $1000~\hbar k$ atom optics, $10^8~\text{atoms/s}$ flux, shot noise limited). At lower frequencies the detector sensitivity is likely limited by systematic errors (e.g., time-varying blackbody radiation or magnetic fields). We assume a density of $0.3~\text{GeV}/\text{cm}^3$ for the $B-L$ field.  The gray shaded regions show bounds from equivalence principle tests using torsion pendula (EP)~\cite{Graham:2015ifn, Wagner_2012} and the MICROSCOPE satellite experiment~\cite{Touboul_2019microscope}.  Potential sensitivities of this detector method to general other dark matter candidates are discussed in~\cite{Graham:2015ifn}.}
\label{fig:BL-DM-sensitivity}
\end{center}
\end{figure}

Second, dark matter that causes accelerations can be searched for by comparing the accelerometer signals from two simultaneous atom interferometers run with different isotopes (${}^{88}$Sr and ${}^{87}$Sr for example)~\cite{Graham:2015ifn}.  This requires running a dual-species atom interferometer, which is well established~\cite{PhysRevLett.113.023005,PhysRevA.88.043615,kuhn2014bose,overstreet2018effective}.  The potential sensitivity of MAGIS-100 to one such dark matter candidate, a $B-L$ coupled new vector boson, is shown in \Fref{fig:BL-DM-sensitivity}.  In general, potential sensitivities to dark matter candidates are shown in~\cite{Graham:2015ifn}.  Note that, compared to existing bounds, MAGIS-100 has the potential to improve the sensitivity to any such dark matter particles with mass (frequency) below approximately \SI{e-15}{eV} (\SI{0.1}{Hz}) by about two orders of magnitude.  Interestingly, the two dark matter searches outlined here are sensitive to similar dark matter candidates, but within complementary mass ranges, extending the coverage of the dark matter parameter space.

Third, dark matter that causes precession of nuclear spins, such as general axions, can be searched for by comparing simultaneous, co-located interferometers using Sr atoms in quantum states with differing nuclear spins.  See~\cite{Graham:2017ivz} for a discussion and potential sensitivities.

\subsection{New forces}

In addition to these dark matter searches, new fundamental particles may also be discovered by searching for new forces~\cite{WalsworthReport}. Ultralight particles that have highly suppressed interactions with Standard Model particles, often dubbed ``dark sectors'', emerge in a variety of beyond-the-Standard-Model frameworks. These theories include forces mediated by particles that can dynamically solve naturalness problems in the Standard Model, such as the strong CP problem (QCD axion~\cite{Moody:1984ba}) and the hierarchy problem (relaxion~\cite{Graham:2015cka}). Such forces can also arise in theories with extra dimensions~\cite{ArkaniHamed:1998nn} as well as super-symmetry~\cite{Dimopoulos:1996kp}.  Due to its high precision, MAGIS-100 can search for these ultra-weak forces, sourced either by the Earth or a test mass. Several of these particles have already been considered above as ultralight dark matter candidates. However, there are alternative ways to search for the presence of these new fields, without necessarily requiring them to be dark matter. In principle there are two ways to do this. First, if the range of the new force is short, it can be observed by modulating the distance between a test mass and the atomic sensor. Second, long range forces sourced by the Earth other than gravity may lead to differential free-fall accelerations between different elements/isotopes. A comparison between atomic sensors made out of different atomic species could reveal the existence of such forces.  MAGIS-100 can perform such a test by performing simultaneous acceleration measurements with two isotopes of Sr.

\subsection{Quantum science}

It is widely recognized that quantum technologies such as quantum computing, quantum simulation, and quantum sensing will play a key role in a variety of scientific applications~\cite{PreskillReport, WalsworthReport}.  MAGIS-100 aims to both develop and employ quantum techniques that are broadly relevant for these applications. The operation of the instrument requires the ability to coherently manipulate atoms with high fidelity, serving as a testbed for quantum control protocols that are widely applicable in quantum information science.  The advanced techniques developed in MAGIS-100 for fundamental physics can also be leveraged for applied inertial sensors including gyroscopes~\cite{gustavson1997precision,Durfee2006}, accelerometers ~\cite{kasevich1991atomic,Peters2001}, and gravity gradiometers~\cite{mcguirk2002sensitive}.

MAGIS-100 can also be used to test quantum mechanics itself.  The detector takes advantage of recent advances in manipulation of atoms using light~\cite{kovachy2015quantum,asenbaum2016phase, Rudolph2020}, as well as long free-fall times~\cite{Dickerson2013}, to realize macroscopic quantum mechanical superposition states.  Atom wave packets in each interferometer are expected to be separated by distances of several meters, improving substantially over previous records~\cite{kovachy2015quantum,asenbaum2016phase} and leading to atomic states that are delocalized on a truly macroscopic scale.  By operating two or more such interferometers simultaneously and comparing their interference patterns, MAGIS-100 can probe whether the coherence of the macroscopic quantum superposition state is maintained at such large length scales.  The instrument can also study whether these macroscopically delocalized quantum states preserve their coherence at long times (up to 9 seconds for a full 100~m launch).  Such measurements have implications for a variety of proposed fundamental decoherence models~\cite{Arndt2014,Bassi2013,Nimmrichter2013,Bassi2017,altamirano2018gravity}. The detector can also potentially search for non-linear corrections to the Schr\"{o}dinger equation~\cite{PreskillReport, weinberg1989testing}. Moreover, MAGIS-100 may be able to measure phase shifts arising from higher order variations in the gravitational potential across the wavefunction \cite{Bertoldi2019,overstreet2021physically}, which were not resolvable in previous work~\cite{asenbaum2016phase} and would probe quantum mechanics in a new regime \cite{overstreet2021physically}.

The detector platform can also eventually incorporate entangled quantum sources to reduce noise, permitting enhanced sensitivity. Spin-squeezed atom sources~\cite{hosten2016measurement,PhysRevLett.116.093602} take advantage of quantum correlations within an atom ensemble to realize a reduction in sensor noise below the standard quantum limit (shot noise).  In addition to improving sensitivity, combining spin-squeezed atom ensembles with large wavefunction delocalization may offer new ways to test quantum mechanics~\cite{engelsen2017bell}.

\pagebreak[3]
\section{MAGIS Experimental Concept}
\label{Sec:MAGISConcept}

MAGIS is a novel quantum sensor concept for ultralight wave-like dark matter searches and gravitational wave detection. Passing gravitational waves cause a small modulation in the distance between objects. Detecting this modulation requires two ingredients:

\begin{description}
\item \textbf{Inertial references:} A pair of objects to act as inertial reference points, separated by some distance (the baseline). Good inertial reference objects must be largely decoupled from the environment and immune to perturbations from non-gravitational forces.

\item \textbf{Clocks:} A means of precisely measuring the separation between the inertial reference objects. This is typically done by measuring the time for light to cross the baseline, via comparison to a precise phase reference (e.g.\ a clock).
\end{description}

In the MAGIS detector design, dilute clouds of freely-falling ultracold Sr atoms play both of these roles, simultaneously acting as inertial references and as precise clocks. Laser light propagates between the two atom ensembles separated by the baseline. It drives transitions between the ground and excited states of the Sr clock transition (${}^1S_0 \rightarrow {}^3P_0$), the same line used in state-of-the-art optical lattice clocks~\cite{campbell2017fermi}. A sequence of short light pulses generates a pair of atom interferometers, one on each end of the baseline~\cite{Hogan2009}.  The timing of the atomic transitions, and thus the time the atoms spend in a superposition of the ground and excited states, depends on the light travel time across the baseline~\cite{graham2013new}. The resulting atom interferometer phase $\phi$ is then proportional to the length $L$ of the baseline: $\phi \,\propto\,\omega_A L/c$, where $\hbar\, \omega_A$ is the energy splitting of the clock transition.  As a result, the differential phase measurement between the two atom interferometers is sensitive to variations in both the baseline $L$ and the clock frequency $\omega_A$ that arise during the light-pulse sequence. A passing gravitational wave modulates the baseline length, while coupling to an ultralight dark matter field can cause a modulation in the clock frequency\footnote{This effect does not require an inhomogeneity in the dark matter field across the baseline, but instead is a result of a time-delayed clock comparison due to the finite light travel time~\cite{arvanitaki2018search}.}. Thus, the MAGIS concept combines the prospects for both gravitational wave detection and dark matter searches into a single detector design, and both science signals are measured concurrently.

It is illustrative to compare the MAGIS concept to laser interferometer detectors such as LIGO, which fundamentally operate using the same ingredients defined above.  In laser interferometers, the mirrors on each end of one of the arms act as the inertial references, while light bouncing between the mirrors serves to sample the light travel time.  These mirrors are decoupled from vibration noise and other non-gravitational perturbations by a vibration isolation system.  In MAGIS, the atom inertial references are in free fall, so they are intrinsically decoupled from many of these noise sources without the need for a sophisticated suspension system.

It is important to emphasize that, in principle, a LIGO-like sensor could detect gravitational waves using a single linear arm if it had access to a sufficiently stable phase reference against which to compare the phase of the returning light.  However, lacking an idealized laser source, a laser interferometer instead uses a second (optimally orthogonal) arm as its phase reference.  The signal is then the phase difference between the two arms, leading to a suppression of laser phase noise in the differential measurement, and therefore one arm can be thought of as acting as the phase reference for the other.  In MAGIS, the atomic proof masses themselves act as the phase reference, encoding the noisy laser phase of each light pulse in the quantum coherence of the atom superposition state.  This noise is then rejected as a common mode in the differential measurement between the atom interferometers on each end of the detector, eliminating the need for an auxiliary baseline to serve as the phase reference.  In this sense, the `active' atomic proof masses used in MAGIS allow for a single-baseline detector geometry which is not possible using `passive' mirror proof masses such as in laser interferometers.

\subsection{Clock atom interferometry}
\label{sec:LightPulseAI}

\begin{figure}[t]
\begin{center}
%\vspace{-10mm}
\includegraphics[width=0.5\textwidth]{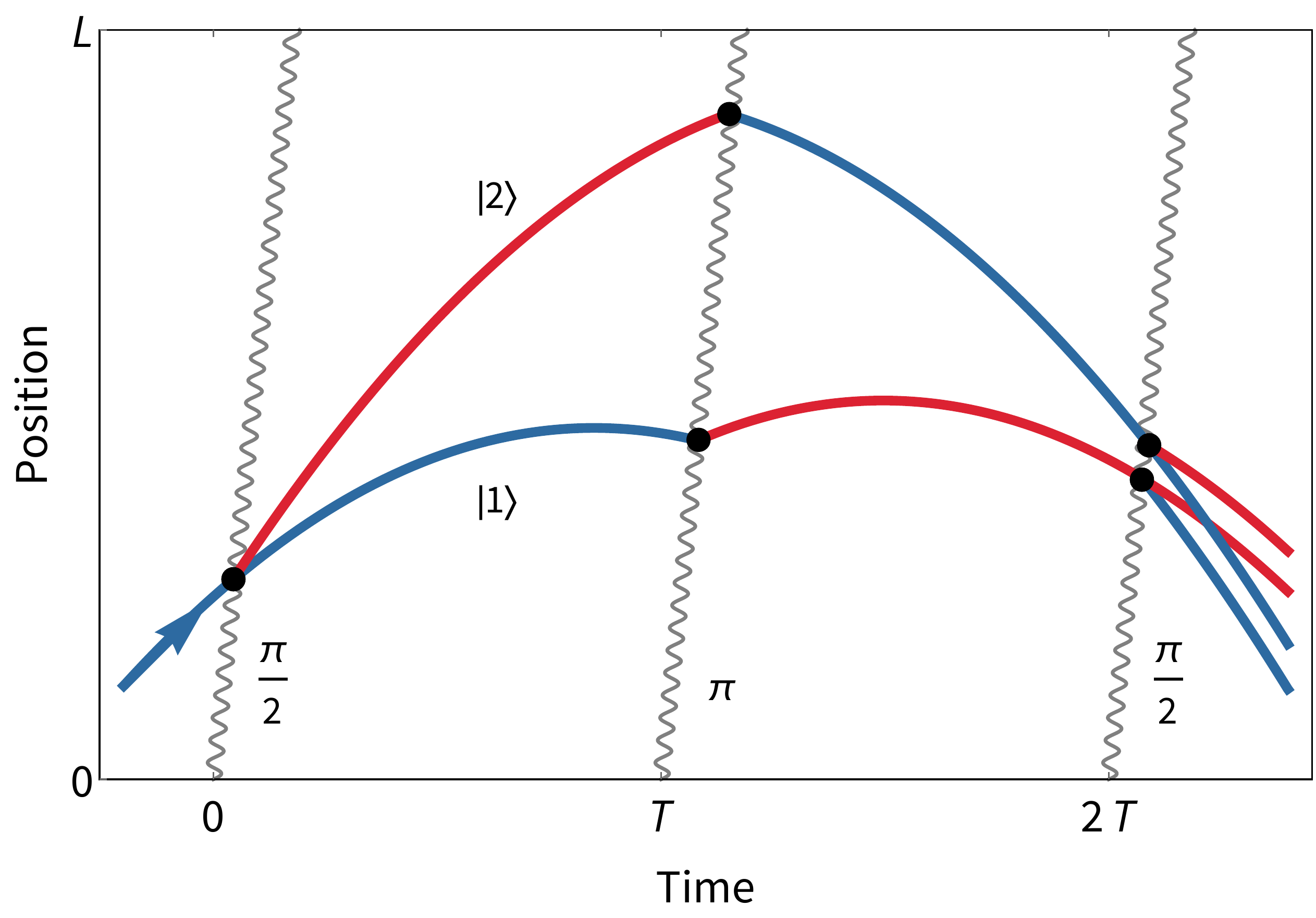}
    \caption{Space-time diagram of a clock atom interferometer. An atom is launched upwards and is freely falling under the influence of gravity. The atom is manipulated by a series of three laser pulses (wavy lines), emitted at times $0$, $T$, and $2T$. The first pulse, with pulse area $\pi/2$, puts the atom in an equal superposition of ground and excited states ($\ket{1}$ and $\ket{2}$). The second pulse, with pulse area $\pi$, flips each state and redirects the two parts of the wavefunction. A third pulse with pulse area $\pi/2$ recombines the two paths, leading to a ground and excited state population whose ratio depends on the relative phase accumulated between the two arms.}
\label{Fig:AIspacetime}
\end{center}
\end{figure}

MAGIS employs light-pulse atom interferometry to coherently split, redirect, and recombine matter waves with laser pulses~\cite{Borde1989,kasevich1991atomic,tino2013atom,berman1997atom,Hogan2009}. The manipulation of matter waves is achieved through the stimulated absorption and emission of photons, driving transitions between two long-lived atomic states. Conventional atom interferometry uses alkali atoms like Rb and Cs, which lack a pair of long-lived states coupled by a suitable optical transition. Instead, a pair of counter-propagating laser beams is used to drive two-photon Raman or Bragg transitions via a third, far-detuned atomic state~\cite{muller2008, kasevich1991atomic}. MAGIS relies on a new variation of light-pulse atom interferometry that takes advantage of long-lived excited states in alkaline-earth-like atoms such as Sr~\cite{graham2013new}. Transitions to these states can be resonantly driven by a single laser beam and are used in some of the world's most precise atomic clocks~\cite{nicholson2015systematic,campbell2017fermi}. Leveraging the advances in this field, the MAGIS detector will employ single-photon transitions on the narrow Sr clock line (${}^1S_0 \rightarrow {}^3P_0$) in a clock atom interferometer~\cite{hu2017atom,Rudolph2020}. \Fref{Fig:AIspacetime} shows a space-time diagram that illustrates the pulse sequence of such an interferometer in a Mach-Zehnder type configuration. This symmetric sequence measures and compares the time each interferometer arm spends in the excited state through the increased phase accumulation rate in the clock state. Thus, a modulation of the light travel time in between the laser pulses changes the relative phase of the two paths, which is then expressed in the population ratio of the output states.

The phase sensitivity of the clock interferometer can be enhanced with additional $\pi$-pulses from alternating directions, in between the beamsplitter pulses. Every new pair of pulses represents an additional measurement of the light travel time across the baseline and linearly increases the phase sensitivity. This enhancement is analogous to the Fabry-Perot cavities employed in laser interferometers such as LIGO, which coherently enhance the sensitivity with each reflection between the mirrors, effectively folding a much larger baseline. In MAGIS, additional $\pi$-pulses can also be applied from a single direction, creating a multi-loop geometry in which the atomic wavepackets oscillate back and forth with some chosen period $T$, resonantly enhancing the sensitivity of the detector at frequency $\sim 1/T$~\cite{Graham:2016plp}. This resonant enhancement comes at the cost of bandwidth, so the pulse sequence needs to be adjusted dynamically to scan over the target frequency range. MAGIS-100 will provide a testbed for developing these advanced atom optics techniques for future generations of MAGIS detectors.

Increasing the space-time area through additional laser pulses is a common tool to boost the sensitivity of atom interferometers~\cite{mcguirk2000,muller2008,muller2009,clade2009,chiow2011,Close:2013,kovachy2015quantum,Mazzoni2015,Kotru2015,Plotkin2018,gebbe2019twin,pagel2019bloch,Rudolph2020}. Recently, several key advances have been made in such large-momentum-transfer (LMT) atom optics techniques. First, LMT atom optics have been successfully incorporated into atom interferometers with long durations of up to several seconds, enabling ultrasensitive atom interferometers in which the atomic wavefunctions are delocalized over macroscopic spatial and temporal scales~\cite{kovachy2015quantum,asenbaum2016phase}. Second, LMT atom optics based on two-photon Bragg transitions have been adapted to dual-isotope atom interferometry~\cite{kuhn2014bose,overstreet2018effective, asenbaum2020atom}.  In one of its operating modes, MAGIS-100 will use such dual-isotope, Bragg-based LMT techniques to search for dark matter (see \Fref{fig:BL-DM-sensitivity} and \ref{Fig:magis-config}). Finally, LMT atom optics have been successfully implemented in clock interferometers (on the $^{1}{S}_0 \rightarrow {}^{3}{P}_1$ transition in Sr), demonstrating state-of-the-art performance~\cite{Rudolph2020}.

The Sr clock transition ($^{1}{S}_0 \rightarrow {}^{3}{P}_0$) offers the potential for dramatic improvement of LMT atom optics compared to conventional atom interferometers. One of the key limitations in scaling up the number of additional pulses is atom loss from residual spontaneous emission. For atom optics based on two-photon transitions, spontaneous emission is suppressed by detuning the counter-propagating lasers from a short-lived excited state. This detuning, and therefore the extent to which spontaneous emission can be reduced, is limited by the available laser intensity. For $^{87}$Sr, the clock transition has a corresponding natural excited state lifetime in excess of $100~\text{s}$~\cite{Boyd2007,muniz2020cavity}. So by contrast, spontaneous emission loss from excited state decay is substantially diminished, while spontaneous scattering from other off-resonant lines is suppressed by terahertz detunings. For the laser intensities available in MAGIS-100, this can support many thousands of LMT pulses before the majority of the atoms are lost to spontaneous emission. Moreover, the increase in total pulse area enabled by clock atom interferometry offers promising potential for improving the fidelities of LMT pulses via optimal quantum control~\cite{Saywell2018}.

Clock atom interferometry with Sr atoms has other advantages over interferometry with alkali atoms. For example, such instruments must use magnetically insensitive $m=0$ states to minimize the effect of magnetic forces on the atoms. However, even the second order Zeeman energy shift in alkali atoms, on the order of kHz/G, requires magnetic field control below the milligauss level for optimal performance. The magnetic susceptibility of the Sr clock transition is approximately 1000 times smaller\footnote{The fermionic isotope has an additional linear susceptibility, but this can be canceled by interrogating atoms with both positive and negative spin projection.}, easing the requirements for magnetic shielding~\cite{taichenachev2006magnetic}.

\subsection{Clock gradiometry}
\label{Sec:Gradiometer}

\begin{figure}
\begin{center}
\includegraphics[width=0.65\columnwidth]{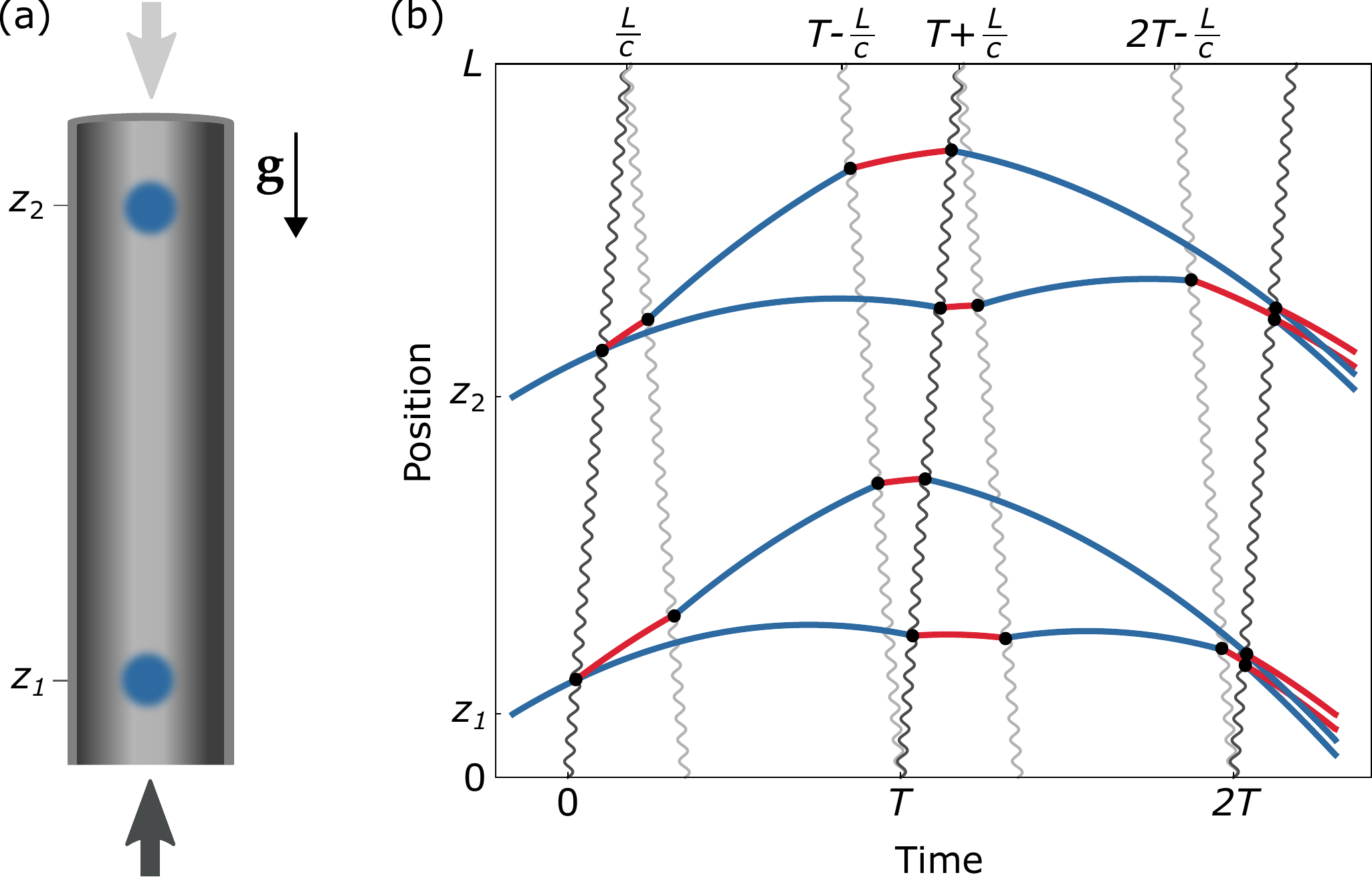}
    \caption{Clock gradiometer. (a) Two dilute clouds of Sr atoms (blue dots) are initially launched from positions $z_1$ and $z_2$, and are freely falling in vacuum under the influence of gravity. Laser light (dark and light gray arrows) propagates between the atoms from either side, creating a symmetric pair of atom interferometers at opposite ends of the baseline. (b) Space-time diagram of the interferometer trajectories based on single-photon transitions between ground (blue) and excited (red) states driven by laser pulses from both directions (dark and light gray). In contrast to \Fref{Fig:AIspacetime}, the pulse sequence shown here features an additional series of $\pi$-pulses (light gray) traveling in the opposite direction to illustrate the implementation of LMT atom optics (here $n=2$).}
    \label{fig:magis-detector-setup}
\end{center}
\end{figure}

In a gradiometer configuration, two identical atom interferometers are run simultaneously on opposite ends of a baseline, using the same laser sources. The phase measured by each interferometer includes an undesirable contribution from laser phase noise. This noise arises from the intrinsic instability of the laser and from the vibration of the delivery optics. However, a comparison of the individual atom interferometer signals yields a differential measurement that enables the cancellation of noise common to both interferometers, such as the laser noise~\cite{mcguirk2002sensitive,Yu2011,graham2013new}. For atom interferometers using two-photon transitions driven by counter-propagating laser beams, laser frequency noise does not exactly cancel due to the asymmetry in the light travel times to the atoms~\cite{PhysRevD.78.042003,Yu2011,graham2013new}. However, in a clock gradiometer such as MAGIS, the laser pulses are derived from a single laser and both interferometers are driven by nominally identical laser pulses (see \Fref{fig:magis-detector-setup}). Thus, clock gradiometry in principle enables superior common-mode rejection of laser frequency noise compared to what is possible with two-photon transitions in a single-baseline configuration\footnote{With two-photon atom optics, it is possible to achieve sufficient rejection of laser frequency noise by using multiple baselines~\cite{tino2013atom,canuel2018MIGA}.}.

The measurement concept described here is closely related to recent proposals to detect gravitational waves and dark matter using two optical lattice clocks separated over a baseline~\cite{kolkowitz2016gravitational, norcia2017role}. Optical lattices circumvent the need to account for phase shifts associated with the motion and the recoil of the atoms. However, in contrast to freely-falling atoms, those trapped in optical lattices do not intrinsically serve as well-isolated inertial references since they are rigidly connected to the sensor frame by the optical lattice trapping potential. Instead, these proposals require an auxiliary inertial reference that can be realized by, for example, placing the optical lattice clocks on drag-free satellites~\cite{kolkowitz2016gravitational}.

\section{MAGIS-100 Detector Design}
\label{sec:magis-100 design}

\begin{figure}
\centering
\includegraphics[width=0.5\columnwidth]{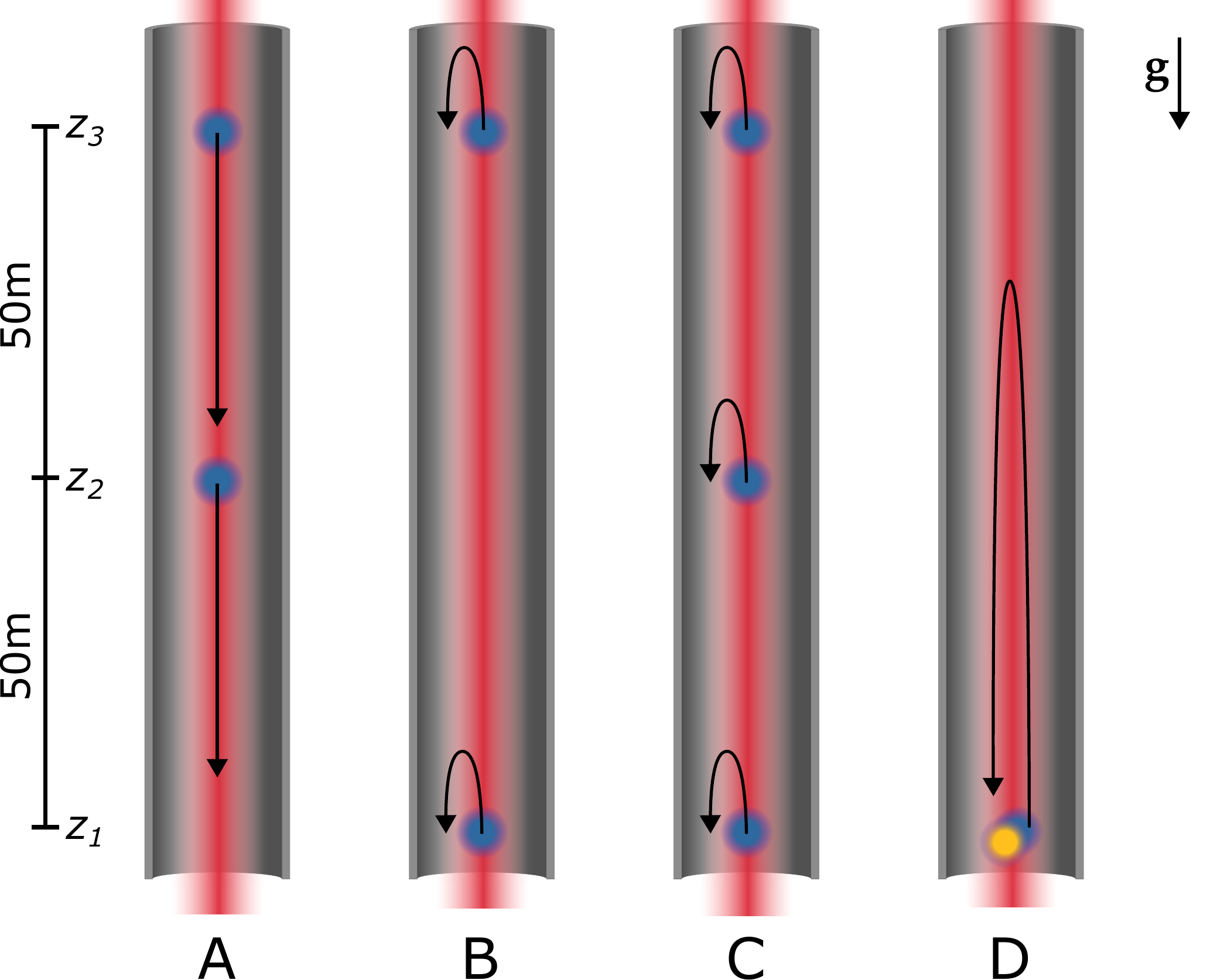}
    \caption{The MAGIS-100 detector features four distinct operating modes (A-D), using the three atom sources that connect to the 100~m vacuum tube at locations $z_1$, $z_2$ and $z_3$. At these locations, atom clouds can be prepared, dropped, launched, and detected. Light pulses (red beam) travel along the vacuum tube in both directions and interact with the atoms (blue clouds) while they are in free fall. \textbf{Mode A:} Maximum gradiometer drop time. Atoms are dropped from locations $z_3$ and $z_2$ over 50~m and are detected at locations $z_2$ and $z_1$, respectively. \textbf{Mode B:} Maximum gradiometer baseline. Atom clouds are launched for several meters from $z_1$ and $z_3$ and then detected at their initial launch positions. \textbf{Mode C:} GGN characterization. All three sources can be used with short launches in order to explore Newtonian noise variation along the baseline. \textbf{Mode D:} Dual-isotope launch. In this alternative dark matter detection mode two Sr isotopes (blue and orange clouds) are simultaneously launched from $z_1$.}
\label{Fig:magis-config}
\end{figure}

The MAGIS-100 detector is a long-baseline atom interferometer, interrogating ultracold atoms in free fall along a 100~m baseline with a vertically propagating laser.  The operation of the detector is cyclic.  In each experimental cycle, an ensemble of atoms is prepared and used to perform light-pulse atom interferometry.  At the end of the interferometer pulse sequence, the resulting matter wave interference pattern is recorded and analyzed to determine its phase. Each detector cycle consists of the following stages: 
\begin{enumerate}
\item Atom ensemble preparation. Atoms are collected and cooled in a multi-step process involving laser cooling, evaporative cooling, and matter wave lensing.  The velocity spread of the ensemble must be sufficiently small to support efficient light-atom interactions and allow for long interferometry times.  To maximize sensitivity, it is important that the cooled ensembles have as many atoms as possible.
\item Shuttle.  The cooled atom cloud is translated from the atom source to the adjoining 100~m interferometry region.  This is implemented using a horizontal optical lattice shuttle. To realize the full MAGIS detector configuration, it is important that two or more atom sources are operated simultaneously (see \Fref{Fig:magis-config}), so that two ensembles of atoms are delivered to the interferometer region at the same time to implement the gradiometer measurement.
\item Launch.  Once in the interferometer region, the atom ensembles are launched vertically and they begin to follow free-fall trajectories.  The launch is implemented by a vertical optical lattice.  The target launch height varies based on the detector mode, as described below.
\item Atom interferometry.  Once the atoms are falling, a series of laser pulses are sent along the vertical axis of the 100~m interferometer region.  These light pulses act as beamsplitters and mirrors for the matter waves.  The number of pulses and their timings depend on the detector mode and may optionally be configured to implement LMT or resonant sequences as needed.
\item Atom detection.  The matter wave interference pattern is imaged using a camera.  The images are formed by shining resonant light on the atoms and then collecting the resulting fluorescence.
\end{enumerate}
\noindent The phase shift extracted from the imaged interference pattern constitutes one data point.  The science signals of interest such as from gravitational waves or dark matter cause time-varying phase shifts in the interferometer. Therefore, the interferometer phase must be sampled at a rate faster than the evolution of the signals (i.e., up to several Hz).  The above preparation cycle must therefore operate at this rate as well.

The MAGIS-100 detector consists of three main components: the 100~m atom interferometer region, the atom interferometry laser system, and the three atom sources.  The interferometer region is where the science measurement is performed and consists of a 100~m vertical vacuum pipe with pressure in the ultra-high vacuum range, along with magnetic shielding and coils along its length to ensure a controlled magnetic field.  The atom interferometry laser system, located at the top of the detector, produces the laser pulses needed for atom interferometry and delivers them to the 100~m interferometer region.  Finally, MAGIS-100 has three atom sources, located at the top, middle, and bottom of the interferometer region.  Each atom source is responsible for producing and delivering a cold ensemble of Sr atoms to the interferometry region at the beginning of each measurement, and for detecting the atoms at the end of the cycle.

The three MAGIS-100 atom sources are distributed along the length of the instrument to enable flexible operation in several different modes (A-D), as illustrated in \Fref{Fig:magis-config}.  These modes are optimized for the various science goals of the experiment.  In mode A, atoms from the upper and middle sources are dropped simultaneously, and fall 50~m before being detected in the middle and lower detection areas, respectively.  This mode allows for a long free-fall time of more than 3~s while maintaining a 50~m baseline length between the atoms.  In mode B, atoms from the bottom and upper sources are simultaneously launched upward for several meters and are later detected at their original locations.  This mode maximizes the baseline to approximately 100~m while still allowing for 1-second-scale interferometer durations.  In mode C, all three sources can also be operated simultaneously in this manner.  Measuring correlations across more than two sources distributed along the baseline in this way can potentially be used to characterize and suppress Newtonian gravity gradient noise (GGN)~\cite{PhysRevD.93.021101} that otherwise likely limits any terrestrial detector at low frequency (see \Sref{sec:seismic-measurements}).  Finally, mode D involves operating simultaneous interferometers using two isotopes of strontium launched together from the bottom atom source.  This mode is indicated for searches for dark matter fields or new forces that couple differentially to the two isotopes, i.e.\ equivalence principle violating effects (see \Sref{sec:DMIntro}). Launching from the bottom atom source also provides the maximum possible free-fall time with launch heights over 50~m, and potentially up to 100~m.

\subsection{Interferometer region}
\label{sec:interferometer region}

\begin{figure}
\centering
\includegraphics[width=1\columnwidth]{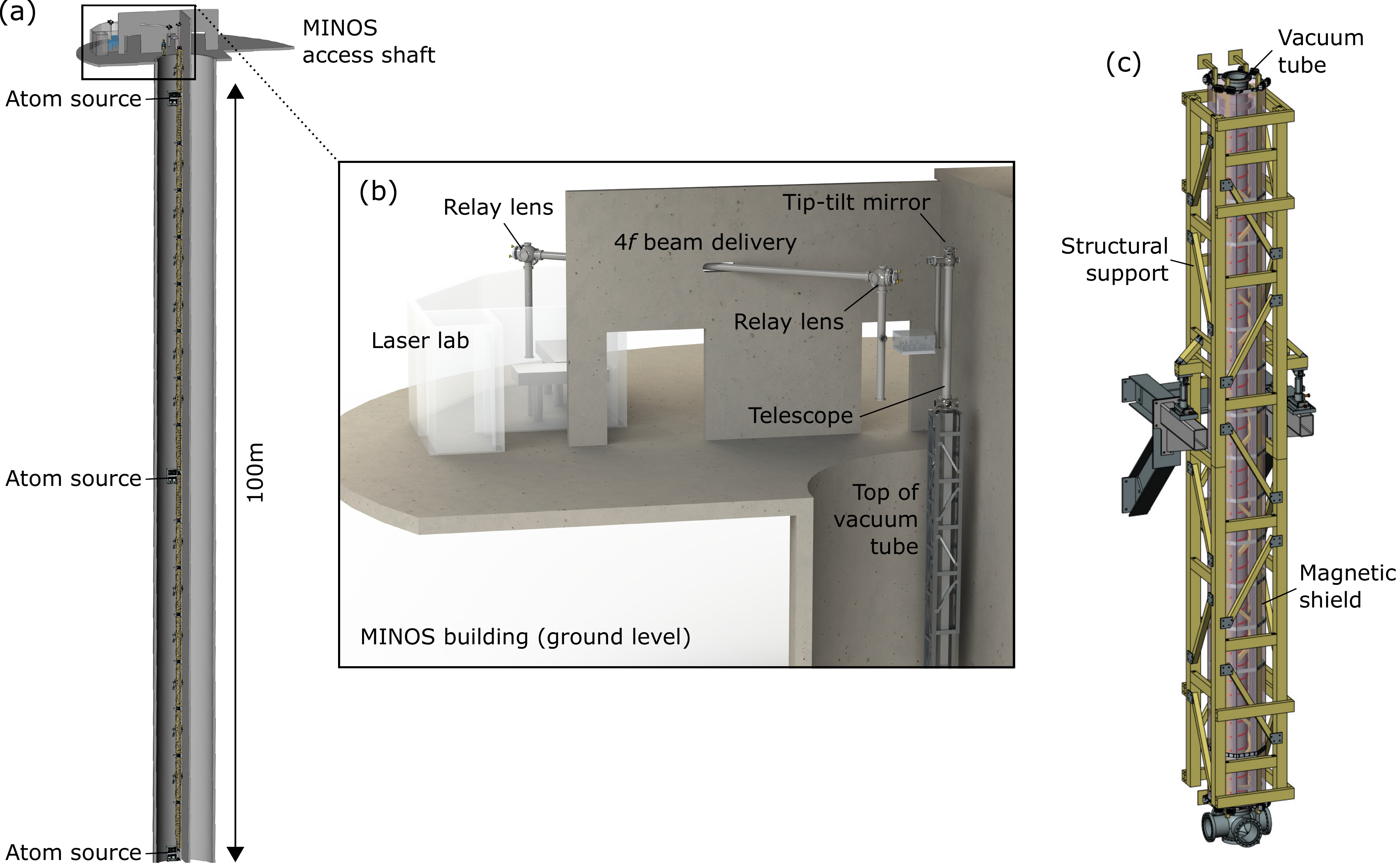}
    \caption{MAGIS-100 detector layout. (a) CAD model of the $100~\text{m}$ interferometer region installed in the MINOS shaft.  The detector is attached to the wall of the $6.7~\text{m}$ diameter underground vertical shaft (shown here in cross section). (b) Close-up of the MINOS building at ground level. The atom interferometry lasers and frequency comb are housed in a temperature-controlled laser lab. Two in-vacuum relay lenses in a $4f$ configuration are used to deliver the interferometry laser beam to the top of the shaft. A short, single-mode optical fiber removes residual pointing jitter and provides initial spatial filtering. Afterwards, the interferometry laser beam undergoes further spatial filtering via in-vacuum, free space propagation and is magnified by a $1:30$ telescope (see \Sref{Sec:AILasers} for details).  (c) CAD model of the MAGIS-100 modular sections. A total of 17 sections span the length of the shaft and are connected end to end. Each 5.3~m long module is mounted to the shaft wall and contains a section of vacuum pipe, vacuum pumps, magnetic shielding, and coils for magnetic field control.}
\label{fig:modules}
\end{figure}

\Fref{fig:modules}(a) is a CAD rendering of the full MAGIS-100 detector, which is to be installed in the existing MINOS underground shaft at Fermilab~\cite{laughton2003construction}.  The central feature of the atom interferometer region is a 100~m vertical vacuum pipe. The targeted pressure in this chamber is in the $10^{-11}~\text{Torr}$ range to permit atom lifetimes of tens of seconds or longer.  This pressure will be maintained with a combination of ion pumps and passive getter pumps placed every \SI{5.3}{m} along the detector.  A mu-metal magnetic shield surrounding the vacuum system is designed to reduce the background magnetic field to below $\sim 1~\text{mG}$.  A set of vertical wire bars inside the shield running the length of the interferometer region produces a horizontal magnetic bias field. This horizontal field orientation is perpendicular to the laser propagation direction, as required by the selection rules for the desired clock transition.

To facilitate assembly and installation, the interferometer region uses a modular design.  The 100~m detector is made up of 17 identical modules, each \SI{5.3}{m} long, attached end to end.  \Fref{fig:modules}(c) is a CAD model of one of these modules. Each module consists of a section of vacuum pipe, magnetic shielding, bias coils, vacuum bakeout hardware, and temperature and magnetic field sensors, as well as a structural support cage that is anchored to the wall.

The modules are joined together with custom vacuum chambers called connection nodes (\Fref{fig:modules}(c), bottom).  The connection nodes have ports where the vacuum pumps are attached, as well as viewports for diagnostic imaging of atoms along the length of the interferometer region.  To allow for thermal expansion during vacuum bakeout of the full system, a short section of vacuum bellows is placed between each module at the connection node.  Furthermore, each vacuum pipe section is clamped to the support cage via the connection node on only one end.  The other end of each pipe section is secured radially by means of a linear bearing assembly that permits cm-scale translation during bakeout.

\begin{figure}
  \centering
  \includegraphics[width=0.75\columnwidth]{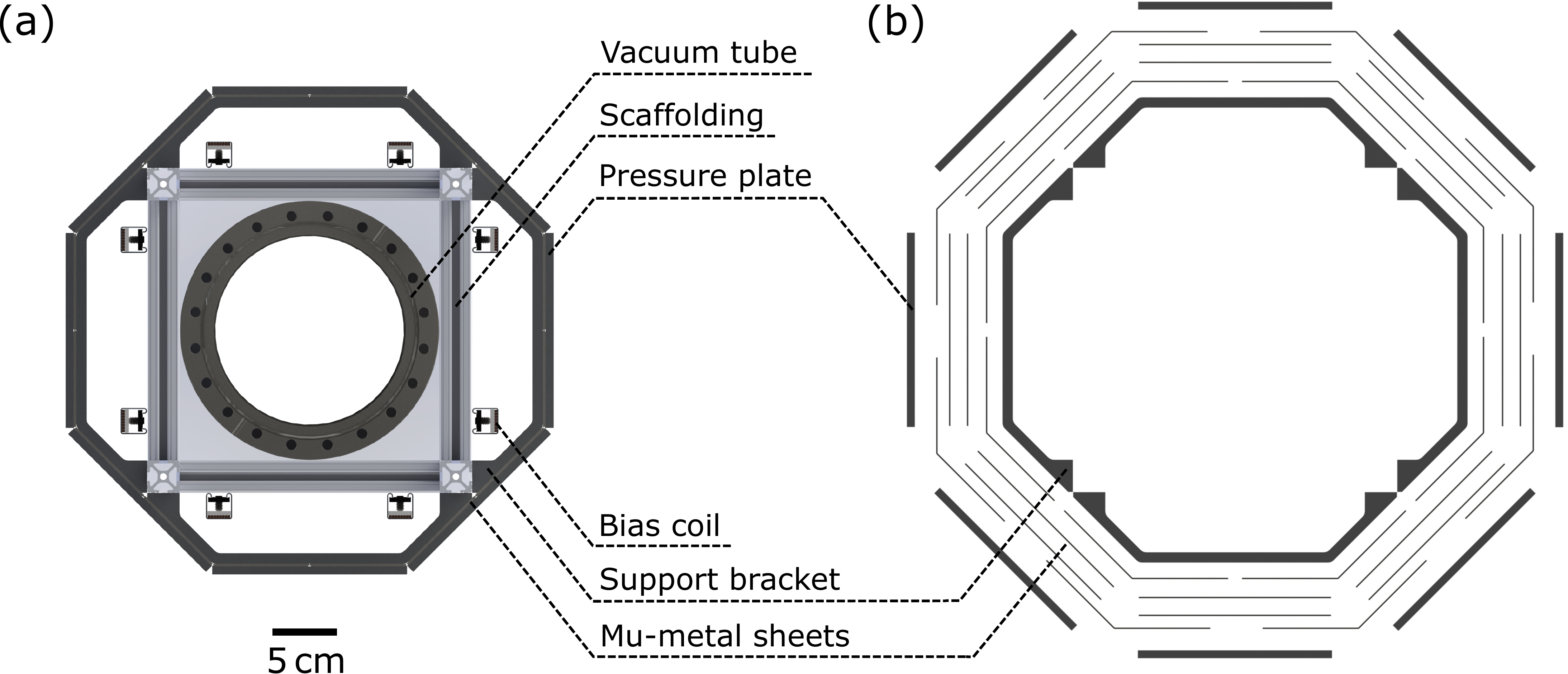}
    \caption{Interferometer region magnetic shield. (a) Cross-sectional view of the interferometer vacuum pipe and surrounding magnetic shield.  The octagonal magnetic shield is supported internally by a square cross-section aluminum scaffolding truss that surrounds the vacuum pipe.  A set of support brackets are attached periodically to this scaffolding, serving as octagonal ribs for the shield.  The shield sheet metal plates attach to these ribs, and pressure plates are applied from the outside to clamp the multiple sheet metal layers together.  Eight vertical bias coil bars, each consisting of magnet wires guided by a channel, attach to the scaffolding inside the shield, providing a uniform transverse magnetic field in the center of the vacuum pipe. (b) Exploded view of the magnetic shield assembly. On each face of the octagon, four sheets of mu-metal are clamped together, with sheets on the corners and on the faces arranged to avoid radial gaps.  The sheets are also staggered vertically (out of the page) to avoid gaps in the axial direction. This pattern of overlapping sheets reduces magnetic field leakage~\cite{wodey2020scalable}.}
\label{fig:Shield}
\end{figure}

The magnetic shield design for MAGIS-100 is a challenge because of its total length, and because of the usual difficulties associated with field leakage in large length-to-diameter ratio shields~\cite{dickerson2012high}.  To improve shield continuity and simplify manufacturing of the mu-metal, the MAGIS-100 shield uses an adaptation of a proven design concept based on a multi-layer stack of overlapping sheets~\cite{wodey2020scalable}.  \Fref{fig:Shield} shows a top-down view of the shield design.  Four overlapping mu-metal sheets are stacked and clamped together to form the single-layer octagon-shaped shield.  The sheets are staggered to avoid radial gaps, reducing field leakage of transverse applied fields.  Similarly, multiple mu-metal sheets are used to cover the length of each 5.3~m modular section, and these sheets are also staggered to maximize overlap and avoid axial gaps.  The magnetic circuit requires good contact between the individual sheets that make up the shield, and this is provided by a set of brackets and corresponding clamping plates that squeeze the sheets together every 30~cm along the shield.  This shield design was validated by extensive 2D and 3D FEA modeling work.

Each module also has a set of vertical magnetic coil bars that run along its length to provide a $1~\text{G}$ horizontal bias field.  In order to maximize the uniformity of the bias field in the center of the vacuum chamber, the positions of the current-carrying bars were numerically optimized to take advantage of image currents in the shield~\cite{bidinosti2014passive}.  A set of four coil bars is sufficient for this purpose, and ensures that the bias field uniformity is compatible with the 1~mG field homogeneity requirement.  For redundancy, the design includes an additional set of four bars oriented to produce a field in the orthogonal direction. Together, these two coil sets can be used to set the bias field in any desired direction in the transverse plane.

\subsection{Atom sources}

The atom sources are responsible for producing the ultracold Sr atom ensembles at the beginning of each cycle of the detector.  MAGIS-100 has three independent atom sources, located at the top, middle, and bottom of the 100-meter interferometer region, facilitating the variety of operating modes shown in \Fref{Fig:magis-config}.  Each of the three atom sources consists of a vacuum chamber that interfaces with the interferometer region, lasers and optics needed to implement the atom cooling and transport, associated control electronics, magnetic coils, an atom imaging system, and an environmental enclosure.

\subsubsection*{Laser system}

\begin{figure}
    \centering
    \includegraphics[width=\textwidth]{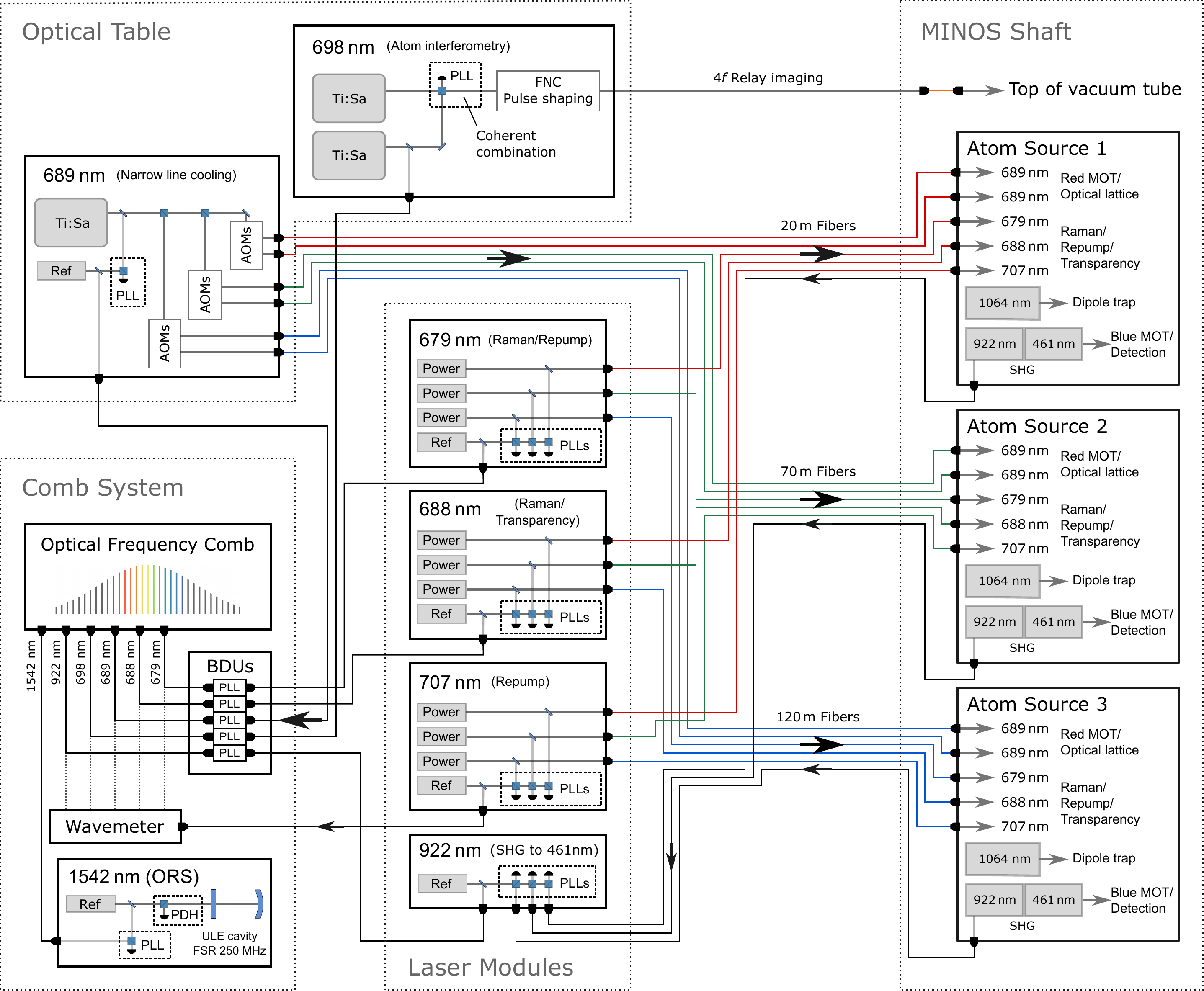}
    \caption{\label{fig:full-laser-schematic}
    Schematic of the MAGIS-100 laser system.  The majority of the laser sources are housed in a temperature-controlled laser lab (left). The cornerstone of the design is a commercial optical frequency comb (Menlo Systems), which is used to stabilize all the lasers needed to operate the detector. One reference laser at each wavelength is stabilized to the comb via a phase-locked loop (PLL).  The comb itself is locked to a cavity-stabilized 1542~nm reference laser (Menlo Systems ORS).  To allow independent frequency tuning of all three atom sources, additional lasers (``Power'') at each wavelength are offset locked to the associated reference laser (``Ref'').  Light is sent to the atom sources in the MINOS shaft (right) via long optical fibers (color-coded red, green, and blue by atom source destination). At 689~nm, a high-power Ti:sapphire laser is offset locked to the reference laser to provide light for three pairs of optical lattice beams for the three sources, as well as laser cooling light.  Pairs of double-passed acousto-optic modulators (AOMs) are used to control the lattice light frequency and amplitude profiles.  The atom interferometry laser system at 698~nm is also stabilized to the frequency comb. The details of this system are illustrated in \Fref{fig:general-laser-system}.  This laser beam is delivered to the top of the shaft via relay imaging (see \Sref{Sec:AILasers}).}
\end{figure}

The MAGIS-100 atom sources require laser light at a variety of wavelengths to serve functions including laser cooling, dipole trapping, atom transport and launch, internal state preparation, and atom detection.  \Fref{fig:full-laser-schematic} is a schematic of the laser system that provides stabilized, frequency-tunable light at all needed wavelengths for the three atom sources.  A commercial optical frequency comb (Menlo Systems FC1500-250-ULN) is used to stabilize all the lasers at the correct frequencies.  The comb itself is stabilized to a commercial optical reference cavity (Menlo Systems ORS) at 1542~nm.  For each required wavelength in the system, an external cavity diode laser (ECDL) at the appropriate wavelength is locked to the comb.  These reference lasers are then used to stabilize additional power lasers dedicated to each individual atom source.  This approach allows for flexible tuning of the light for the independent atom sources, while ensuring sufficient power for a high signal-to-noise ratio error signal for the comb locks.

All lasers at a given wavelength are grouped into common rack-mounted laser frames where the beat note measurements are made to generate the feedback signals for each of the phase-locked loops.  The light from the power lasers is then fiber-coupled and delivered to the three atom sources.  The entire laser system is located in a temperature-controlled laser room at the top of the MINOS shaft.  Long optical fibers ($\sim 100~\text{m}$) are required to send the light from the laser room down to the location of the atom sources installed in the shaft.

The primary cooling and trapping light for Sr is at 461~nm (``blue MOT'').  To avoid the high losses of blue light over long optical fibers, about 800~mW of 461~nm light is produced locally inside each atom source frame by a commercial source via second harmonic generation of a high power 922~nm diode laser (Toptica TA-SHG).  To stabilize this light, a small amount of 922~nm carrier light is sent from each atom source over optical fibers to the comb-stabilized 922~nm reference laser, where beat notes are formed to provide feedback for the phase-locked loops.  The flexibility of these offset locks supports independent, dynamic tuning over hundreds of MHz for each atom source, allowing for the preparation of any of the Sr isotopes (or mixtures).  Light at 461~nm is also used for fluorescence imaging in each atom source, as described below.

Optimal cooling performance on the 461~nm line benefits from repumper light at 679~nm and 707~nm~\cite{boyd2007high}.  These wavelengths are also needed during imaging of the atom interference pattern to return atoms in the excited clock state back to the ground state prior to fluorescence detection.  The 679~nm reference laser is comb-stabilized since this wavelength is also needed for coherent manipulations, including state preparation and Bragg transitions.  
The 707~nm light is strictly needed for repumping, so it has reduced frequency noise requirements compared to all the other wavelengths.  For simplicity, the 707~nm reference laser is stabilized using a wavemeter rather than a comb lock.

Light at 689~nm is needed for narrow-line laser cooling (``red MOT'' phase).  The red cooling light for all three atom sources is derived from a common Ti:sapphire laser offset locked to the comb-stabilized 689~nm reference diode laser.  In addition to cooling, this 689~nm light is also used for the optical lattices.  For each of the three atom sources, two independently tunable beams are delivered over optical fibers to the atoms, supplying the counter-propagating light needed to form the optical lattice potential.  Each lattice beam is controlled in frequency and amplitude using a double-passed acousto-optic modulator.  The differential lattice beam phase is stabilized by a local measurement in each atom source to account for noise introduced in the long fibers.  The same pair of lattice beams are reused for both the vertical lattice launch and horizontal lattice shuttle sequences, as described below.

Light at 688~nm is used in conjunction with 689~nm and 679~nm to allow for coherent three-photon Raman transitions to the $^3P_0$ state during state preparation in some detector modes, as described below.  The 688~nm light is also available to serve as a transparency beam during evaporative cooling~\cite{PhysRevLett.110.263003}.

Each atom source contains a 100~W dipole trap laser system at a wavelength of 1064~nm.  This light is used to form a crossed dipole trap for evaporative cooling.  The trap curvature and depth are dynamically controlled by a reflective spatial light modulator (SLM).  The use of such dynamic optical potentials can be used for rapid and efficient evaporation to high phase space densities in alkaline-earth-like elements~\cite{PhysRevA.93.043403}.  Here, the SLM imprints a phase pattern onto the beam upon reflection, which then yields the desired intensity distribution at the position of the atom trap.  The SLM can also be used to  shape the trapping potential to implement matter wave lensing sequences to collimate the atomic ensemble prior to interferometry~\cite{kovachy2015matter}.

\subsubsection*{Ultracold atom production}

\begin{figure}
  \centering
  {\includegraphics[width=0.98\textwidth]{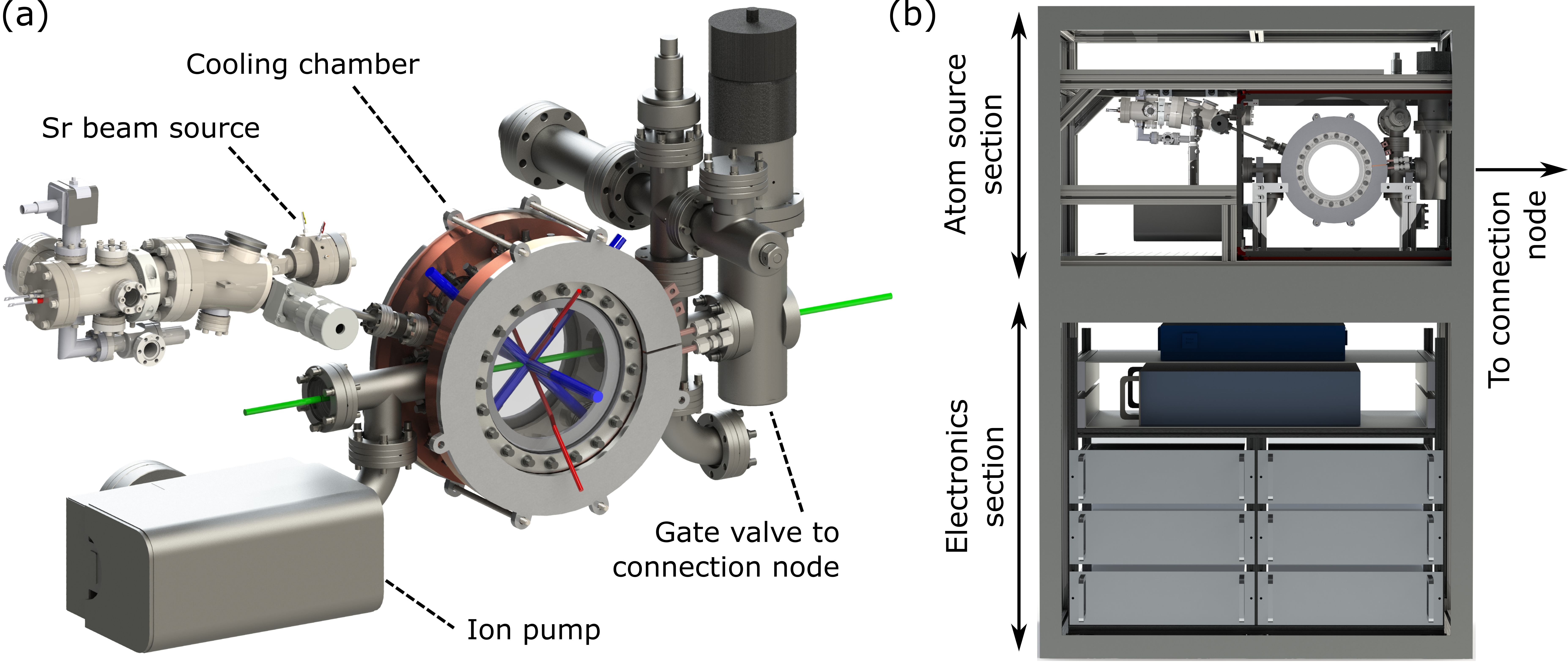}}
    \caption{Atom source CAD model. (a) The atom source vacuum assembly. The Sr beam source produces a beam of atoms that enters the main cooling chamber where atoms are trapped and cooled in a MOT.  Quadrupole magnetic coils (copper orange) surround the large vacuum viewports of the cooling chamber to help produce the MOT.  The 3D MOT beams are shown for reference (blue), as well as other beams for the crossed dipole trap (red), and shuttle optical lattice (green).  An aluminum frame supports the system. (b) Atom source enclosure (access panel doors omitted for interior viewing).  An aluminum, NEMA-rated enclosure ($0.6~\text{m}\times 1~\text{m}\times 0.7~\text{m}$) contains the vacuum system in (a) as well as accompanying rack-mounted electronics below. The enclosure provides a light-tight, temperature-controlled environment that protects the atom source from water and debris. It also houses the 461~nm and 1064~nm lasers for the MOT and dipole trap, respectively, in extendable drawers below the vacuum system.}
\label{fig:AtomSources}
 \end{figure}

The atom ensembles are produced in the cooling vacuum chamber, shown in \Fref{fig:AtomSources}(a).  A commercial beam source (AOSense) consisting of a Sr oven, Zeeman slower, and 2D MOT, delivers an atomic beam to the cooling chamber through a 10~cm long differential pumping stage with a conically expanding aperture (from 3~mm to 7~mm).  During the loading stage, the atoms are captured in a 3D MOT using the 461~nm blue transition.  The atom source also includes a set of quadrupole magnetic coils centered around the cooling chamber for producing the MOT, as well as bias trim coils along three axes to null the background magnetic field.

After reaching temperatures in the millikelvin range, the atoms are then transferred to a red MOT using the narrow linewidth 689~nm intercombination transition and are further cooled to a temperature of a few microkelvin.  Next, evaporative cooling in a crossed optical dipole trap is used to lower the ensemble temperature to the nanokelvin range~\cite{PhysRevA.71.011602,PhysRevA.79.061406,PhysRevA.93.043403}.  As a final step, a matter wave lensing sequence is used to collimate the atoms, a technique that has been demonstrated to produce effective temperatures as low as \SI{50}{pK}~\cite{kovachy2015matter,Rudolph2016Thesis}.

The MINOS shaft varies in temperature throughout the year and suffers from substantial groundwater seepage.  To protect the atom source hardware from water damage and improve temperature stability, each atom source is built inside a custom aluminum enclosure (see \Fref{fig:AtomSources}(b)).  The atom source enclosure accommodates the vacuum system, the laser delivery optics, and support electronics, as well as high power laser sources at 461~nm and 1064~nm. A dedicated closed-cycle chiller provides local water cooling and temperature stabilization for the interior space.  Removable panels on all sides allow access during maintenance.

When the atom sources are installed in the shaft, access for any repairs or optics realignment is challenging, and nominally should be infrequent.  The atom source design is therefore targeted to be robust against misalignment.  A number of mitigations and diagnostics will be used both to minimize drift and to allow for remote realignment.  Where practical, monolithically machined, compact optomechanics will be used, as well as fixed mirror mounts without extra degrees of freedom wherever possible.  Temperature drift will be minimized thanks to the temperature-stabilized enclosure and water-cooled optics breadboards.  Power monitoring beam samplers will be installed after all optical fibers and in other important areas.  For critical alignments, motorized mirror mounts will be installed, along with position sensitive beam monitors.  The atom source enclosure interior will include temperature sensors and a magnetometer for remote monitoring.

Each atom source attaches to the 100~m vacuum region at specialized connection node vacuum chambers, shown in \Fref{fig:LaunchLattice}(a) and \Fref{fig:ImagingRender}. The atom source connection node plays several roles.  In addition to joining together two modules of the 100~m interferometer vacuum system as described in \Sref{sec:interferometer region}, the connection node vacuum chamber is used for the optical lattice launch and final atom state preparation.  It also provides optical access for detection of the atom interference pattern.  These functions are each described in more detail below.

\subsubsection*{Optical lattice shuttle and launch}\label{Sec:launch}

After cooling, the atom ensemble is translated from the atom source chamber to the connection node using a horizontally moving optical lattice shuttle~\cite{schmid2006long}. The optical lattice light is blue-detuned from the 689~nm transition to reduce spontaneous emission loss during the transport~\cite{PhysRevA.82.013615, browaeys2019current}.  

\begin{figure}
  \centering
  {\includegraphics[width=0.95\textwidth]{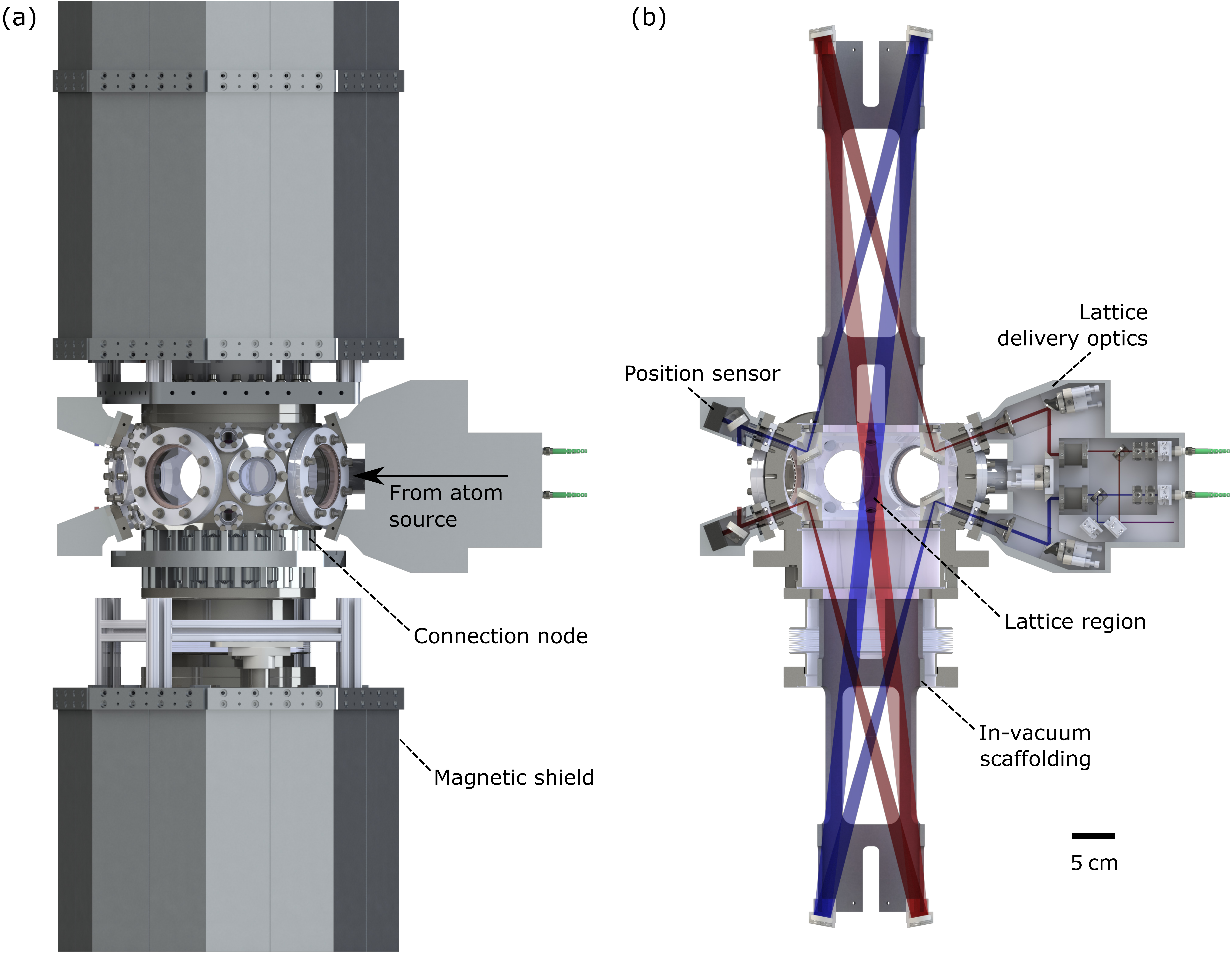}}
    \caption{Connection node and launch lattice optics CAD models. (a) At the bottom of each of the MAGIS-100 modules, below the module's magnetic shield, is a connection node vacuum chamber.  The connection node joins adjacent MAGIS-100 modules and serves as vacuum pump attachment point.  Trim coils are used to compensate for the field inhomogeneity caused by the gap in the magnetic shield.  Specialized connection nodes (one shown here) are used to attach the three atom sources. These connection nodes include in-vacuum optics for the lattice launch, and also have high optical access for atom detection.  Cold atoms are translated to the center of the connection node chamber from the atom source by a horizontal shuttle lattice, and are then launched upward by the vertical lattice before interferometry. (b) Cutaway view showing the in-vacuum lattice scaffolding and delivery optics, as well as the two lattice beam paths (red and blue) which intersect in the center of the chamber to form the diamond-shaped launch lattice region. The blue-detuned 689~nm lattice laser beams are delivered from a compact, stainless steel module containing beam shaping optics and piezo-actuated mirrors. The two beams are directed to a pair of concave mirrors supported by the in-vacuum scaffolding above and below the connection node, which collimate and reflect the light towards the center of the chamber at a shallow angle with respect to vertical. The two lattice beams exit the connection node and are then incident on a pair of position sensitive detectors, which provide feedback to the piezo-actuated delivery mirrors to stabilize the lattice.}
\label{fig:LaunchLattice}
\end{figure}

Once the atoms reach the center of the connection node chamber, the same shuttle lattice light is redirected with an optical fiber switch to form the vertical launch lattice. The atoms are loaded into this lattice and accelerated upwards into the interferometer region. To achieve the launch velocities needed for the desired free-fall times, the two laser beams forming the lattice must overlap over a sufficient vertical distance.  However, a simple vertical counter-propagating lattice aligned along the 100~m interferometer axis is not ideal, since it prevents the three atom sources from operating independently and makes it difficult to launch more than one atom ensemble at a time when operating at higher interferometer sampling rates.  To avoid this, the lattice launch light for each atom source is instead delivered locally, with the two beams intersecting at a shallow angle such that they overlap for about 10~cm in the center of the connection node.

As is shown in \Fref{fig:LaunchLattice}, the lattice light is directed into the connection node vacuum chamber using piezo-actuated mirrors mounted inside a monolithic stainless steel housing attached to the outside of the connection node.  The light is subsequently steered to the center of the chamber at the appropriate angle by a set of in-vacuum mirrors.  An in-vacuum aluminum scaffolding extending 50~cm above and below the connection node chamber supports a set of concave mirrors that collimate the lattice beams and provide the necessary shallow intersection angle.  This lattice delivery and in-vacuum scaffolding approach is designed  to provide good temperature and vibrational isolation for the optics, while also minimizing the gap in the magnetic shield at the connection node.

To adjust the atom launch angle with respect to gravity, the angle of the lattice can be fine-tuned using the piezo-actuated mirrors outside the vacuum chamber.  In particular, one actuated mirror is positioned such that the lattice can be tilted about the connection node center with minimal translation.  On the other side of the connection node, position sensitive detectors are used to sample the lattice light after it exits the vacuum chamber to monitor for lattice beam pointing drift. These position measurements provide feedback to the piezo actuators on the input mirrors to stabilize the position and angle of the lattice. Additionally, we monitor the beat note between the two lattice beams and apply feedback to remove differential phase noise between the beams introduced by the long delivery fibers.

\subsubsection*{State preparation}

After atoms have been cooled to a sufficiently low temperature, they must be prepared in the proper internal state for the operation mode in question.  For clock interferometry on the 698~nm transition, ensembles of $^{87}$Sr atoms will be prepared in the $m_F=\pm9/2$ stretched states~\cite{Takamoto2006}. The choice to operate using both $m_F=+9/2$ and $m_F=-9/2$ states allows for simultaneous, co-located interferometers with opposite-sign linear Zeeman coefficients.  Taking the average of the phase of these two $m_F$ interferometers then suppresses magnetic field dependent phase shifts (up to the quadratic Zeeman response), while the phase difference between them can provide a magnetic field measurement that may be monitored or used for additional corrections.

Alternatively, in another detector mode, Bragg transitions at 679~nm will be used to implement the atom interferometry.  This transition offers advantages for simultaneous interferometry with two isotopes, including reduced spontaneous emission and less sensitivity to isotope shifts (see \Sref{Sec:AILasers}).  To operate in this mode, the atoms must first be transferred to the ${}^3P_0$ state.  However, the bosonic isotopes have no natural coupling on the
${}^1S_0 \rightarrow {}^3P_0$ transition, so instead the atoms will be transferred via a coherent multi-photon Raman process, with ${}^3P_1$ and ${}^3S_1$ as intermediate levels (in practice, using a four-level STIRAP sequence such as in~\cite{sola1999optimal}).

\subsubsection*{Atom detection and phase extraction}
\begin{figure}
\centering
\includegraphics[width=0.8\textwidth]{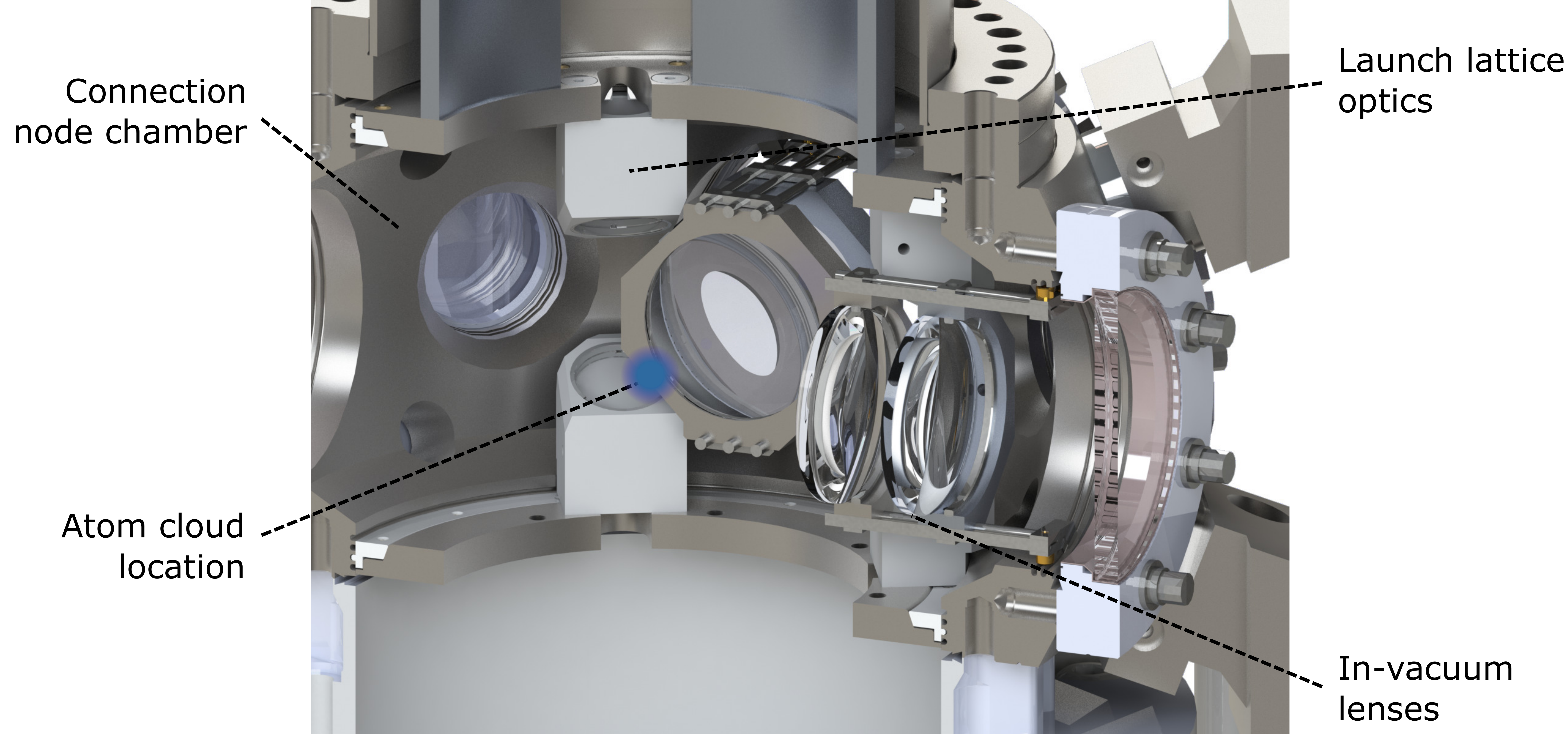}
    \caption{Cutaway view of the detection area inside the connection node vacuum chamber. In-vacuum lenses provide improved light collection efficiency with a large solid angle when detecting dilute clouds (e.g., for long interferometer durations).  The blue ball represents the location of the atom cloud during imaging.  In-vacuum imaging lenses on the orthogonal axis are visible behind the atom cloud. An additional set of four unobstructed viewports around the perimeter of the chamber are also available for imaging with external optics. Also shown are the launch lattice input mirrors, as well as part of the lattice in-vacuum scaffolding (see \Fref{fig:LaunchLattice}).}
\label{fig:ImagingRender}
\end{figure}

The science signal for the detector is encoded in the relative phase between the interferometers.  In order to achieve high phase sensitivity, the detection system for MAGIS-100 is designed to be limited by atom shot noise.  The phase of each interferometer can be extracted from the image of the atom interference pattern acquired using fluorescence imaging on the 461~nm line.  As shown in \Fref{fig:ImagingRender}, the connection node chamber has several large viewports surrounding the atoms to support imaging from multiple directions.  The primary cameras are positioned along two perpendicular axes intersecting at the center of each connection node so that the cloud can be imaged from both directions simultaneously.  To perform detection using phase shear readout, one of the interferometer pulses will be given a small additional tilt relative to the other pulses using the tip-tilt mirror system (see~\Fref{fig:RetroRender}), producing spatial fringes across the atom cloud~\cite{Sugarbaker2013}.  To supplement this detection strategy, two additional viewports are fitted with a set of in-vacuum lenses that are positioned close to the center of the chamber to maximize light collection efficiency (see \Fref{fig:ImagingRender}).  The in-vacuum lenses are well-suited for measuring the total light emitted by the two interferometer ports in situations where the atom density or total atom number may be low, such as after a long free-fall time or extreme LMT sequences.

Numerical simulations of the detection  process and associated sources of noise have been used to inform the design of the imaging system and the required camera specifications. Moreover, these simulations facilitate comparisons of various image analysis techniques.  The simulation translates the distribution of atoms at the end of an interferometer into a simulated image, accounting for the photon scattering rate, realistic imaging times, the diffusion of the atoms during the imaging process, the geometrical acceptance of the optics, the geometry of the pixels in the camera, the quantum efficiency of the pixels, and the camera readout noise.

\begin{figure}[t]
\begin{center}
\includegraphics[width=0.55\textwidth]{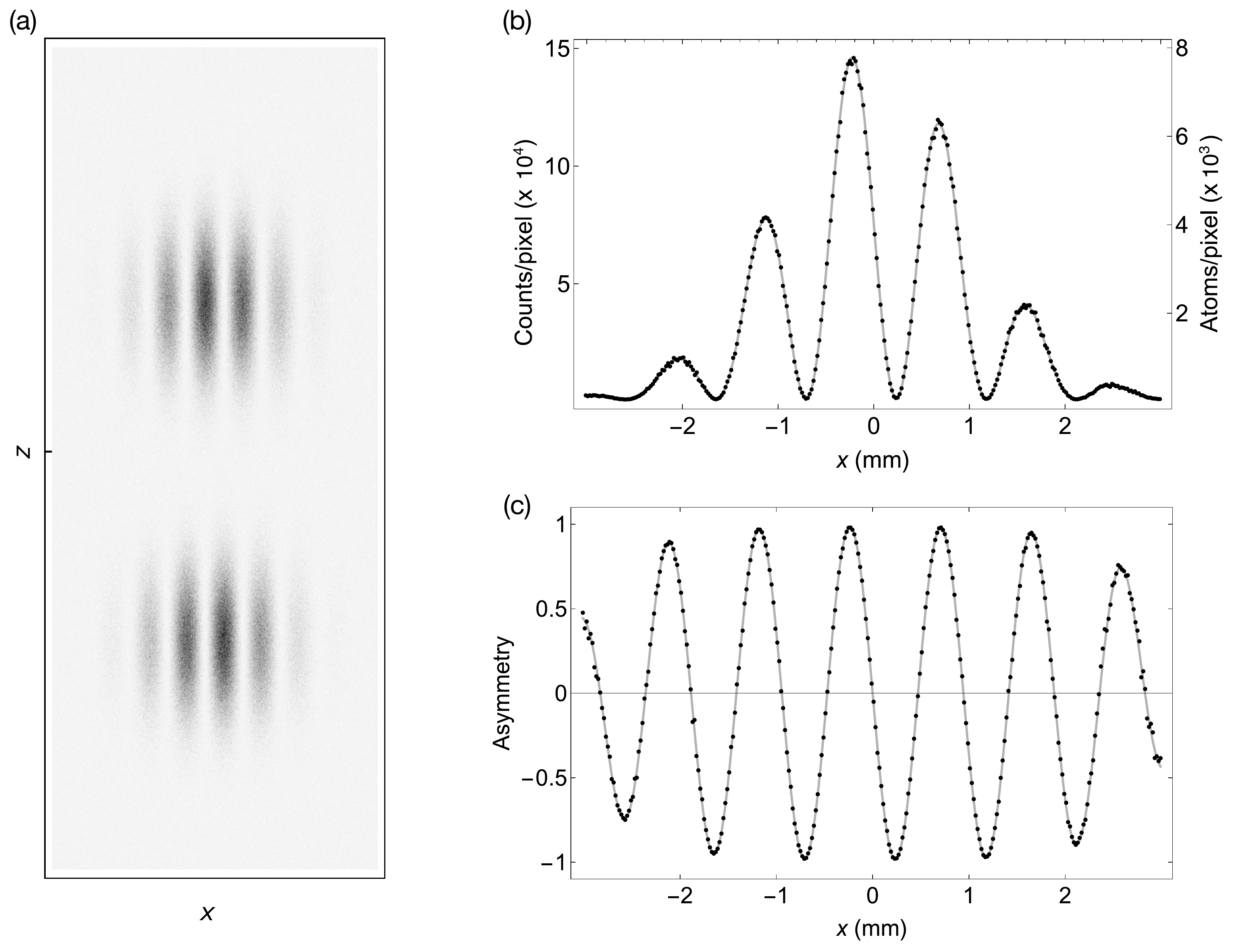}
    \caption{Simulated image of an atom interference pattern in the detection region at the end of a MAGIS-100 interferometer. (a) The simulated image shows the two detected sub-populations corresponding to the two output ports of the interferometer.  The fringes are the result of the phase shear readout technique. (b) The $x$-projection of the upper-half of the pixel plane which contains the image associated with the upper of the two output ports. (c) The two-port asymmetry constructed from the $x$-projections of the two ports. The curves in (b) and (c) panels result from fitting the simulated data to obtain the phase associated with the interference pattern.}
    \label{fig:simulation}
\end{center}
\end{figure}

A representative example of a simulated image is shown in \Fref{fig:simulation}(a). The image is in the ($x$,$z$)-plane, where $z$ points downward along the 100~m interferometer axis. At the location of detection, the atoms are distributed in two sub-populations corresponding to the two output ports, port A and port B.  For each port, the sub-distribution is modelled as a 2D-Gaussian envelope with a sinusoidal fringe pattern imprinted in the $x$-direction. \Fref{fig:simulation}(b) shows the upper half of the image (port A) with counts binned along the $z$-axis, showing the interference fringes along $x$. \Fref{fig:simulation}(c) shows the two-port asymmetry distribution constructed from the difference of the port A and port B fringe distributions. The curves overlaid on \Fref{fig:simulation}(b) and \ref{fig:simulation}(c) are from fits to these distributions, which are used to determine the phase associated with the interference pattern.  The fits to the simulated images yield a phase precision approaching the atom shot noise limit of 1~mrad, as expected for the assumed cloud size of $N=10^6$ atoms.  These simulation results help validate the imaging system design and detection protocol. However it is important to note that the detection noise simulated here does not account for other sources of interferometer phase noise such as those discussed in \Sref{Sec:systematics}.

\subsection{Atom interferometry laser system}

\label{Sec:AILasers}

The MAGIS-100 atom interferometry laser system is designed to deliver a laser beam with high optical power, a stable absolute frequency, a spatial mode with minimal wavefront aberrations, and stable and dynamically adjustable pointing. In order to realize LMT atom interferometry based on the 698\;nm clock transition of strontium, the laser system provides 8\;W of CW power at this wavelength. This high power enables a Rabi frequency of several kHz on the 698\;nm clock transition with a cm-scale laser beam waist.  It is beneficial to have sufficient power to reach Rabi frequencies at this level in order to minimize $\pi$-pulse inefficiencies arising from nonzero detunings, which can be caused by laser frequency noise or the Doppler spread of the atom cloud.  For a small detuning $\delta$, the $\pi$-pulse transfer efficiency is $P_e \approx 1- \frac{\delta^2}{\Omega^2}$, where $\Omega$ is the Rabi frequency~\cite{metcalf2007laser}.  To further minimize pulse inefficiencies arising from laser detuning errors, the absolute frequency of the laser will be stabilized to within $\sim 10$\;Hz of the clock transition resonance using the optical frequency comb (\Fref{fig:full-laser-schematic}).  In principle this level of laser stability allows for up to $1000 \hbar k$ atom optics with less than one percent total loss. Auxiliary spectroscopic measurements will be incorporated to correct for long-term drifts.  Pulse efficiency can be further enhanced via composite pulses~\cite{Butts2013}, adiabatic rapid passage~\cite{Kotru2015}, or optimal quantum control~\cite{Saywell2018}.

\begin{figure}
    \centering
    \includegraphics[width=1.0\textwidth]{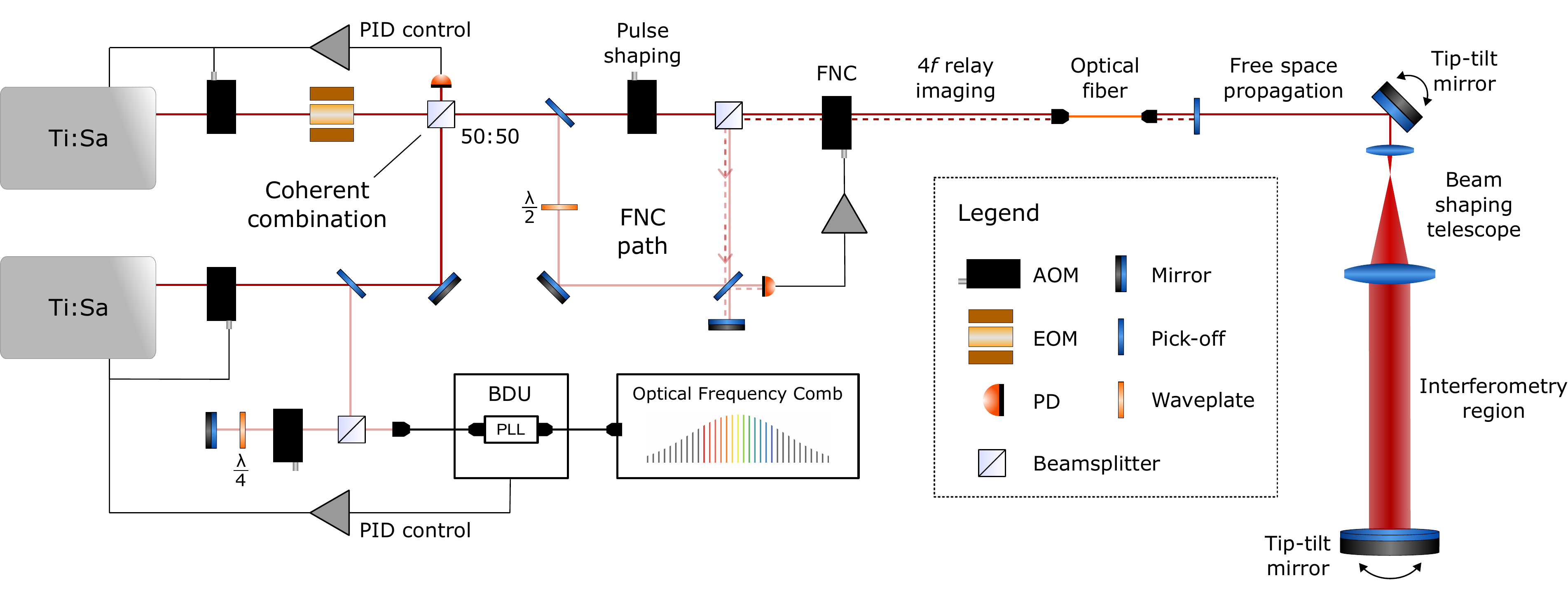}
    \caption{Simplified conceptual schematic of the MAGIS-100 interferometry laser system. On the left, two Ti:sapphire (Ti:Sa) lasers are coherently combined. Acousto-optic modulators (AOMs) are used for frequency control and pulse shaping, and an electro-optic phase modulator (EOM) is used to generate an error signal for the coherent combination.  The laser is referenced to a frequency comb (see \Fref{fig:full-laser-schematic}) with an offset controlled by a double-passed AOM.  This enables rapid frequency tuning over $200~\text{MHz}$, which is required to account for Doppler shifts of the atoms.  A path for generating a continuous fiber noise cancellation (FNC) error signal bypasses the AOM used for temporal pulse shaping of the main beam.  After being relay imaged and sent through a short optical fiber, the laser light is guided by a tip-tilt mirror through a 1:30 telescope and into the interferometry region. A retroreflecting mirror is positioned at the bottom for rotation compensation and to allow pulses of light to travel in both directions. In an alternate operating mode, the two lasers are separately offset locked to the comb (instead of coherently combined) to drive AC-stark-shift-compensated Bragg transitions~\cite{kovachy2015quantum} at 679\;nm for dual-isotope dark matter searches.}\label{fig:general-laser-system}
\end{figure}

\Fref{fig:general-laser-system} shows a conceptual schematic of the interferometry laser system.  In order to achieve the desired power, two commercial Ti:sapphire lasers (M Squared SolsTiS) are coherently combined on a 50:50 beamsplitter.  One laser is stabilized to the frequency comb (see \Fref{fig:full-laser-schematic}), while the second laser is phase locked to the first in order to achieve the coherent combination. The lasers are mounted on an optical table inside a temperature-controlled laser lab.  The laser lab location (see \Fref{fig:modules}) is dictated by building infrastructure constraints, and is approximately 10~m away from the top of interferometer region.  As shown in \Fref{fig:modules}, the laser beam is relay-imaged from the laser lab to the top of the shaft via two in-vacuum lenses in a $4f$ configuration. In-vacuum relay imaging is used instead of a long optical fiber to avoid loss from stimulated Brillouin scattering~\cite{Kobyakov2010}. The relay-imaged beam passes through a short, single-mode optical fiber which provides initial spatial filtering and removes any residual pointing jitter from earlier  optics. A feedback loop will actively stabilize the laser power after the fiber, and fiber noise cancellation~\cite{Ma1994} will be used compensate for any frequency noise introduced by the fiber. The laser beam's spatial mode is then further cleaned via in-vacuum, free space propagation so as to provide a uniform laser wavefront at the location of the atoms, as discussed below.  All subsequent optics that manipulate the beam must be high-quality in order to minimize any beam perturbations introduced after this spatial filtering.  An in-vacuum telescope then expands the laser beam to a waist of $\sim 1$\;cm before interaction of the light with the atoms.  \Fref{fig:modules} illustrates this beam delivery path.  The laser beam is retro-reflected by a mirror at the bottom of the 100-meter-tall vacuum pipe.  Piezo-controlled tip-tilt mirrors at the top and bottom of the shaft provide precise and dynamical laser pointing control, as described below~\cite{Lan2012,Dickerson2013,Sugarbaker2013,asenbaum2020atom}.

To implement LMT pulse sequences, sequential laser pulses are applied from alternating directions. Although the atom optics laser is delivered only from the top, pulses traveling from the bottom to the top can also be realized because of the retro-reflection mirror at the bottom of the tube. This mirror ensures that all laser pulses propagate in both directions. Whether an atom absorbs/emits an upward or downward traveling pulse is determined by setting the frequency of the pulse, since the two cases have opposite Doppler shifts.

In addition to driving single-photon transitions on the 698\;nm clock transition, the same laser system can be readily reconfigured to implement single-photon transitions on the intermediate linewidth 689\;nm $^{1}S_0$ $\rightarrow$ $^{3}P_1$ transition~\cite{Rudolph2020} or AC-stark-shift-compensated~\cite{kovachy2015quantum} two-photon Bragg transitions~\cite{Mazzoni2015} on the broad 679 nm $^{3}P_0$ $\rightarrow$ $^{3}S_1$ transition.  For a single atom source, Bragg transitions or single-photon transitions on the 689\;nm line can be used effectively for overlapped dual-species experiments (such as searches for time-varying new forces).  Bragg or single-photon 689\;nm atom optics are needed when comparing two isotopes of Sr in MAGIS-100, as the 698\;nm clock transition is naturally coupled only in the fermionic $^{87}\text{Sr}$ isotope~\cite{boyd2007high}.

\subsubsection*{Laser wavefront aberrations}

\label{Sec:WavefrontAberrationsMainText}

Aberrations in the atom optics laser beam result in laser phase shifts that depend on the position of the atom cloud with respect to the aberrations.  Since the atom ensembles are at different heights, aberrations can vary from one atom ensemble to the other due to diffraction, leading to unwanted differential phase shifts that can be a source of noise if either the initial atom kinematics (position, velocity, and temperature) or the wavefront aberrations vary in time. These effects have been extensively analyzed in the context of an atomic gravitational wave detector~\cite{hogan2011atomic} as well as more generally~\cite{Wicht2005,Gibble:2006,Schkolnik:2015,Zhou:2016,Karcher_2018,Bade:2018}, and a study of higher order corrections to this previous treatment is presented in \ref{appendix:wavefront}.  Further details can be found in \Sref{Sec:WavefrontAnalysis}.

Spatially resolved detection of the atom interferometer phase at different points within the atom cloud is a powerful tool for mitigating phase errors from wavefront imperfections, providing in situ information about wavefront perturbations~\cite{Dickerson2013,Sugarbaker2013}.  Such in situ wavefront characterization enables temporal variations in the wavefront to be tracked~\cite{hogan2011atomic}.  Moreover, spatially resolved detection also allows phase shifts arising from the coupling of initial atom kinematics to wavefront aberrations to be determined and corrected for by measuring the initial cloud position, velocity and temperature for each experimental shot~\cite{Dickerson2013,Sugarbaker2013,kovachy2015quantum,asenbaum2016phase,overstreet2018effective}.  In correcting for these phase shifts, it is possible to discriminate between initial position and initial velocity fluctuations by making an independent position measurement prior to the start of the interferometer.  This position measurement can be done, for example, by taking an absorption image of the cloud that only destructively measures a small fraction of the atoms.

\label{Sec:FreePropModeCleaning}

In addition, as discussed above, MAGIS-100 will use free space propagation of the interferometry laser beam to reduce laser wavefront aberrations. As the beam propagates in free space, aberrations imprinted on the profile of the beam by imperfect optical components diffract away from the propagation axis, filtering the spatial mode of the beam. To guide the design of the free space spatial filter region, we study the filtering of beam aberrations in MAGIS-100 by numerically evaluating the Rayleigh-Sommerfeld diffraction integral~\cite{Shen2006}. Approximate analytic scaling relationships are generated by evaluating the diffraction integral in the limit of paraxial optics~\cite{Siegman1986}.  The numerical simulations show good agreement with the analytical solutions in a variety of test cases, including a localized defect simulated by an absorptive screen, circular apertures, and intensity/phase gratings.

\begin{figure}
\centering
\includegraphics[width=0.85\textwidth]{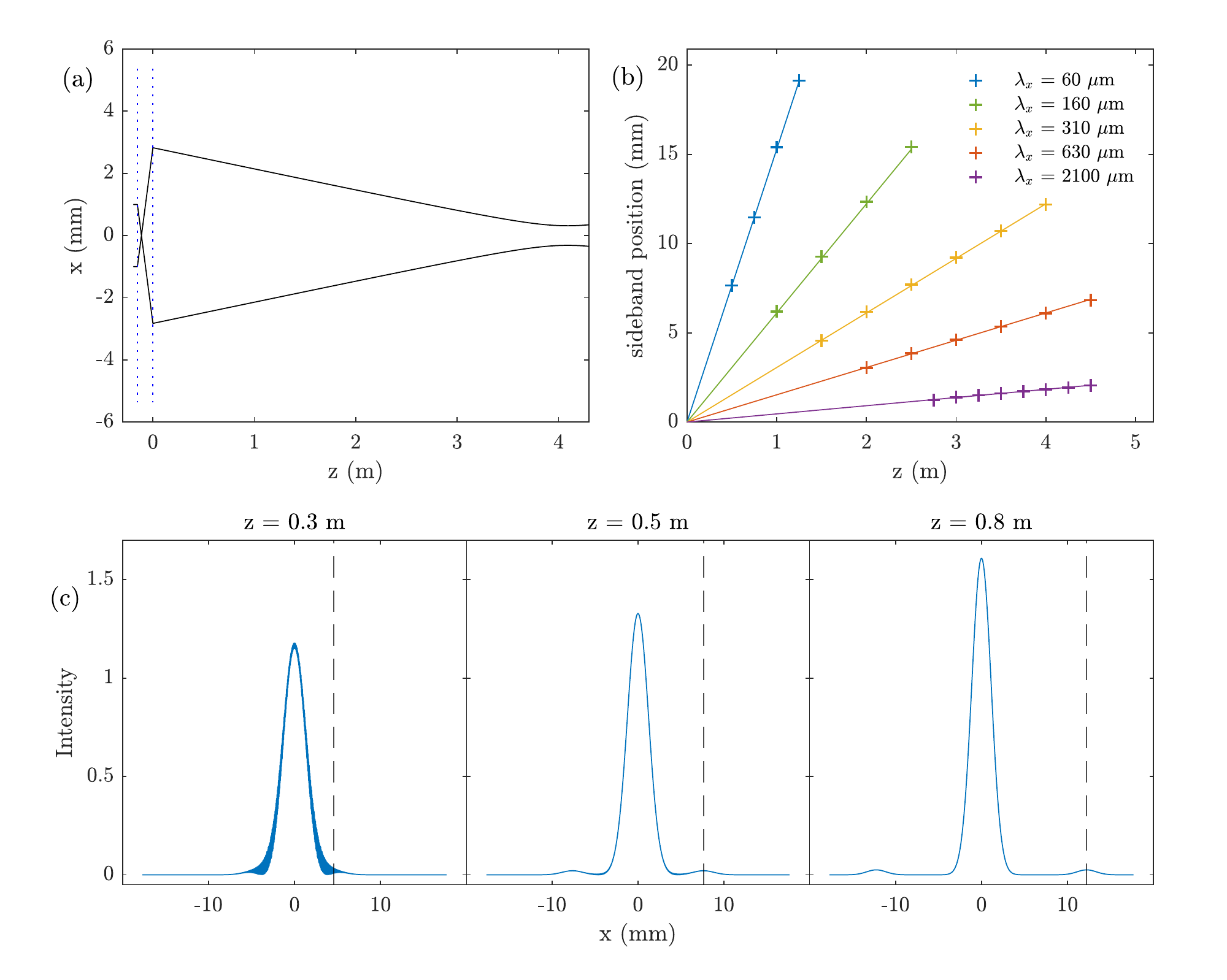}
    \caption{Simulation results demonstrating spatial filtering via free space propagation. (a) A pre-vacuum telescope (lens positions at dotted lines) generates a beam that focuses down to a waist of 300~$\mu$m in 4~m (before the 1:30 magnifying telescope). (b) Perturbations with different transverse wavelength $\lambda_x$ diffract out of the laser beam at different rates. Numerical results (indicated by +) for the positions of the diffracted side-peaks arising from the perturbation vs. distance for various values of $\lambda_x$ are compared to the expected linear scaling with propagation distance (solid lines). (c) Numerical simulation (blue) of the focusing beam in (a) with a sinusoidal intensity perturbation ($\lambda_x=60\ \mu$m) introduced right after the pre-vacuum telescope. The vertical axis is normalized to a peak intensity of 1 right after the pre-vacuum telescope.  From left to right are intensity cross-sections of beams propagated for 0.3 m, 0.5 m and 0.8 m respectively, illustrating the diffraction of the side-peaks out of the main beam as the propagation distance increases. The analytical positions of the diffracted side-peaks are denoted by the dashed lines in (c).}
\label{side}
\end{figure}

\begin{figure}
\centering
\includegraphics[width=0.95\textwidth]{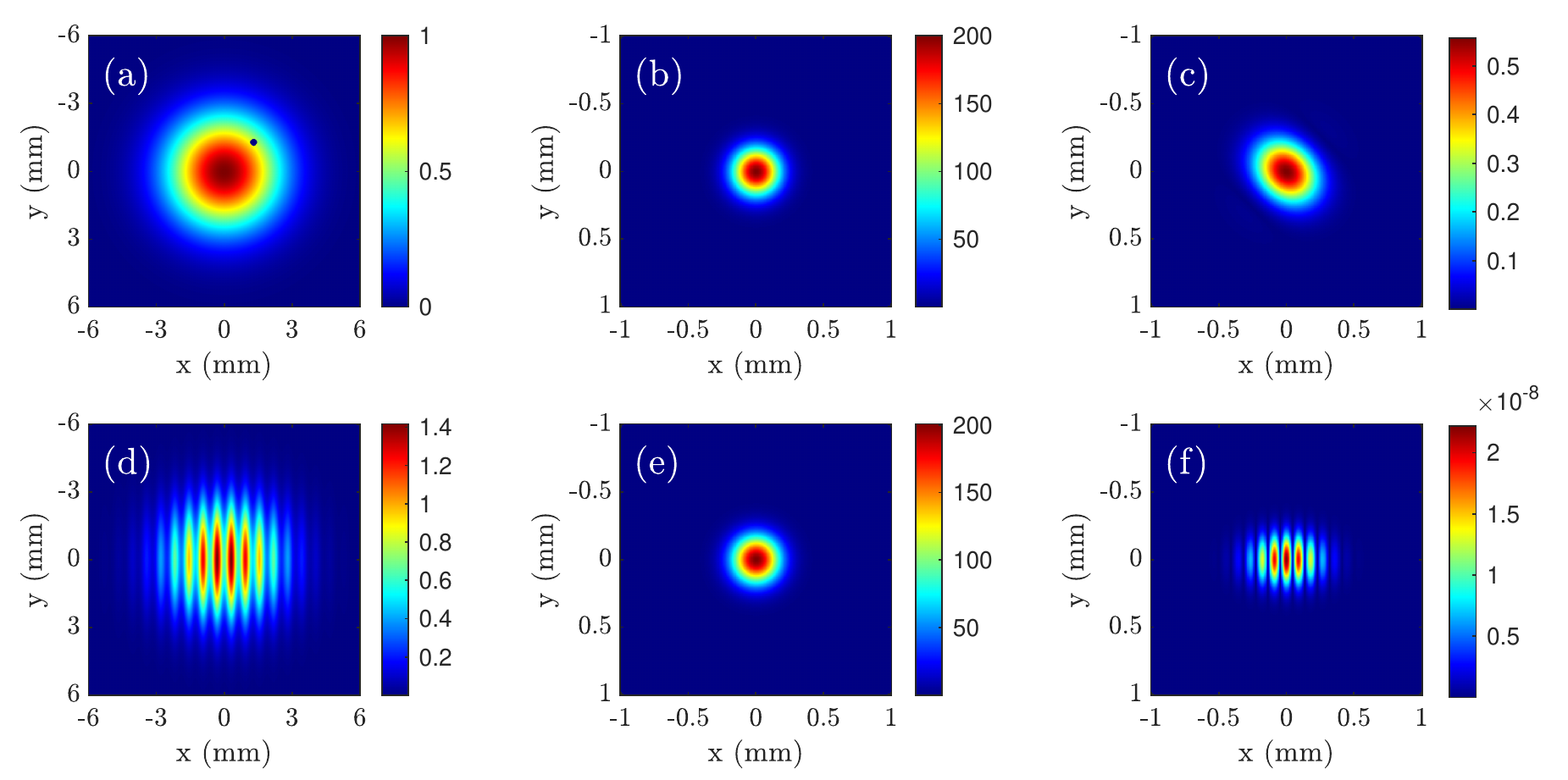}
    \caption{Simulation results illustrating spatial mode cleaning of the laser beam profile via free space propagation for two example beam perturbations.  The color indicates the local beam intensity (arbitrary units).  As in \Fref{side}, an initial Gaussian beam is focused down to 300~$\mu$m after 4~m propagation. (a) Intensity of initial beam incident on a small absorptive screen and (b) after propagation. (c) Difference between (b) and the ideal Gaussian beam after propagation. (d) Intensity of initial beam incident on an intensity grating and (e) after propagation. (f) Difference between (e) and the ideal Gaussian beam after propagation.  The small residual difference in (f) is from numerical effects related to the finite grid size used in the simulation.}
\label{fbcomp}
\end{figure}

Imperfect optics can imprint phase and amplitude aberrations onto the profile of the laser beam.  These beam perturbations can be decomposed into spatial frequency components, and the effect of each Fourier component can be evaluated individually. A sinusoidal phase or amplitude aberration of the laser beam with transverse spatial frequency $k_x=2\pi/\lambda_x$ will diffract out of the main beam into side-peaks, where $\lambda_x$ denotes the transverse wavelength of the perturbation.  As shown in \Fref{side}, these side-peaks correspond to copies of the main beam propagating at an angle $\lambda/\lambda_x$ from the main beam's propagation axis (here the $z$-axis)~\cite{Siegman1986}. Assuming a perturbation generated at $z=0$, the radial offset $r$ of these diffracted side-peaks is therefore $r=z \lambda/\lambda_x$. Significant cleaning of the beam mode will occur when the radial offset of the side-peaks is greater than the 300~$\mu$m waist size of the interferometry laser beam prior to the 1:30 telescope, indicating that the side peaks have significantly diffracted out of the main beam (\Fref{side}). For example, free propagation before the telescope of a distance 4~m effectively filters Fourier components of transverse wavelengths less than $\lambda_x \approx 8.8$~mm. The telescope will subsequently magnify by a factor of 30 any remaining perturbations that have not completely diffracted out of the beam.  This magnification is beneficial because larger scale wavefront perturbations are generally less of a concern for the interferometer (see \Sref{Sec:WavefrontAnalysis}).  To complement this Fourier-based analysis, \Fref{fbcomp} shows the effect on the wavefront of two example beam aberrations.  We also find that spherical aberration from the 1:30 magnifying telescope appears to be negligible.

\subsubsection*{Tip-tilt mirrors and rotation compensation} \label{Sec:TipTiltMirrors}

Tip-tilt mirrors at the top and bottom of the shaft will be used to dynamically control the pointing direction of the interferometry laser beams. These mirrors can be used to mitigate the effect of the rotation of the Earth, which leads to Coriolis forces that cause velocity-dependent interferometer phase shifts. The effect of the Coriolis force can be compensated by using the mirrors to counter-rotate the atom optics laser beam, opposite the direction of the rotation of the Earth~\cite{Hogan2009, Lan2012, Dickerson2013}. Additionally, the ability to dynamically tune the beam angle enables phase shear readout~\cite{Sugarbaker2013}, which can be beneficial for low-noise atom interferometer phase extraction. Alternatively, multi-loop interferometers with highly suppressed sensitivity to rotations can be used in order to alleviate the need for rotation compensation and the corresponding beam deflection~\cite{dubetsky2006atom, hogan2011atomic}.  Rotation compensation can also be used in conjunction with multi-loop interferometry.

\begin{figure}
\centering
\includegraphics[width=0.8\textwidth]{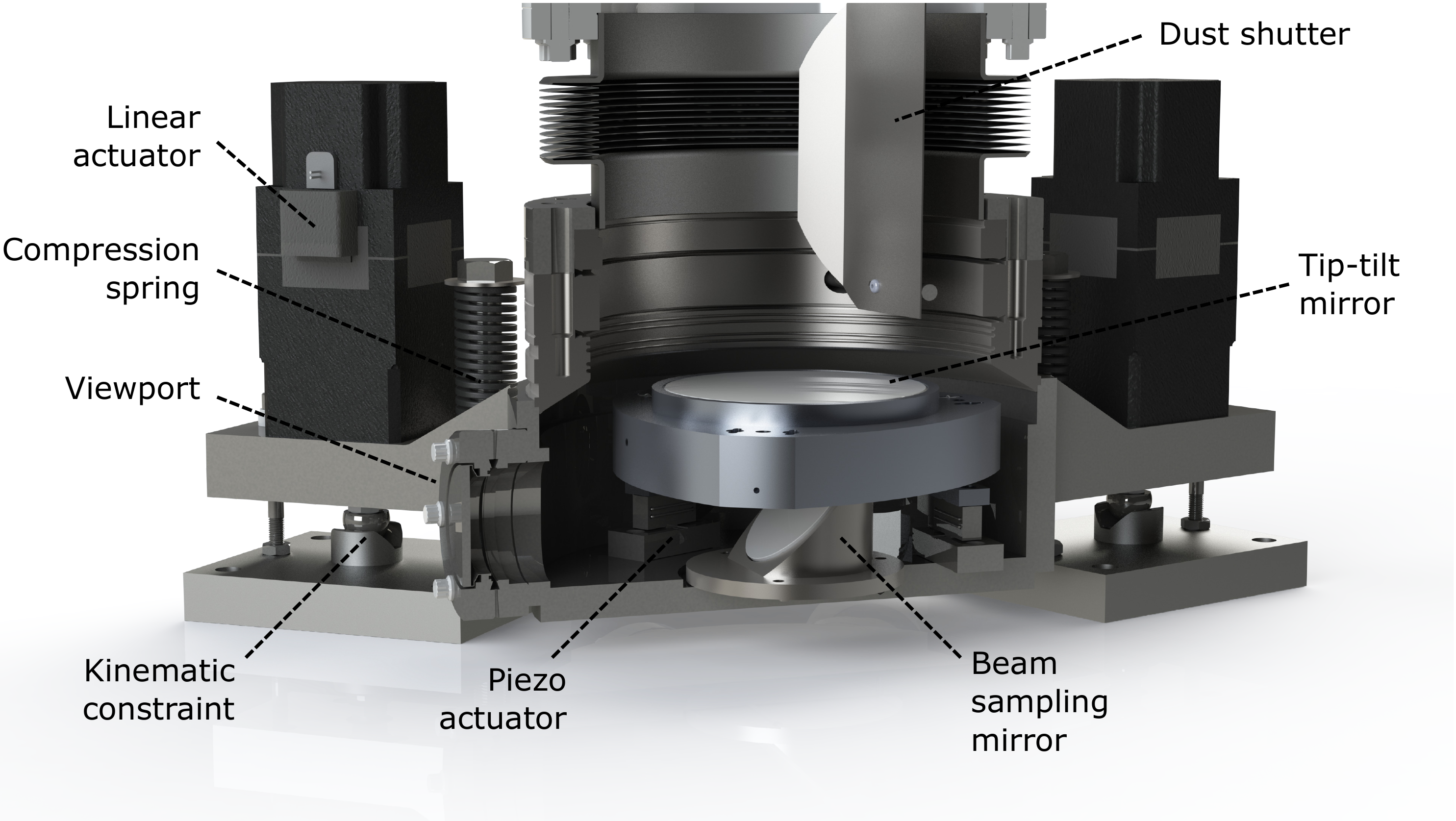}
    \caption{A cutaway view of the tip-tilt mirror system CAD model at the bottom of the 100~m interferometer region.  The mirror angle is controlled by three in-vacuum piezoelectric actuators, together allowing 1.3~mrad angle tuning range.  The inclination of the vacuum chamber itself can be coarsely adjusted by several degrees using three stepper motor linear actuators mounted to the outside of the chamber.  A set of three compression springs, one near each stepper motor, compensates for the force of atmospheric pressure on the chamber and preloads the linear actuators with positive axial force.  An in-vacuum shutter flap is situated above the mirror to protect its surface from dust during initial bake-out.  The mirror angle can be directly measured from the bottom using light reflected off the back surface of the mirror into a position sensitive detector, enabling closed-loop control.}
\label{fig:RetroRender}
\end{figure}

\Fref{fig:RetroRender} shows the design of the bottom tip-tilt mirror stage for MAGIS.  The mirror is mounted in a tripod configuration, with three piezoelectric-actuated stages under the mirror to control the two angles and the offset height of the assembly.  The piezo actuators are kinematically mounted to the underside of the custom mirror mount plate using bearings and springs to allow the plate to rotate freely as the piezos extend.  The mirror and piezoelectric actuator stages are mounted inside a vacuum chamber connected to the bottom of the 100~m interferometer region.  This avoids the need for a vacuum viewport, which could otherwise cause aberrations on the interferometer laser beam.  To cope with the limited tuning range of the piezo actuators, coarse tuning is provided by a set of three external electromechanical linear actuators.  A short section of vacuum bellows allows these linear actuators to articulate the entire tip-tilt stage vacuum chamber with respect to the 100~m vacuum pipe during initial alignment (the piezo actuators alone are sufficient for dynamic control during atom interferometry).

\section{Detector Systematics}
\label{Sec:systematics}

In addition to the science signals of interest, atom interferometers are sensitive to many other effects that can act as noise backgrounds.  It is important to understand the impact of these effects on interferometer phase in order to develop a suitable detector design.  The theoretical tools discussed in~\cite{Hogan2009,Antoine2003,bongs2006high} provide a framework for carrying out such an analysis.

Background modeling for MAGIS-100 is based on extensive, previously published analysis of noise sources in atom interferometry. The relevant sources of noise for MAGIS-100 include laser frequency noise~\cite{graham2013new}, laser wavefront aberrations~\cite{Wicht2005,Gibble:2006,hogan2011atomic, Dickerson2013, Schkolnik:2015, Zhou:2016, Karcher_2018, Bade:2018,Hensel2021}, seismic vibration~\cite{graham2013new,asenbaum2016phase}, Coriolis effects arising from Earth's rotation~\cite{Dickerson2013, Sugarbaker2013,dimopoulos2008atomic,hogan2011atomic}, laser pointing jitter~\cite{Dickerson2013, Sugarbaker2013,hogan2011precision, hogan2011atomic,graham2017mid}, AC Stark shifts~\cite{kovachy2015quantum}, variations in initial cloud kinematics~\cite{Dickerson2013, Sugarbaker2013,overstreet2018effective,dimopoulos2008atomic, hogan2011atomic}, mean field shifts~\cite{swallows2012operating, Hensel2021}, variations in magnetic fields~\cite{hogan2011atomic, graham2017mid,falke201187sr,dickerson2012high}, and blackbody radiation shifts~\cite{graham2017mid,falke201187sr,haslinger2018attractive}.  These effects are all well-understood, and many of them have been studied preliminarily in a 10-meter scale apparatus~\cite{Dickerson2013,asenbaum2016phase,Sugarbaker2013,kovachy2015quantum,overstreet2018effective}. MAGIS-100 benefits from the use of a gradiometer configuration in which differential phase measurements are made between atom interferometers separated over a baseline, which substantially suppresses many of these noise sources as a common mode.  Many long baseline atom interferometry groups have carried out similar analysis and modeling for their specific experiments. They include MIGA~\cite{canuel2018MIGA}, the proposed ELGAR~\cite{Canuel2019ELGAR}, ZAIGA~\cite{ZAIGA2020}, and the AION program~\cite{Badurina_2020}.

This section describes anticipated noise sources for MAGIS-100 and some of the strategies used in the detector design to minimize their impact.  This noise analysis translates into experimental requirements, as summarized in \Tref{tab:parameters}.  An important aspect of the MAGIS research program is experimentally studying these noise sources and the associated mitigation strategies in a long-baseline atom interferometry configuration.  Modeling and analysis of many of the background contributions has been carried out in the context of atomic gravitational wave detector configurations similar to MAGIS-100 and are directly applicable~\cite{dimopoulos2008atomic,hogan2011atomic,graham2013new,graham2017mid}.  

The MAGIS-100 experiment is designed to be sensitive to time-varying gravitational wave or dark matter signals in the frequency band $\sim$\;0.1 -- 10\;Hz. The relevant backgrounds are those that temporally vary in this frequency band.  This frequency selectivity eases a number of requirements.  For instance, any backgrounds that lead to constant phase offsets or long-term phase drifts would not affect the potential science signals.  MAGIS-100 initially aims for $100 \hbar k$ atom optics and a phase resolution of  $10^{-3}~\text{rad}/\sqrt{\text{Hz}}$ for a differential measurement between interferometers separated by baselines of up to 100\;m.  Research and development efforts aim to boost instrument sensitivity via increased momentum splitting and improved phase resolution (resulting from higher atom flux and/or squeezed atom sources) and to further reduce the influence of the corresponding technical noise sources discussed here.

\begin{table}[t]
\centering
\caption{Summary of key experimental parameters and target values of MAGIS-100 (initial) (see \Tref{table:futurevision}).  Spectral densities are taken to be in the $\sim 0.1-3$ Hz frequency band of interest. Note that the cloud kinematics can either be stabilized to below the target values or measured each shot at the target uncertainty.}\label{tab:parameters}
\small
\begin{tabular}{@{}llp{8.05cm}}  
 \toprule
 Parameter & Target Value & Primary Driving Factors \tabularnewline
 \midrule
 LMT atom optics & $n=100$ & Increase sensitivity to science signals \tabularnewline
 Phase resolution& $10^{-3}~\text{rad}/\sqrt{\text{Hz}}$ & Increase sensitivity to science signals \tabularnewline
 Frequency noise/drift & $<10$ Hz & Increase pulse transfer efficiency (\Sref{Sec:AILasers}) \tabularnewline
 Per shot position uncertainty &10\;$\mu$\text{m}$/\sqrt{\text{Hz}}$  & \multirow{2}{*}{Coupling to wavefront aberrations (\Sref{Sec:WavefrontAnalysis})}\tabularnewline
 Per shot velocity uncertainty & 10\;$\mu$\text{m/s}$/\sqrt{\text{Hz}}$  & \tabularnewline
 \multirow{2}{*}{Laser wavefront variation} & \multirow{2}{*}{5\;mrad\textsuperscript{*}} & Coupling to cloud kinematic and laser pointing jitter (\Sref{Sec:WavefrontAnalysis} and \Sref{Sec:PointingJitter})\tabularnewline
 Laser intensity stabilization &$0.1 \%/\sqrt{\text{Hz}}$ & AC Stark shifts (\Sref{Sec:ACStark})\tabularnewline
 Laser pointing stability &$30~\text{nrad}/\sqrt{\text{Hz}}$ & Coupling to wavefront aberrations (\Sref{Sec:PointingJitter})\tabularnewline
 Magnetic field uniformity & $1~\text{mG}$ (rms) & Clock frequency shifts \tabularnewline
 \bottomrule
 \addlinespace[2pt]
 \multicolumn{3}{l}{\textsuperscript{*}\footnotesize{at transverse length scales $\lesssim 3~\text{mm}$}}
 \end{tabular}
\end{table}

\subsection{Laser phase noise} 

\label{Sec:FreqNoise}

The atom-laser interactions are dependent on the phase of the laser at each of the interaction points, so laser technical noise can affect the atom interferometer signal.  The multiple atom ensembles in the gradiometer are subject to the same laser pulses, so this noise effect is expected to be common-mode suppressed to a significant degree.  Single-photon atom optics on the clock transition~\cite{hu2017atom} will be employed to realize the necessary level of laser noise rejection~\cite{graham2013new}.  Specifically, the residual noise $\delta \phi_{\text{freq}}$ in the interferometer phase arising from laser frequency noise has the leading contribution (term 3 from Table 1 in \cite{graham2013new}) 
\begin{equation}
    \delta \phi_{\text{freq}} \sim \left(10^{-13} ~ \text{rad}/\sqrt{\text{Hz}}\right) \left(\frac{n}{100} \right) \left(\frac{\Delta v}{100~\mu\text{m}/\text{s}}\right) \left(\frac{\delta f}{10~\text{Hz}/\sqrt{\text{Hz}}}\right) \left(\frac{\Delta \tau}{100~\mu\text{s}}\right) \nonumber
\end{equation}
related to the finite duration $\Delta \tau$ of each laser pulse, the velocity difference $\Delta v$ between the two atom clouds in the gradiometer, the beamsplitter momentum $n \hbar k$, and the amplitude spectral density $\delta f$ of laser frequency noise.

For all conceivable experimental parameters, this source of $\delta \phi_{\text{freq}}$ is negligibly small. However, the practical degree of common-mode cancellation of laser noise that can be expected in a clock gradiometer has not been fully established experimentally. MAGIS-100 will study this question in detail and will investigate the possibility of other mechanisms of differential laser noise imprinting.

\subsection{Laser wavefront aberrations}

\label{Sec:WavefrontAnalysis}

The influence of laser wavefront aberrations and associated mitigation strategies, including in situ wavefront measurements and spatial filtering of the atom optics laser beam via free propagation, are discussed in \Sref{Sec:WavefrontAberrationsMainText}.  The effect of wavefront aberrations on the phase of an atom interferometer has been modeled in detail~\cite{Wicht2005,Gibble:2006,hogan2011atomic,Schkolnik:2015,Zhou:2016,Karcher_2018,Bade:2018}. This interferometer phase response analysis can be combined with in situ wavefront measurements and measurements of atom cloud kinematics in MAGIS-100 to correct for backgrounds arising from wavefront aberrations.  \ref{appendix:wavefront} expands upon the treatment in~\cite{hogan2011atomic} with a full perturbative Fourier analysis.  This new analysis goes beyond previous work to also include effective momentum kicks that arise from the wavefront aberration phase varying spatially in the transverse plane. The calculation is taken to second order in the aberration amplitude $\delta$ (in radians), and gives the same result at first order as previous calculations, as expected. 

We consider here the phase shift size due to terms up to second order in $\delta$ at different scales of LMT. This example considers a laser with wavelength $\lambda = \SI{698}{nm}$ ($k = 2 \pi/\lambda$); vertical height $H$ spanned by the interferometer; duration between the interferometer beamsplitter and mirror $T$; and a transverse atom trajectory fluctuation of size $\Delta x$ during the interferometer, which could result from shot-to-shot variations in initial atom positions and/or velocities.  It is reiterated here that the only phase shifts that contribute noise to the detector are those that temporally vary within the target frequency band.  To first order in $\delta$, the noise in the interferometer phase resulting from the interplay of wavefront aberrations and initial kinematic fluctuations is
\begin{align}
\delta\phi_{wf,1} &\sim n \delta  k_{t} \Delta x \sin{\left[\frac{H k_t^2}{2 k} \right]} \nonumber \\
&\sim \qty(\SI{5e-4}{rad\per\sqrt{\hertz}})\qty(\frac{\delta}{0.005})\qty(\frac{n}{100})\qty(\frac{k_t}{(\SI{3}{mm})^{-1}})\qty(\frac{\Delta x}{10\,\mu\text{m}/\sqrt{\text{Hz}}})\qty(\frac{\sin{\left[H k_t^2/2k \right]}}{0.3}). \nonumber
\end{align}
For second order in $\delta$, the leading contribution is 
\begin{align}
\delta\phi_{wf,2} &\sim \delta^2\frac{k_t^2 n^2 T \hbar}{m} k_t \Delta x \sin{\left[\frac{H k_t^2}{2 k} \right]} \nonumber \\
&\sim \qty(\SI{3e-8}{ \tfrac{\text{rad}}{\sqrt{\text{Hz}}} })\qty(\frac{\delta}{0.005})^2\qty(\frac{n}{100})^2\qty(\frac{k_t}{(\SI{3}{mm})^{-1}})^3 \qty(\frac{\Delta x}{10~\tfrac{\mu\text{m}}{\sqrt{\text{Hz}}}}) \qty(\frac{\sin{\left[H k_t^2/2 k \right]}}{0.3}), \nonumber 
\end{align}

\noindent assuming $T=1\;\text{s}$. These phase shifts are suppressed to the extent that the local wavefront perturbation experienced by a given atom remains constant throughout the entire interferometer. For simplicity, here we assume that the dominant cause of imperfect suppression is due to the evolution of the wavefront perturbations between the various vertical positions of the atom-light interactions\footnote{We find in this case that this assumption only leads to a factor of $\sim 3$ suppression compared to the situation in which the wavefronts are completely uncorrelated between different pulses.}.

The $(nk_t)^2$ dependence at second order in $\delta$ comes from the kinetic energy of the momentum kicks from the transverse component of the local wave vector of the perturbed laser field.  The additional $k_t \Delta x$ factor in both terms reflects the rate at which the phase shift varies with $\Delta x$, which increases proportionally with $k_t$.  

As discussed in \Sref{Sec:FreePropModeCleaning}, wavefront perturbations with a spatial scale of $1/k_t\lesssim 3~\text{mm}$ (after magnification by the 1:30 telescope) will be minimized via spatial filtering and by the use of high-quality in-vacuum optics.  To reduce wavefront-induced phase noise, it will be important to either control or measure displacements arising from initial kinematic offsets at the level of $10~\mu\text{m}$ or better on each experimental shot.  For the anticipated 1\;mm cloud sizes and $10^6$ atoms in MAGIS-100, measurements at this level are compatible with the atom shot noise limit.

There are additional strategies to further mitigate aberration-induced phase noise in the interferometer.  For example, as the number of LMT enhancement pulses increases, the total laser interaction time can become a substantial fraction of a second.  Longitudinal averaging of the imprinted wavefront perturbation due to the motion of the atoms during this interaction time is expected to significantly reduce the size of the aberration-induced noise.  To be conservative, this averaging has not been included in the phase error estimates here.  In addition, spatially resolved detection including point source interferometry~\cite{Dickerson2013} and phase shear readout~\cite{hogan2011atomic} techniques allows the atoms to be used as an in-situ probe of the wavefront, providing an additional tool for mitigating aberration-induced phase noise~\cite{tino2013atom}.  For example, the beam profile can be controllably translated across the atom ensemble using the tip-tilt delivery mirrors, allowing the aberrated wavefront to be imprinted on the cloud with different transverse offsets.  Phase noise in the target frequency band can also be caused by time-varying wavefront aberrations~\cite{hogan2011atomic}.  In-situ wavefront characterization can be used to measure such temporal variations and account for their effects in post-processing~\cite{Dickerson2013,Sugarbaker2013,hogan2011atomic}.

\subsection{Seismic vibration}

Ground vibrations imprint phase noise on the interferometry laser pulses due to vibrations of the delivery optics.  This phase noise impacts the detector in the same way as intrinsic laser noise (see above) and also cancels to a high degree as a common mode~\cite{graham2013new,asenbaum2016phase}.  For a velocity mismatch $\Delta v$ between the two clouds in the gradiometer, there is a residual phase shift 
\begin{equation}
    \delta \phi_{\text{vibration}} \sim \left(10^{-8} ~ \text{rad}/\sqrt{\text{Hz}}\right) \left(\frac{n}{100} \right) \left(\frac{\Delta v}{100~\mu\text{m}/\text{s}}\right) \left(\frac{T}{1~\text{s}}\right) \left(\frac{\delta a}{10^{-4} \text{m}/\text{s}^2/\sqrt{\text{Hz}}}\right) \nonumber
\end{equation}
associated with vibration of the critical beam steering optics with amplitude spectral density $\delta a$.  Here $T$ is the duration between the interferometer beamsplitter and mirror interactions, and all other parameters are as defined above~\cite{graham2013new}.  We find that $\delta \phi_{\text{vibration}}$ is below the fundamental detection noise floor for typical experimental parameters, and $\delta a$ can be reduced as needed by adding modest vibration isolation to critical optical elements.  Seismic vibrations are also associated with moving mass in the vicinity of the detector.  This can lead to additional background noise due to the gravitational coupling of the moving objects to the atom ensembles (see the discussion of gravity gradient noise in \Sref{sec:seismic-measurements}).

\subsection{Laser pointing jitter}

\label{Sec:PointingJitter}

Uncontrolled pointing jitter of the laser causes phase shifts in the interferometer since it changes the position of the atom with respect to the laser wavefronts. The behavior of these phase shifts is well-understood~\cite{Dickerson2013,Sugarbaker2013,hogan2011atomic}.  Pointing jitter in the target frequency band for the science signal ($\sim$\;0.1 -- 10\;Hz) can act as a noise background.  Let $\delta \Phi$ denote the amplitude spectral density for pointing jitter in this band.  If the pointing jitter originates from optics near one of the atom ensembles, the laser wavefronts are tilted and the beam is displaced, leading to a transverse displacement by a distance $L \delta \Phi$ at the far interferometer (and much less for the near interferometer), where $L \approx 100~\text{m}$ is the baseline length.  For MAGIS-100, the dominant source of noise in this situation arises from the coupling of this displacement to laser wavefront aberrations.  With the wavefront aberrations parameterized as above, the level of background noise from laser pointing jitter is 
\begin{equation}
    \delta \phi_{\text{Pointing}} \sim \left(6 \times 10^{-4} ~ \text{rad}/\sqrt{\text{Hz}}\right) \left(\frac{n}{100} \right) \left(\frac{\delta}{0.005}\right) \left(\frac{\delta \Phi}{30~\text{nrad}/\sqrt{\text{Hz}}}\right) \left(\frac{k_t}{\left(3~\text{mm} \right)^{-1}}\right). \nonumber
\end{equation}
Laser pointing will be monitored using split photodetectors (see~\cite{graham2017mid} for a more detailed discussion), and feedback to control the laser direction can be used if needed.  Also, spatially resolved detection of the atomic interference pattern can be leveraged to provide measurements of the pointing jitter on each shot~\cite{Sugarbaker2013,tino2013atom}.

\subsection{AC Stark shifts}

\label{Sec:ACStark}

Off-resonant light causes an AC Stark shift of the atomic line.  For example, even for resonant excitation of the clock transition in Sr ($\lambda =$ 698~nm), there is an AC Stark shift due to off-resonant coupling to the other atomic levels. In particular, off-resonant coupling to the $^ 1 S_0\rightarrow \!\,^1P_1$ (461~nm) transition shifts the energy of the ground ($^ 1 S_0$) state, and off-resonant coupling to the $^ 3 P_0\rightarrow \!\,^3S_1$ (679~nm) transition shifts the energy of the excited clock ($^ 3 P_0$) state~\cite{boyd2007high}.  This energy shift can cause a phase shift in an interferometer that mimics the target signal to the extent that it varies spatially and fluctuates in time in the target frequency band.  Intensity fluctuations of the laser are a dominant source of this, so laser intensity control can reduce the effect.  

For interferometer sequences in MAGIS-100 with $n \hbar k$ beamsplitters, phase backgrounds from AC Stark shifts are at the level of 
\begin{equation}
    \delta\phi_{\text{AC}} \sim \left(6 \times 10^{-4} ~ \text{rad}/\sqrt{\text{Hz}}\right) \left(\frac{n}{100} \right) \left(\frac{\delta (\Delta I)/I}{10^{-5}/\sqrt{\text{Hz}}}\right) \nonumber
\end{equation}
where $\delta (\Delta I)/I$ is the amplitude spectral density in the target frequency band for the fractional fluctuation of the differential laser intensity (averaged over the atom ensembles) between the two interferometers. $\delta (\Delta I)/I \sim 10^{-5}/\sqrt{\text{Hz}}$ can be realized, for example, with a $1\%$ spatial intensity variation between the two interferometers and laser intensity stabilization at the level of $0.1\%/\sqrt{\text{Hz}}$, both of which are readily achievable~\cite{Takahashi2008}.  Transverse-position-dependent AC Stark shifts can also couple to initial atom kinematic jitter in a manner analogous to wavefront aberrations (discussed above).  For a given intensity/wavefront perturbation amplitude, phase errors from AC Stark couplings of this type will generally be smaller than the corresponding wavefront-induced phase errors.  For such AC Stark couplings, analogous mitigation strategies can be applied as in the wavefront case.  In situ measurements of intensity perturbations can be performed by implementing spatially resolved detection combined with short duration interferometers in a superposition of different internal states, while leaving on a long, Doppler-detuned laser pulse to imprint the AC Stark profile.

\subsection{Rotations and gravity gradients}

\label{Sec:CoriolisandGG}

Shot-to-shot fluctuations in the atom trajectory can couple to rotations and gravity gradients, leading to time-dependent phase errors~\cite{Hogan2009,Antoine2003,berman1997atom}.  Multi-loop interferometers provide a way to cancel phase shifts from the coupling of initial kinematics to gravity gradients or rotations while preserving the time-varying dark matter or gravitational wave signal~\cite{dubetsky2006atom, hogan2011atomic}.  As an illustrative example, a three-loop interferometer configured to cancel out leading order phase shift contributions from rotations and gravity gradients is considered~\cite{dubetsky2006atom}.  In such an interferometer, the coupling of higher-order rotation/gravity gradient phase shifts to initial kinematic jitter can still be a source of noise.  The dominant such noise term arises from a cross-coupling between rotations, gravity gradients, and initial atom velocity and has the form $\delta \phi_\text{RGGV} = \left( 17/3 +4\sqrt{2} \right) n k \Delta v_x \Omega_y T_{zz} T^4$.  Here, the $x$, $y$, and $z$ axes are defined so that $z$ is normal to Earth's surface and Earth's rotation vector lies in the $yz$ plane.  $\Delta v_x$ denotes the shot-to-shot jitter in the atom cloud velocity along the $x$ axis (or the accuracy to which this jitter can be measured on each experimental shot if post-processing corrections are implemented), $\Omega_y$ is the $y$ component of Earth's rotation vector, $T_{zz}$ is the vertical gravity gradient, and $T$ is the duration between the first beamsplitter and mirror interactions. The associated noise has the magnitude 
\begin{equation}
    \delta \phi_\text{RGGV} \sim \left(2 \times 10^{-5} ~ \text{rad}/\sqrt{\text{Hz}}\right) \left(\frac{n}{100}\right) \left(\frac{\Delta v_x}{10~\mu\text{m}/\text{s}/\sqrt{\text{Hz}}}\right) \left(\frac{T}{1\;\text{s}}\right).  \nonumber
\end{equation}
Adding additional loops to the interferometer would further suppress phase shifts from the cross-coupling of rotations, gravity gradients, and initial atom velocity.

Gravity gradients can also affect the pulse transfer efficiency. In configurations involving atom sources on each end of the long baseline, the Earth’s gravity gradient causes the velocities of the vertically separated atom clouds to evolve differently, resulting in different Doppler detunings.  For $T=1$\;s and a baseline length of $L = 100$\;m, the maximum magnitude of the Doppler detuning arising from this effect is approximately $2\pi \times 200$\;Hz.  For typical Rabi frequencies of $2 \pi \times 3$\;kHz, this corresponds to an inefficiency per pulse of $\sim 0.5 \%$, which is consistent with atom interferometers with $\sim\!\! 100 \hbar k$ atom optics.  MAGIS-100 can further mitigate this effect by employing composite pulses~\cite{Butts2013}, adiabatic rapid passage~\cite{Kotru2015}, or optimal quantum control~\cite{Saywell2018} to dramatically reduce the impact of detunings on pulse efficiency.  Alternatively, atom clouds at different heights can be purposefully offset in velocity by a controlled amount to allow them to be independently and resonantly addressed by different spectral components~\cite{dimopoulos2008atomic} of the clock laser beam, which can be generated by driving an AOM at multiple frequencies.

\subsection{Mean field shifts}

Atom-atom interactions cause an energy shift of the clock transition proportional to the atomic density.  This can cause a systematic error in an interferometer if the two arms occupy different atomic states (with different mean field shifts), or if the density is asymmetric.  The detector is sensitive to any time-varying mean field shifts, which may arise if the density fluctuates shot-to-shot.  This background can be suppressed by using ensembles with low density (after matter wave lensing), by employing sequences that use symmetric internal states for the two arms, and by using symmetric beam splitting sequences~\cite{Ahlers2016,overstreet2018effective}.

For the fermionic isotope $^{87}$Sr, the density-dependent mean field shift has been measured to be $\sim 10^{-11}~\frac{\text{Hz}}{\text{atoms}/\text{cm}^3}$~\cite{swallows2012operating, campbell2009probing}.  The mean field shift arises when atoms are inhomogeneously excited on the clock transition, as otherwise atom-atom interactions are suppressed by Pauli blocking~\cite{swallows2012operating}.  Therefore, the times at which mean field shifts will affect phase evolution in MAGIS-100 are during the atom optics pulses.

To estimate this effect in MAGIS-100, we assume there are $\sim 10^6$ atoms per shot in the interferometer, with an ensemble volume of $\sim 1~\text{mm}^3$.  Conservatively assuming the same mean field shift as measured in~\cite{swallows2012operating, campbell2009probing} (in actuality, MAGIS-100 will likely have a smaller excitation inhomogeneity),  and assuming a $\pi$-pulse duration of $\Delta \tau$, phase backgrounds from mean field shifts are 
\begin{equation}
    \delta\phi_{\text{MF}} \sim \left(5 \times 10^{-5} ~ \text{rad}/\sqrt{\text{Hz}}\right)\left(\frac{n}{100}\right) \left(\frac{\Delta \tau}{200~\mu s}\right)\left(\frac{\delta (\Delta N_A)/N_A}{0.01}\right), \nonumber
\end{equation}
where $\delta (\Delta N_A)/N_A$ is the shot-to-shot fluctuation of the fractional atom population asymmetry between the two interferometer arms (the relevant fluctuations are those in the target frequency band). Additionally, closely matching the densities of the two separated interferometers would allow the mean field phase shift to be further suppressed as a common mode, and highly robust beamsplitters and mirrors using composite pulses along with advanced quantum control~\cite{levitt2007composite,Butts2013,Dunning2014,Berg2015,Saywell2018,Saywell_2020biselective,Saywell_2020} can dramatically reduce fluctuations in the population asymmetry. 

\subsection{Magnetic fields}

The clock energy levels of Sr shift in response to magnetic fields.  Time-varying magnetic fields can cause systematic frequency shifts that behave like a gravitational wave or ultralight dark matter  signal.  As discussed in \Sref{sec:interferometer region}, magnetic shielding will be employed to reduce the influence of stray fields in the interferometer region, and the local field will be monitored with commercial magnetometers.  Using multiple sequential transitions, the atom interferometry sequence can be designed so that both arms of the interferometer spend most of the time in the ground state, reducing the differential magnetic phase shifts.  For $\,^{87}\text{Sr}$, a co-magnetometer can be realized by simultaneously operating two interferometers using states with opposite magnetic field response, suppressing the linear response to magnetic fields and allowing the magnetic field dependent phase shift to be measured and subtracted.  The residual quadratic Zeeman shift coefficient is $-0.23~\text{Hz}/\text{G}^2$~\cite{falke201187sr}, implying that for a bias field $B_0$ and a fluctuating component of the magnetic field $\delta B$, the associated phase background is 
\begin{equation}
    \delta\phi_{\text{mag}} \sim \left(1 \times 10^{-3} ~ \text{rad}/\sqrt{\text{Hz}}\right)\left(\frac{B_0}{1\;\text{G}}\right) \left(\frac{\delta B}{1 \;\text{mG}/\sqrt{\text{Hz}}}\right) \left(\frac{T}{1\;\text{s}}\right). \nonumber
\end{equation}
For a bias field of $B_0 = 1$\;G, maintaining fluctuations in the magnetic field at a level below $\delta B = 1 \;\text{mG}/\sqrt{\text{Hz}}$ requires a fractional stability in the bias field current of below $0.1\%/\sqrt{\text{Hz}}$ in the target frequency band, which is readily achievable.  Fluctuations in Earth's magnetic field in the target frequency band are expected to be at the level of $\sim 1\;\mu\text{G}/\sqrt{\text{Hz}}$~\cite{ringler2020magnetic}, and so can likely be neglected.  Other sources of stray time-dependent magnetic fields, such as from local current carrying wires, are still being investigated.

Spatial curvature of the magnetic field (i.e., a nonzero second spatial derivative) can produce force gradients on the atoms, which can lead to time-varying phase shifts via coupling to initial kinematic fluctuations in a manner analagous to the case of gravity gradients.  As discussed in \Sref{Sec:CoriolisandGG}, the use of multi-loop interferometers can substantially suppress such phase shifts.  For a bias field of $1~ \text{G}$ and a magnetic field curvature $\sim 1\;\text{mG}/\text{m}^2$, the associated force gradient from magnetic field variations is $\sim 6$ orders of magnitude smaller than the force gradient from Earth's gravitational field.

\subsection{Blackbody radiation shifts.} Blackbody  radiation causes  an  energy shift of the atomic energy levels.  This can result in phase noise  in the interferometer if the temperature of the vacuum tube varies in time in the target frequency band.   For  the  strontium  clock  transition, the  blackbody  shift  has  a  temperature  coefficient of $-2.3~\text{Hz}~\left(\frac{T_{\text{system}}}{300~\text{K}}\right)^4$, where $T_{\text{system}}$ is the temperature of the apparatus~\cite{falke201187sr}.  It is important to note that apparatus temperature drifts will naturally occur at frequencies much lower than the target frequency band.  For a three-loop interferometer with $T$ as defined above and a temperature oscillation at frequency $\omega_{\text{Temp}}$, the interferometer phase response to temperature variations at low frequency $\omega_{\text{Temp}}$ is suppressed by a factor of $\left(\omega_{\text{Temp}} T\right)^2$.  For example, for a temperature oscillation of amplitude $1\;$K and period 1 hour, the associated interferometer noise is at the level of $\sim 1 \times 10^{-6} ~ \text{rad}/\sqrt{\text{Hz}}$ for $T= 1$~s, which would be negligible.

\subsection{Timing jitter}

A timing-jitter induced asymmetry $\delta T$ in the duration of the different free propagation zones of the interferometer, in combination with any velocity mismatch $\Delta v$ between the two clouds in the gradiometer, leads to interferometer phase noise of magnitude~\cite{graham2013new}
$$\delta \phi_{\text{timing}} \sim \left(10^{-4} ~ \text{rad}/\sqrt{\text{Hz}}\right) \left(\frac{n}{100} \right) \left(\frac{\Delta v}{100~\mu\text{m}/\text{s}}\right)  \left(\frac{\delta T}{1~\text{ns}/\sqrt{\text{Hz}}}\right).$$ The necessary timing stability can be achieved with stable pulse generators.

\subsection{Background gas index of refraction}  

The index of refraction from background gas in the pipe modifies the optical path length associated with the baseline.  Noise $\delta \eta$ in the index of refraction $\eta$ therefore leads to a spurious strain signal $\delta h_{\text{index}} = \delta \eta$~\cite{dimopoulos2008atomic}.  The index of refraction of air is $\eta \sim 1 + 3 \times 10^{-4} \left(\frac{P}{760\;\text{Torr}} \right)\left(\frac{300\;\text{K}}{T_{\text{system}}} \right)$, where $P$ is the pressure and $T_{\text{system}}$ is the temperature of the system.  The spurious strain signal associated with index of refraction variation due to temperature fluctuation $\delta T_{\text{system}}$ with a period of 1 hour (for $T=1\;$s) is $$\delta h_{\text{index}} = \delta \eta \sim \left(4 \times 10^{-26}/\sqrt{\text{Hz}}\right) \left(\frac{P}{10^{-11}\;\text{Torr}} \right)\left(\frac{300\;\text{K}}{T_{\text{system}}} \right)\left(\frac{\delta T_{\text{system}}}{1~\text{K}} \right).$$  The spurious strain signal associated with index of refraction variation due to fractional pressure fluctuation $\delta P/P$ in the frequency band of interest is $$\delta h_{\text{index}} = \delta \eta \sim \left(4 \times 10^{-21}/\sqrt{\text{Hz}}\right) \left(\frac{P}{10^{-11}\;\text{Torr}} \right)\left(\frac{300\;\text{K}}{T_{\text{system}}} \right)\left(\frac{\delta P/P}{0.001/\sqrt{\text{Hz}}} \right).$$  For reference, it is noted that the same index of refraction effects apply to LIGO, which maintains ultra-high vacuum of $10^{-8}$ to $10^{-9}$ Torr~\cite{Abbott2009}.

Additionally, the index of refraction of the ultracold atom clouds themselves will induce a phase shift on the laser beam when it passes through a cloud.  For atom cloud density $\rho$, transition linewidth $\Gamma$, laser detuning $\Delta$ from the transition resonance, on-resonance saturation parameter $s_0$, and on-resonance scattering cross section $\sigma_0 = 3 \lambda^2/(2 \pi)$ the atom cloud modifies the real part of the index of refraction by an amount $\delta \eta \sim \rho \sigma_0 \lambda/(4 \pi) \left[(-2 \Delta/\Gamma)/(1 + (2 \Delta/\Gamma)^2+s_0)\right]$~\cite{metcalf2007laser, meppelink2010phase}.  Since the effect vanishes on resonance, we conservatively assume the light is detuned by approximately the Rabi frequency ($\Delta = 2 \pi \times 1$\;kHz), yielding interferometer phase noise from this effect of magnitude
$$\delta \phi_{\text{cloud}} \sim \left(10^{-7} \text{rad}/\sqrt{\text{Hz}}\right) \left(\frac{n}{100} \right)\left(\frac{N_{\text{a}}}{10^{6}} \right)\left(\frac{r_{\text{c}}}{1\;\text{mm}} \right)^{-2}\left[\left(\frac{\delta N_{\text{a}}/N_{\text{a}}}{0.01/\sqrt{\text{Hz}}} \right)^2 + 4 \left(\frac{\delta r_{\text{c}}/r_{\text{c}}}{0.01/\sqrt{\text{Hz}}} \right)^2 \right]^{1/2}$$
where $N_{\text{a}}$ is the atom number in a given experimental shot, $r_{\text{c}}$ is the atom cloud radius, $\delta N_{\text{a}}/N_{\text{a}}$ is the fractional shot-to-shot fluctuation in the atom number, and $\delta r_{\text{c}}/r_{\text{c}}$ is the fractional shot-to-shot fluctuation in the cloud radius.  If needed, $\delta N_{\text{a}}/N_{\text{a}}$ and $\delta r_{\text{c}}/r_{\text{c}}$ can be measured on each shot and associated phase shifts can be subtracted in post-processing.  Phase shifts from the effect of far-off-resonant transitions on the index of refraction are smaller than the value given here.

\subsection{Gravity gradient noise}
\label{sec:seismic-measurements}

\begin{figure}[t]
    \centering
    \includegraphics[width=\textwidth]{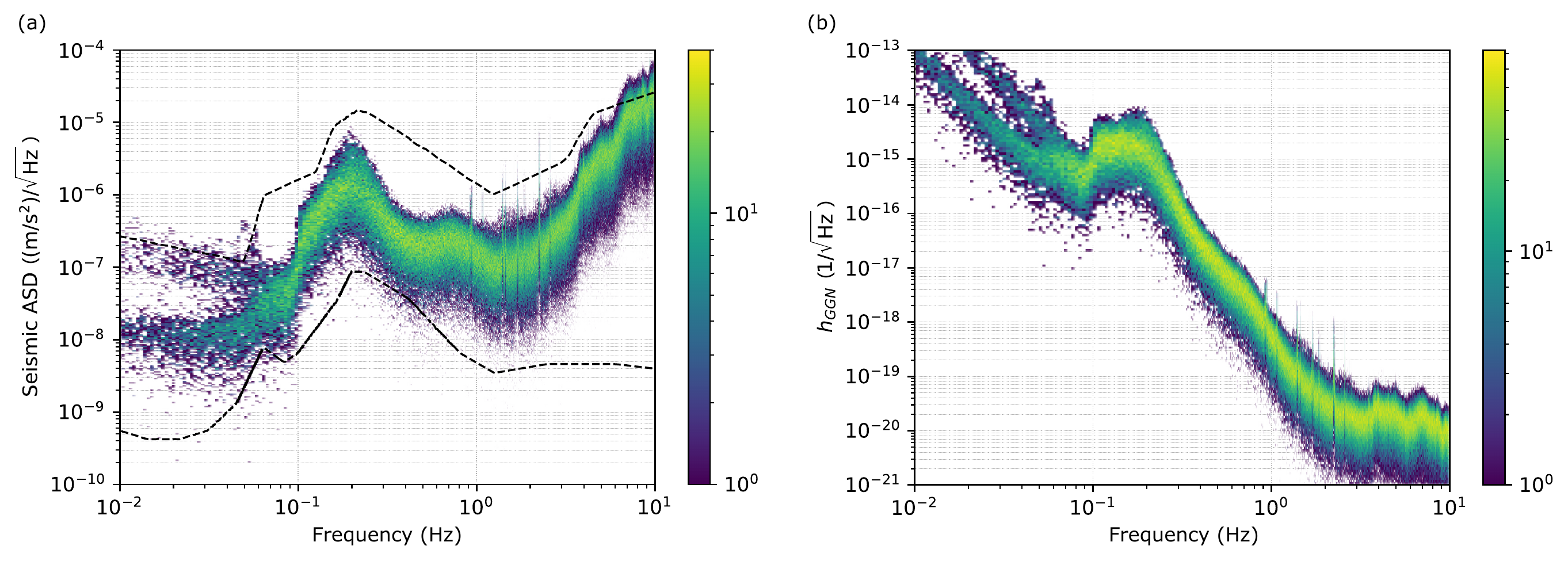}
    \caption{(a) Seismic acceleration amplitude spectral density for the surface of the MINOS shaft at Fermilab over a period of 30 days. Dashed black lines represent the NLNM and NHNM. (b) Inferred GGN strain amplitude spectral density modeled from surface displacement measurements. The color bar represents amplitude counts per frequency bin.  All data was acquired with a sampling rate of \SI{50}{Hz}.}
    \label{fig:seismic-spectra}
\end{figure}

Newtonian gravity gradient noise (GGN) is an important source of background noise for any terrestrial gravitational wave detector, including MAGIS~\cite{HarmsGGN}.  This noise source is caused by local seismic activity exerting gravitational forces on the detector proof masses. GGN likely imposes a practical limit on the sensitivity at low frequencies for Earth-based gravitational wave detectors, and is one of the primary motivations for space-based detectors.  One of the goals of MAGIS-100 is to study GGN and evaluate possible mitigation strategies to extend the detector sensitivity to lower frequencies.

Despite the fact that the freely-falling atom proof masses in MAGIS are decoupled from the direct effect of ground vibration, seismic noise can still couple to the atoms via gravity.  Local seismic activity is associated with density fluctuations in the earth surrounding the detector, and this moving mass produces a time-varying gravitational force on the atoms.  The resulting relative acceleration of the two proof masses on either side of the baseline is indistinguishable from the strain caused by a gravitational wave.

One possible mitigation is to use more than two test masses along the baseline to attempt to distinguish GGN from a gravitational wave~\cite{PhysRevD.93.021101}.  Since the dynamic gravitational potential from seismic waves is sourced locally, its magnitude can vary non-linearly with position across the baseline.  On the contrary, an incoming gravitational wave is well-described as a plane wave, leading to a highly linear response as a function of proof mass separation.  Therefore, using three or more atom proof masses spaced uniformly along the baseline, the detector may be able to differentiate GGN from the desired gravitational wave signal by measuring the curvature and other higher non-linear deviations.  If successful, this concept could allow the GGN to be measured and then subtracted from the signal channel.

To evaluate this strategy, MAGIS-100 can operate using three test masses (mode C in \Fref{Fig:magis-config}).  An important part of studying this detector mode will be to correlate the measurements made by the atom interferometers with independent local vibration measurements.  An initial survey was performed using a seismometer to measure the ground motion at several relevant locations around the detector site.  Data was collected near the surface of the MINOS shaft, in the location of the laser lab, and at the bottom of the 100~m shaft.  \Fref{fig:seismic-spectra}(a) shows the surface seismic acceleration amplitude spectral density for measurements made over a period of two months. For reference, the lower and upper dashed black lines are the New Low Noise Model (NLNM) and New High Noise Model (NHNM), respectively~\cite{Peterson:1993,mcnamara2006seismic}.  It is worth noting that seismic noise can vary widely by site, with some sites significantly quieter (such as MIGA~\cite{Junca2019}).  In considering a site for a full km-scale detector, it will be important to choose such a low noise site, such as SURF~\cite{Harms_2010}.  MAGIS-100 will also characterize the influence of time-varying gravity gradients that may arise from other sources, such as variations in the local water table or water flow.

Using this acceleration spectrum, we can estimate the amplitude of GGN at the MAGIS-100 site. Following~\cite{HarmsGGN}, in \Fref{fig:seismic-spectra}(b) we calculate the inferred strain amplitude spectral density for GGN for the MAGIS-100 detector geometry. GGN at this level would set the ultimate noise floor of MAGIS-100 below about 1~Hz (see \Fref{fig:gw-sensitivity-100}).  MAGIS-100 will serve as a testbed for developing strategies to manage this noise, with the goal of extending the detector range to lower frequencies.  Such strategies will be critical for maximizing the science reach of a future full-scale detector.

\section{Conclusions and Outlook}
\label{Sec:Conclusions}

MAGIS-100 aims to advance quantum science, perform sensitive searches for ultralight dark matter, and serve as a pathfinder experiment for mid-band gravitational wave detection.  It is anticipated that MAGIS-100 will pave the way for future atom interferometric sensors with even longer baselines and even greater sensitivities to science signals of interest. The pursuit of these goals is made possible by several key features.  Clock atom interferometry in principle supports a high degree of common-mode laser noise cancellation using only a single-baseline, and also offers the potential for dramatically enhanced LMT atom optics.  Another benefit of the MAGIS concept is the dual-purpose nature of science signal searches. With a single detection mode, searches can be carried out for ultralight dark matter as well as gravitational waves simultaneously. In addition, the frequency response of MAGIS-100 can be tuned by adjusting the pulse sequences, without the need for any hardware changes.  This allows the detector to rapidly switch between resonant and broadband detection modes.  MAGIS-100 also features multiple atom sources along its baseline to study and characterize Newtonian gravity gradient noise.

MAGIS-100 intends to make significant progress toward establishing the requirements and demonstrating the feasibility of a kilometer-scale detector. The technical design of MAGIS-100 benefits from a foundation laid by many years of research in the areas of atom interferometry and optical atomic clocks. Substantial research and development efforts are needed to advance the detector technology further, to a level that supports strain measurements in the scientifically interesting range below $10^{-20}/\sqrt{\text{Hz}}$. In order to reach this level of sensitivity, advances in LMT atom optics, as well as development of high-flux atom sources that incorporate spin squeezing for enhanced phase resolution, are expected to play a critical role. Another important part of this effort will be to continue to develop methods to reduce technical noise sources to a level consistent with the target sensitivity. A promising site for a future follow-on detector MAGIS-km is the SURF Laboratory in South Dakota, which already contains kilometer-scale vertical access shafts. Additional work is needed to better understand the challenges of further increasing the detector baseline and to assess the suitability of the SURF site itself.

As an increasing number of large-scale atom interferometers aimed at gravitational wave and dark matter detection~\cite{canuel2018MIGA,Canuel2019ELGAR,Badurina_2020,ZAIGA2020} are built around the world, a promising direction is to explore establishing a network of these detectors.  Studying how to optimally leverage a global network of such observatories has been a substantial focus of the AION research program~\cite{Badurina_2020}.  These terrestrial instruments may also ultimately pave the way for space-based detectors with even greater sensitivity.  In anticipation of this, initial studies of the requirements of and scientific motivation for space-based, long-baseline atom interferometers have already begun~\cite{graham2017mid,loriani2019atomic,abou2020aedge}.

\ack

We thank Philippe Bouyer, Ernst Rasel, Naceur Gaaloul, Dennis Schlippert, \'{E}tienne Wodey, Stefan Seckmeyer, Florian Fitzek, and Chris Overstreet for valuable discussions.  We also thank Menlo System GmbH, M Squared Lasers, and AOSense, Inc. for useful design inputs.  This project is funded in part by the Gordon and Betty Moore Foundation Grant GBMF7945.  This document was prepared by the MAGIS collaboration using the resources of the Fermi National Accelerator Laboratory (Fermilab), a U.S. Department of Energy, Office of Science, HEP User Facility. Fermilab is managed by Fermi Research Alliance, LLC (FRA), acting under Contract No. DE-AC02-07CH11359.  This work is supported in part by the U.S. Department of Energy, Office of Science, QuantiSED Intitiative.  We acknowledge support from Quantum Technologies for Fundamental Physics (QTFP), a joint program from UKRI’s Science and Technologies Facilities Council (STFC) and the Engineering and Physical Sciences Research Council (EPSRC).  We acknowledge additional support from the Laboratory Directed Research and Development program at SLAC National Accelerator Laboratory under contract DE-AC02-76SF00515.  Jeremiah Mitchell acknowledges support from the Kavli Foundation.  Benjamin Garber is supported by the Department of Defense (DoD) through the National
Defense Science \& Engineering Graduate (NDSEG) Fellowship Program.  Jonah Glick acknowledges support from an NSF Quantum Information Science and Engineering Network (QISE-NET) Graduate Fellowship. Peter Graham and Savas Dimopoulos acknowledge support from the Gordon and Betty Moore Foundation Grant GBMF7946.  Jonathon Coleman acknowledges additional support from the Universities Research Alliance (URA) and the Royal Society, UK.  Northwestern acknowledges additional support from Office of Naval Research grant number N00014-19-1-2181 and National Institute of Standards and Technology grant number 60NANB19D168.  Northern Illinois acknowledges support from the University President's Office and the Office for Research and Innovation.

\appendix
\addtocontents{toc}{\fixappendix}

\newpage

\section{Laser Wavefront Aberration Phase Shifts}
\label{appendix:wavefront}

We consider here a treatment of higher order effects of wavefront perturbations in atom interferometry that will become increasingly important as LMT atom optics continue to improve. The influence of wavefront perturbations in atom interferometry has been explored in, for example~\cite{Wicht2005,Gibble:2006,hogan2011atomic,Schkolnik:2015,Zhou:2016,Karcher_2018}.  The perturbed laser wavefront is imprinted on the atom's wavefunction whenever momentum is transferred to the atom, leading to errors in the atom interferometer phase shift proportional to the amplitude of the wavefront perturbation.  Laser wavefront perturbations also modify the momentum transferred to the atom by each laser pulse~\cite{Wicht2005,Gibble:2006,Bade:2018}.  The influence of modifications to the longitudinal momentum transfer have recently been studied and measured~\cite{Bade:2018}. Here, we extend this work to perform a full atom inteferometer phase shift calculation including the influence of wavefront-perturbation-induced transverse momentum  kicks, and also including wavefront-perturbation-induced longitudinal momentum  kicks in a more general context in which the longitudinal laser phase gradient experienced by the atoms varies from pulse-to-pulse.  Including these effects leads to contributions to the atom interferometer that scale quadratically with the wavefront perturbation amplitude and with the number $n$ of beamsplitter momentum kicks.  These higher order effects become important for the range of LMT values MAGIS-100 aims to explore.

Here we discuss the interferometer phase response to a laser beam perturbation. Consider a Fourier component of the laser beam profile with amplitude $\delta$ and transverse spatial frequency $k_x$.  The beam is treated paraxially, which means that the wavevector, $\vec{k}$, is nearly parallel to the longitudinal axis down the interferometer region. The coordinate system is defined so that the $xy$-plane is transverse to the laser propagation and the $z$-axis is in the direction of propagation, which is vertical.   As an illustrative example, a perturbation Fourier component $k_x$ along the $x$-axis is considered.  An analogous treatment applies for a Fourier component along a general axis in the $xy$-plane.  For the phase calculation, the equations of motion of the atoms are solved by expanding in a power series and keeping terms up to the third order in time.  The calculations presented here are further simplified by assuming the contributions from Earth's rotation will be suppressed and mitigated through the implementation of rotation compensation techniques. 

The laser beam field takes the general form
\begin{equation}
  \label{eq:1}
  E(x,z) = u(x,z)e^{ikz}.
\end{equation}
For an initial perturbation Fourier component $\delta\cos(k_{x}x)$ in the laser field, the initial field and the field at some farther distance are given by multiplying the perturbation by the corresponding paraxial propagator~\cite{Siegman1986}: 
\begin{subequations}
  \begin{align}
    u(x,0) &= 1 + \delta\cos(k_{x}x), \label{eq:2a}\\
    u(x,z) &= 1 + \delta\cos(k_{x}x)e^{-i \frac{k_{x}^{2}}{2k}z}.
    \label{eq:2b}
  \end{align}
\end{subequations}
Equation (\ref{eq:2b}) corresponds to phase perturbation
\begin{align}
    \phi_{w} &= -\delta\cos(k_{x}x)\sin(\frac{k_{x}^{2}}{2k}z). \label{eq:4b}
\end{align}
Local spatial gradients of $\phi_{w}$ result in additional contributions to the local wavevector $\vec{k}$
\begin{subequations}
  \begin{align}
    \Delta k_{x} &= \pdv{\phi_{w}}{x} = k_{x}\delta\sin(k_{x}x)\sin(\frac{k_{x}^{2}}{2k}z), \label{eq:Deltakx}\\
    \Delta k_{z} &= \pdv{\phi_{w}}{z} = -\frac{k_{x}^{2}}{2k}\delta\cos(k_{x}x)\cos(\frac{k_{x}^{2}}{2k}z). \label{eq:Deltakz}
  \end{align}
\end{subequations}
A laser pulse delivers a momentum kick to an atom equal to $\hbar \vec{k}$, where $\vec{k}$ is the local wave vector at the atom's location~\cite{Wicht2005}.  The perturbations $\Delta k_{x}$ and $\Delta k_{z}$ therefore cause the atoms to receive additional momentum kicks proportional to the perturbation size $\delta$.

We now calculate the effect of this beam perturbation on the phase of a Mach-Zehnder interferometer with $n \hbar k$ beamsplitters.  A full phase shift calculation~\cite{Hogan2009} including the additional momentum kicks arising from the wavefront perturbation (the $\Delta k$ kicks)
yields the results shown in Tables \ref{tab:delta-firstorder} and \ref{tab:delta-secondorder}. This calculation differs from previous results~\cite{hogan2011atomic}, because it includes transverse and longitudinal kicks caused by each Fourier component of the wavefront aberration. The first order terms in $\delta$ (shown in \Tref{tab:delta-firstorder}) arise from the aberrated phase of the laser at each of the atom-laser interaction points, while the second order terms (shown in \Tref{tab:delta-secondorder}) can be thought of as arising from the recoil kinetic energy induced by the momentum kicks $\Delta k_x$ and $\Delta k_z$.  These recoil shifts scale as $\hbar^2 \Delta k_i^2/2m$, where $i=x,z$ for the $x$ and $z$ momentum kicks respectively. Terms 1 and 2 in \Tref{tab:delta-secondorder} are associated with the $\Delta k_x$ kick as they scale with $k_x^2$, and terms 3 and 4 are associated with the $\Delta k_z$ kick as they scale with $(k_x^2 /2k)^2$, consistent with Equations~\ref{eq:Deltakx} and \ref{eq:Deltakz}. The second order terms from the longitudinal ($z$)  kicks are therefore generally smaller than those from the transverse ($x$) kicks by a factor of $\sim (k_x/k)^2$.

\begin{table}
  \caption{First order in $\delta$ phase shift terms.  Here $x_i$ and $z_i$ are the atom's initial transverse and longitudinal position, respectively. Similarly, $v_x$ and $v_z$ are the initial transverse and longitudinal velocities.}
  \label{tab:delta-firstorder}
  \centering
  \begin{tabular}{cc}
    \toprule
    Term & Phase shift \tabularnewline
    \midrule
    1 & $-n \delta \cos (k_x x_i ) \sin \left(\frac{k_x^2 z_i}{2 k}\right)$ \tabularnewline
    2 & $-n \delta \cos (k_x x_i + k_x v_x T ) \sin \left(\frac{g k_x^2 T^2}{4 k}-\frac{k_x^2 T v_z}{2 k}-\frac{k_x^2 z_i}{2 k}\right)$ \tabularnewline
    3 & $n \delta \cos (k_x x_i + 2 k_x v_x T ) \sin \left(\frac{g k_x^2 T^2}{k}-\frac{k_x^2 T v_z}{k}-\frac{k_x^2 z_i}{2 k}-\frac{k_x^2 n T \hbar }{2 m}\right)$ \tabularnewline
    4 & $-n \delta \cos (k_x x_i + k_x v_x T ) \sin \left(\frac{g k_x^2 T^2}{4 k}-\frac{k_x^2 T v_z}{2 k}-\frac{k_x^2 z_i}{2 k}-\frac{k_x^2 n T \hbar }{2 m}\right)$ \tabularnewline
    \bottomrule
  \end{tabular}
\end{table}

\begin{table}
\centering
  \caption{Representative phase shift terms at second order in $\delta$.}
  \label{tab:delta-secondorder}
  \begin{tabular}{cc@{}}
    \toprule
    Term & Phase shift \\
    \midrule
    &  \\
    1 & $\frac{\hbar k_{x}^2 n^2 T}{2 m} \delta^2  \sin (k_{x} x_{i} ) \sin \left(\frac{k_{x}^2 z_{i}}{2 k}\right) \sin (k_{x} x_{i} + k_{x} v_{x} T ) \sin \left(\frac{g k_{x}^2 T^2}{4 k}-\frac{k_{x}^2 T v_z}{2 k}-\frac{k_{x}^2 z_{i}}{2 k}\right)$ \\ 
    & \\
    2 & $-\frac{\hbar k_{x}^2 n^2 T}{m} \delta^2 \sin (k_{x} x_{i} ) \sin \left(\frac{k_{x}^2 z_{i}}{2 k}\right) \sin (k_{x} x_{i} + 2 k_{x} v_{x} T ) \sin \left(\frac{g k_{x}^2 T^2}{k}-\frac{k_{x}^2 T v_z}{k}-\frac{k_{x}^2 z_{i}}{2 k}-\frac{k_{x}^2 n T \hbar }{2 m}\right)$ \\ 
    & \\
    3 & $-\frac{\hbar k_{x}^4 n^2 T }{8 k^2 m} \delta^2 \cos (k_{x} x_{i} ) \cos \left(\frac{k_{x}^2 z_{i}}{2 k}\right) \cos (k_{x} x_{i} + k_{x} v_{x} T) \cos \left(\frac{g k_{x}^2 T^2}{4 k}-\frac{k_{x}^2 T v_z}{2 k}-\frac{k_{x}^2 z_{i}}{2 k}\right)$ \\ 
    & \\
    4 & $\frac{\hbar k_{x}^4 n^2 T }{4 k^2 m} \delta^2 \cos (k_{x} x_{i} ) \cos \left(\frac{k_{x}^2 z_{i}}{2 k}\right) \cos (k_{x} x_{i} + 2 k_{x} v_{x} T ) \cos \left(\frac{g k_{x}^2 T^2}{k}-\frac{k_{x}^2 T v_z}{k}-\frac{k_{x}^2 z_{i}}{2 k}-\frac{k_{x}^2 n T \hbar }{2 m}\right)$ \\ 
    & \\
    \bottomrule
  \end{tabular}
\end{table}

\clearpage

\printbibliography

\end{document}